\documentclass{cernrep}
\usepackage{texnames}
\usepackage[T1]{fontenc}

\usepackage{wrapfig}
\usepackage{bbm}
\usepackage{color}
\newcommand\eqn[1]     {Eq.~(\ref{#1})}
\newcommand\eqns[2]    {Eqs.~(\ref{#1}) and~(\ref{#2})}

\newcommand\fig[1]     {Fig.~{\ref{#1}}}
\newcommand\figs[2]    {Figs.~{\ref{#1}} and ~\ref{#2}}

\newcommand\sect[1]    {Sect.~{\ref{#1}}}

\newcommand\tab[1]     {Table~\ref{#1}}

\newcommand\gs{g_{\rm s}}
\newcommand\as{\alpha_{\rm s}}

\newcommand{\im}{{\rm Im}}
\newcommand{\ri}{{\rm i}}
\newcommand{\rc}{{\rm c}}
\newcommand{\rd}{{\rm d}}
\newcommand{\re}{{\rm e}}
\newcommand{\rA}{A}
\newcommand{\rF}{F}
\newcommand{\rL}{{\rm L}}
\newcommand{\rR}{{\rm R}}

\newcommand\bom[1]     {{\mbox{\boldmath $#1$}}}
\newcommand{\bt}{{\bom t}}
\newcommand{\bA}{{\bom A}}

\newcommand{\bJ}{{\bom J}}
\newcommand{\bT}{{\bom T}}

\newcommand{\LO}{{\scriptscriptstyle \rm LO}}
\newcommand{\NLO}{{\scriptscriptstyle \rm NLO}}
\newcommand{\msbar}{$\overline{\mathrm{MS}}$}
\newcommand{\aand}{\!\!\!\!\!\!\!\!&&}
\newcommand{\ldot}{\!\cdot\!}
\newcommand{\lp}{\left}
\newcommand{\rp}{\right}
\newcommand{\la}{\left\langle}
\newcommand{\ra}{\right\rangle}

\newcounter{exercise}
\newcommand{\del}{\partial}
\newcommand{\tr}{\mathrm{Tr}}

\renewcommand{\d}{\mathrm{d}}

\newtheorem{exe}{Exercise}[section]

\begin{document}
\title{QCD for collider experiments}
\author{Z. Tr\'ocs\'anyi\thanks{Z.Trocsanyi@atomki.hu.}}
\institute{Department of Physics and MTA-DE Particle Physics Research
Group, University of Debrecen, Debrecen, Hungary}
\maketitle

\begin{abstract}
These lectures are intended to provide the theoretical basis of
describing high-energy particle collisions at a level appropriate to
graduate students in experimental high energy physics. They are supposed
to be familiar with quantum electrodynamics, the concept of Feynman
rules, Feynman graphs and computation of the cross section in quantum
field theory.
\end{abstract}

\section*{}

\emph{ When you measure what you are speaking about and express it in numbers,
you know something about it, but when you cannot express it in numbers,
your knowledge is of a meagre and unsatisfactory kind. }

\hfill Lord Kelvin

\section{QCD as quantum field theory of the strong interaction}

The Lagrangian of the quark-gluon field based on a non-abelian gauge
symmetry was first proposed in Ref.~\cite{Fritzsch:1973pi} forty years
ago. The paper discussed the advantages of the colour-octet picture.
Since then an immense amount of research lead to a lot of interesting
results and a deep understanding of the strong interaction based on
this quantum field theoretical description of chromo-dynamics, QCD.
Today we are convinced that QCD is the correct description of the
strong interaction, yet we still lack a complete and satisfactory
solution. In such a situation one may set two goals: (i) either an
ambitious one: solve QCD, or (ii) a more pragmatic one: develop tools
for modeling particle interactions in high energy collider experiments.
In these lectures we go for the second one.

\begin{figure}[ht]
\begin{center}
\includegraphics[width=0.49\textwidth]{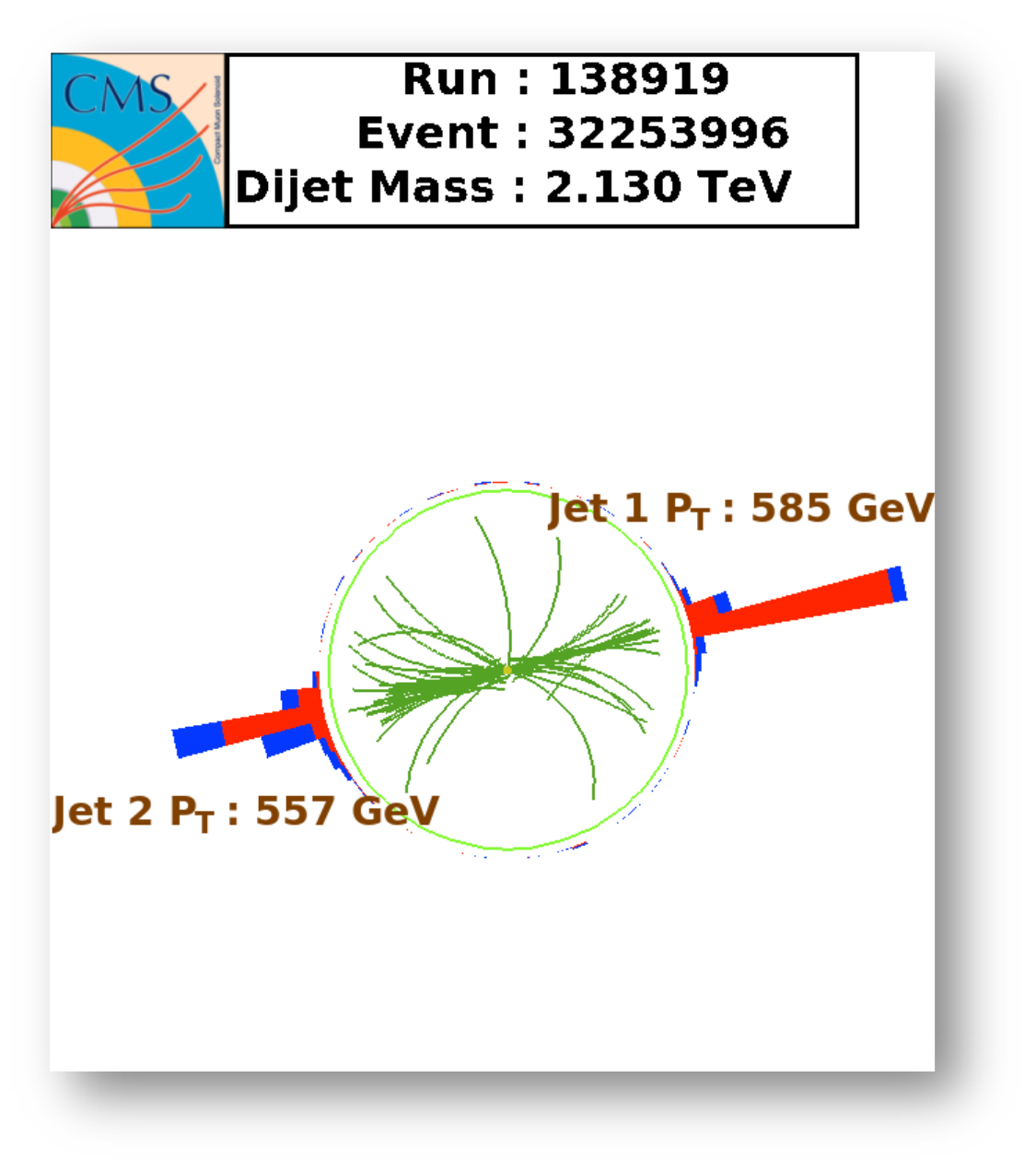}
\includegraphics[width=0.49\textwidth]{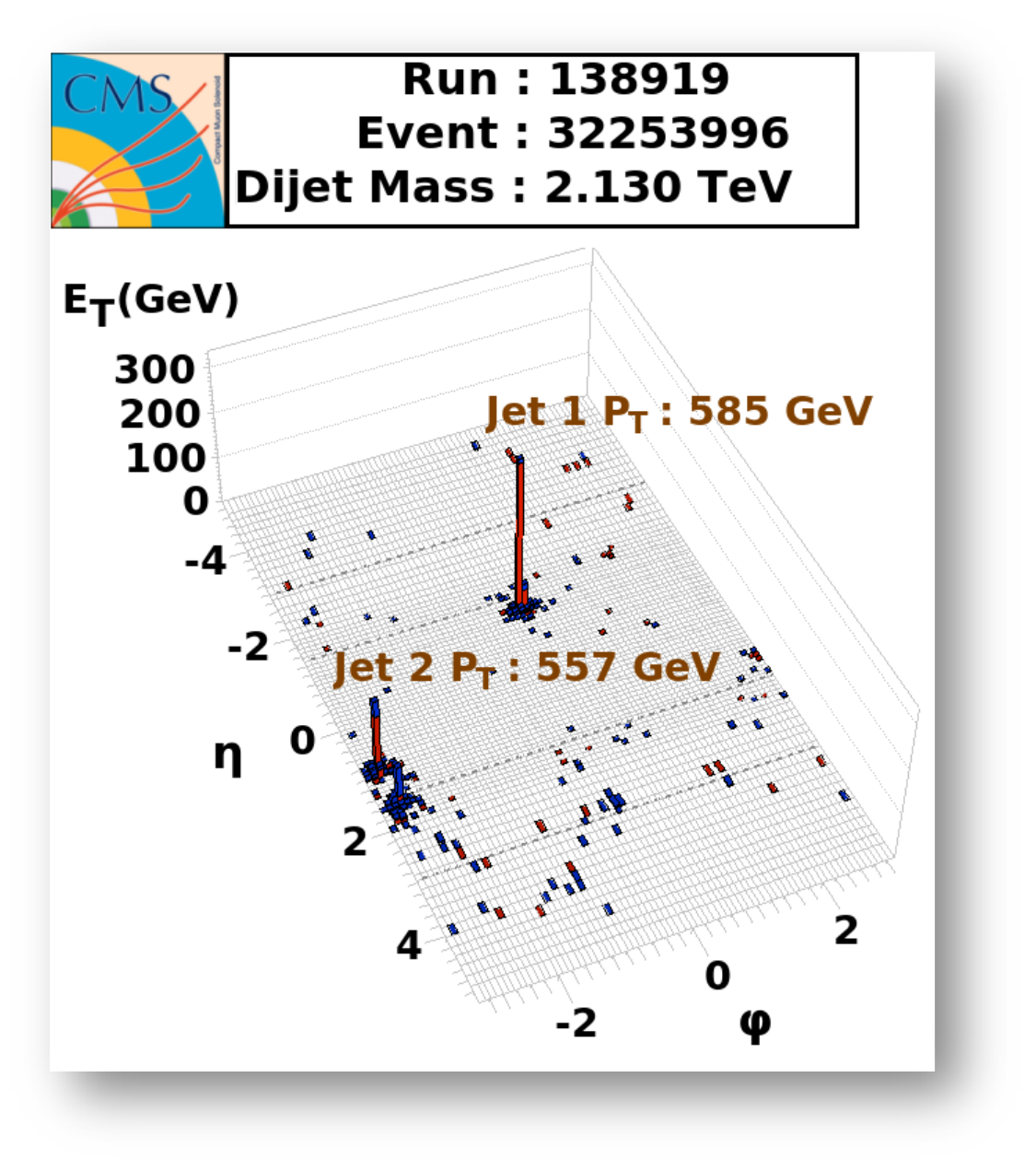}
\caption{A violent proton-proton collision resulting in two hard jets.
Left: tracks and energy deposits in the calorimeters,
right: energy deposits represented by towers in the
pseudorapidity--azimuthal angle plane.}
\label{fig:CMSevents}
\end{center}
\end{figure}

Our aim is to understand high-energy particle collisions quantitatively
from first principles. Examples of such events recorded by the CMS
experiment at the LHC are shown in Fig.~\ref{fig:CMSevents}. In these
events kinematic characteristics of particles, such as energy and
momentum, are collected. Analyzing many such events, we can produce
distributions of kinematic variables, for instance, differential
distribution of the inclusive jet cross section with respect to
pseudorapidity, ${\rm d} \sigma/{\rm d} \eta$. There is a long way from
the QCD Lagrangian to making predictions for such distributions, full
of difficulties. I clearly do not expect students in high energy
experimental physics to be able to solve those difficulties. Instead I
would like to explain the consequences of solving the difficulties,
because an incomplete understanding of these consequences can easily
lead to false interpretation of correct measurements.

\subsection{The QCD Lagrangian}
 
The quantum field theory (QFT) of the strong interactions is a part of
the Standard Model (SM) of elementary particle interactions. The SM is 
based upon the principle of local gauge invariance. The underlying 
gauge group is 
\[SU(3)_{\rc}\,\times\,SU(2)_{\rL}\,\times\,U(1)_{Y}\,,\] 
where c stands for ``colour'', L for ``left'' (or ``weak isospin'') 
and \emph{Y} for ``hypercharge''. As we concentrate on QCD, which is
based on $SU(3)_{\rc}$ gauge symmetry, we can write the Lagrangian as 
\begin{equation}
{\cal L}_{\rm QCD} = {\cal L}_{\rm QCD}^{(0)} + {\cal L}_{\rm sources}\,,
\label{eq:LQCD}
\end{equation}
where
\begin{equation}
{\cal L}_{\rm QCD}^{(0)} =
 {\cal L}_{\rm C} + {\cal L}_{\rm GF} + {\cal L}_{\rm G}\,,
\label{eq:LQCD0}
\end{equation}
while the electroweak sector (based on the
$SU(2)_{\rL}\,\times\,U(1)_{Y}$ symmetry) act as sources.
In \eqn{eq:LQCD0} ${\cal L}_{\rm C}$ is the classical Lagrangian, while
${\cal L}_{\rm GF}$ is the gauge-fixing term. The last piece is the ghost
Lagrangian, absent if we use physical gauges, which will be our choice.

To find the classical Lagrangian, one starts with the Lagrangian of free
Dirac fields,
\begin{equation}
{\cal L}_q^{(0)}(q_f,\,m_f) =
\sum_{k,l=1}^{N_c} \bar{q}_f^k\: ({\rm i} \gamma_\mu \partial^\mu -m_f)_{kl}\: q_f^l
\,,
\end{equation}
where the $\gamma_\mu$ matrices satisfy the Clifford algebra,
\begin{equation}
\lp\{\gamma^{\mu},\gamma^{\nu}\rp\}\,=\,2g^{\mu\nu}\,,\qquad
\lp\{\gamma^{\mu},\gamma^5\rp\}\,=\,0\,.
\label{eq:Clifford}
\end{equation}

The matter field content is dictated by the electroweak sector.
The fermion fields are called {\em quark fields}: $q_{f}^{k}$ with
masses $m_f$ and $f\,=\,1,\dots,n_{f}$, where $n_{f}$ is the number of
different flavours. The quark fields also have an additional degree of
freedom: colour, labelled by $k$, that can take $N_\rc$ values,
$k\,=\,1,\dots,N_{\rc}$. The precise matter content is shown in
\tab{tab:quarks}.
\begin{table}
\begin{center}
\caption{\label{tab:quarks}
The six quark flavours as dictated by the electroweak sector. Their baryon
number is $B=1/3$. Each quark flavour comes in three colours, not shown.
For light flavours (u, d and s) the mass values are not without
controversy and still under investigation. For flavours c and b the
mass values are running \msbar\ quark masses at 2\,MeV (see definition
below), while for t it is the pole mass.}
\begin{tabular}{|c|c|c|c|c|c|c|}
\hline
\hline
 $f$ & 1 & 2 & 3 & 4 & 5 & 6 
\\
\hline
\hline
 $q_f$ & u & d & s & c & b & t \\
\hline
$m_f$ & $\approx 3$\,MeV & $\approx 6$\,MeV &  $\approx 100$\,MeV &
1.2\,GeV & 4.2\,GeV & 172.6\,GeV \\ 
\hline
\hline
\end{tabular}
\end{center}
\end{table}

If we apply a transformation $q^{k}\,\to\,q'^{k}\,=\,U_{kl}q^{l}$, with 
\[
U_{kl}=\exp\lp\{\ri\sum_{a=1}^{N_{\rc}^{2}-1}\,t^{a}\,\theta^{a}\rp\}_{kl} 
\equiv\exp\lp\{\ri\bt\cdot\bom{\theta}\rp\}_{kl}
\,,
\] 
where $\theta^{a}\,\in\,\mathbb{R}$, then the Lagrangian of free Dirac
fields remains invariant,
$\mathcal{L}_{q}^{(0)}\lp(q\rp)=\mathcal{L}_{q}^{(0)}\lp(q'\rp)$. 
The $\lp(t^{a}\rp)_{kl}$ are $N_{\rc}\times N_{\rc}$ matrices
that constitute the fundamental representation of the generators 
$T^{a}$ (called colour-charge operators), which satisfy the Lie algebra:
\begin{equation}
\lp[T^{a},\,T^{b}\rp]=\ri\,f^{abc}\,T^{c} \,,\quad 
\textrm{with normalizaiton} \quad
\mathrm{T_R}(T^a T^b) = T_{\rm R} \delta^{ab}\,.
\label{liealg} 
\end{equation}
For $SU(3)$ the matrices $t^{a}$ are the Gell-Mann matrices (see \eg
\cite{Salam:2011bj}). 

Next we ask the question if we can make $\mathcal{L}_{q}^{(0)}\lp(q\rp)$
invariant under local $SU(N_{\rc})$ transformations. The answer is yes,
we can through the following steps:
\begin{enumerate} 
\item 
Introduce $A_{\mu}^{a}$ coloured vector field with the following
transformation property under $SU(N_{\rc})$ transformations:
\begin{equation*} 
\bt\cdot\bA_{\mu}\longrightarrow\,\bt\cdot\bA'_{\mu}= 
U\lp(x\rp)\bt\cdot\bA_{\mu}U^{-1}\lp(x\rp)+ 
\frac{\ri}{g_{s}}(\partial_{\mu}U\lp(x\rp))U^{-1}\lp(x\rp), 
\end{equation*} 
where 
$U\lp(x\rp)= \exp\lp\{\ri\bt\cdot\bom{\theta}(x)\rp\}$. 
\item 
{Replace $\partial_{\mu}\delta_{kl}$ with 
$D_{\mu}\lp[A\rp]_{kl}= \partial_{\mu}\delta_{kl}+ 
\ri g_{s}\lp(\bt\cdot\bA_{\mu}\rp)_{kl}$. This covariant derivative
$D_{\mu}\lp[A\rp]_{kl}q_l\lp(x\rp)$ transforms the same way as the
quark field $q_k\lp(x\rp)$.} 
\item 
{Introduce a kinetic term 
\begin{equation*} 
\mathcal{L}_g(A)\,=\,-\frac{1}{4}F_{\mu\nu}^{a}\lp[A\rp]F^{a\mu\nu}\lp[A\rp], 
\end{equation*} with the non-abelian field strength $F_{\mu\nu}^{a}$ given by 
\begin{equation*} 
F_{\mu\nu}^{a}\lp[A\rp]= 
\partial_{[\mu}A^{a}_{\nu]} 
-g_{s}\underbrace{f^{abc}A_{\mu}^{b}A_{\nu}^{c}}_{ 
``\bA_{\mu}\times\bA_{\nu}"} 
\,, 
\end{equation*} 
so the Lagrangian contains cubic and quartic terms of the gauge field.}
\end{enumerate} 
The constants $f^{abc}$ are the structure constants of the Lie algebra. 
The structure constants are completely antisymmetric and are related to
the adjoint representation of the generators $F^a_{bc}$ by
$F^a_{bc} = -\ri f^{abc}$.

Thus we find that the gauge boson field, called {\em gluon field}, is a
consequence of the local $SU(N_\rc)$ (gauge) invariance. The classical
Lagrangian of QCD is a sum of interacting Dirac Lagrangians for spin
1/2 fermion fields and a Lagrangian of a gauge field,
\begin{equation}
{\cal L}_{\rm C} = {\cal L}_f + {\cal L}_g =
\sum_{f=1}^{n_f} {\cal L}_q(q_f,\,m_f) + {\cal L}_g(A)
\label{eq:Lcl}
\,,
\end{equation}
where
\begin{equation}
{\cal L}_q(q_f,\,m_f) =
\sum_{k,l=1}^{N_c} \bar{q}_f^k\: ({\rm i} \gamma_\mu D^\mu[A] -m_f)_{kl}\: q_f^l
\,.
\end{equation}
The gluon field is also coloured and self-interacting. In fact, these
self-interactions are the sources of the main difference between QED
and QCD. We shall see that as a result, QCD is a `perfect theory' in
the sense that it is asymptotically free. Furthermore, {\em among
quantum field theories in $d=4$ dimensions only non-Abelian gauge
theories are asymptotically free} (see discussion after
\eqn{eq:beta0graphs}).  It is also plausible that the self-interactions
are the sources of {\em colour confinement}, \ie the colour neutrality
of hadrons, but we do not have a proof based on first principles.  

It is clear that there is an unprecedented large number of degrees of
freedom we have to sum over when computing a cross section:
\begin{enumerate}
\item spin and space-time as in any field theory, not exhibited above,
\item flavour, which also appears in electroweak theory, and colour,
which is specific to QCD only.
\end{enumerate}
As a result computations in QCD are rather cumbersome. During the last
two decades a lot of effort was invested and great progress was made
to find ``simple'' ways of computing QCD cross sections and to automate
the computations. 

\vspace{-4pt}
\noindent\rule{\textwidth}{1pt}

\begin{exe} 
Show that in QED the covariant derivative transforms the same way as the 
field itself, i.e., if $f(x) \to U(x) f(x)$ then 
$D_\mu f(x) \to U(x) D_\mu f(x)$, where $D_\mu=\del_\mu+\ri\,e\,A_\mu$. 
\end{exe} 
 
\begin{exe} 
Show that in QED 
\begin{equation*} 
\left[D_\mu,D_\nu\right]=\ri\,e\,F_{\mu\nu}, 
\end{equation*} 
 where $D_\mu=\del_\mu+\ri\,e\,A_\mu$. 
\end{exe} 
 
 
\begin{exe} 
Show that the generators of a special unitary group are traceless 
and hermitian. 
\end{exe} 
 
 
\begin{exe} 
The generators in the fundamental representation of SU(2) 
are the Pauli matrices divided by two: 
\begin{equation*} 
t^a_f=\frac{\sigma^a}{2},\qquad\sigma_1=\left(\begin{array}{cc}0&1\\1&0\end{array}\right),\quad 
\sigma_2=\left(\begin{array}{cc}0&-i\\i&0\end{array}\right),\quad 
\sigma_3=\left(\begin{array}{cc}1&0\\0&-1\end{array}\right) 
\end{equation*} 
The adjoint representation of a group is defined as 
\begin{equation*} 
(t_A^b)_{ac}=\ri f^{abc} 
\,.
\end{equation*} 
Compute the generators in the adjoint representation of SU(2).

We define the constant $T_{\rm R}$ for a representation $R$ by the condition 
\begin{equation*} 
\tr[t^a_{\rm R}t^b_{\rm R}]=\delta^{ab}T_{\rm R} 
\,.
\end{equation*} 
Compute this constant from the explicit form of the fundamental
($T_{\rm F}$) and the adjoint ($T_{\rm A}$) representation. 

The quadratic Casimir $C_2(R)$ of a representation $R$ is defined by 
\begin{equation*} 
C_2(R)\mathbbm{1}=\sum\limits_a t^a_R t^a_R 
\,.
\end{equation*} 
Compute the quadratic Casimir in the fundamental ($C_{\rm F}$) and
the adjoint ($C_{\rm A}$) representation of $SU(2)$ using the explicit
form of the representation matrices.
\end{exe} 
 
 
\begin{exe} 
 
Show that in SU(N) gauge theories 
$$\left[D_\mu,D_\nu\right]=\ri gF_{\mu\nu}^a T^a \quad \mbox{with} \quad 
F_{\mu\nu}^a = \partial_{\mu} A_{\nu}^a -\partial_{\nu} A_{\mu}^a - 
gf^{abc} A_\mu^b A_\nu^c  \,. 
$$ 
 
\end{exe} 
 
 
\begin{exe} 
 
Show that $F_{\mu\nu}^a$ transforms according to the 
adjoint representation of SU(N): 
$$ F_{\mu\nu}^a \to F_{\mu\nu}^a - f^{abc} \theta^b F_{\mu\nu}^c\,.$$ 
 
\end{exe} 
 
\noindent\rule{\textwidth}{1pt}

\subsection{Feynman rules} 
 
The Feynman rules can be derived from the action,
\[
S={\rm i}\int\!{\rm d}^{4}x(\mathcal{L}_{f}\,+\,\mathcal{L}_{g})
\equiv S_{0}\,+\,S_{I}\,,\quad \mathrm{where}\quad
S_{0}={\rm i}\int\!{\rm d}^{4}x\mathcal{L}_{0}\,,\quad\mathrm{and}\quad
S_{I}={\rm i}\int\!{\rm d}^{4}x\mathcal{L}_{I}.
\]
In this decomposition $\mathcal{L}_{0}$ contains the terms bilinear in
the fields and $\mathcal{L}_{I}$ does all other terms, called interactions.
The gluon propagator $\Delta_{g,\mu\nu}(p)$ is the inverse of the
bilinear term in $A_\mu$.  In momentum space we have the condition 
(we suppress colour indices as these terms are diagonal in colour space)
\begin{equation} 
\Delta_{g,{\mu\nu}}\lp(p\rp)\,
\ri\lp[p^{2}g^{\nu\rho}\,-\,p^{\nu}p^{\rho}\rp]=\delta^{\rho}_{\mu}\,. 
\end{equation} 
However, 
\begin{equation} 
\lp[p^{2}g^{\nu\rho}-p^{\nu}p^{\rho}\rp]p_{\rho}=0 
\,, 
\end{equation} 
which means that the inverse does not exist, the matrix
$\lp[p^{2}g^{\nu\rho}\,-\,p^{\nu}p^{\rho}\rp]$ is not invertible. We
can exploit gauge invariance to rewrite the classical Lagrangian in a
physically equivalent form (action remains the same) such that
$\Delta_{g,{\mu\nu}}$ exists, which is called gauge fixing. This
amounts to imposing a constraint on $A_\mu$ by adding a term to the
Lagrangian with a Lagrange multiplicator (like in classical mechanics).
For example, the {\em covariant gauges} are defined by requiring
$\partial_{\mu}A^\mu\lp(x\rp)=0$ for any $x^\mu$. Adding 
\begin{equation*} 
\mathcal{L}_{\rm GF}= 
-\frac{1}{2\lambda}\lp(\partial_{\mu}A^{\mu}\rp)^{2},\qquad 
\lambda\,\in\,\mathbb{R}, 
\end{equation*} 
to $\mathcal{L}$, the action $S$ remains the same.  The bilinear term
becomes in this case 
\[\ri\lp(p^{2}g^{\nu\rho}-\lp(1-\frac{1}{\lambda}\rp)p^{\nu}p^{\rho}\rp),\] 
with inverse 
\begin{equation*} 
\Delta_{g,{\mu\nu}}\lp(p\rp)= 
-\frac{\ri}{p^{2}}\lp[g_{\mu\nu}-(1-\lambda)\frac{p_{\mu}p_{\nu}}{p^{2}}\rp] 
\,. 
\end{equation*} 
Of course, physical results must be independent of $\lambda$. It is
customary to choose $\lambda=1$ (called covariant Feynman gauge).

In covariant gauges unphysical degrees of freedom (longitudinal and
time-like polarizations) also propagate. The effect of these
unwanted degrees of freedom is cancelled by the ghost fields (coloured
complex scalars with Fermi statistics). We do not elaborate the details 
of these fields as the unwanted degrees of freedom and the ghost fields
can be avoided by choosing {\em axial (physical) gauges}, which is our
choice. The axial gauge is defined with an arbitrary, but fixed vector 
$n^{\mu}$, different from $p^{\mu}$:
\begin{equation*} 
\mathcal{L}_{\rm GF}=-\frac{1}{2\lambda}\lp(n^{\mu}A_{\mu}\rp)^{2}, 
\end{equation*} 
which leads to 
\begin{equation*} 
\Delta_{g,{\mu\nu}}\lp(p,n\rp)= 
-\frac{\ri}{p^{2}}\lp(g_{\mu\nu} 
-\frac{p_{\mu}n_{\nu}+n_{\mu}p_{\nu}}{p\,\ldot\,n}+ 
\frac{\lp(n^{2}+\lambda\,p^{2}\rp)p_{\mu}p_{\nu}}{(p\,\ldot\,n)^{2}}\rp). 
\end{equation*} 
Since $p^{2}=0$, we have: 
\begin{equation*} 
\Delta_{g,{\mu\nu}}\lp(p,n\rp)\,p^{\mu}=0
\,,\qquad
\Delta_{g,{\mu\nu}}\lp(p,n\rp)\,n^{\mu}=0 
\,.
\end{equation*} 
Thus, only 2 degrees of freedom propagate (transverse ones in the
$n^{\mu}+p^{\mu}$ rest-frame). A usual choice is $n^{2}=0,\;\lambda=0$,
called {\em light-cone gauge}. The price we pay by choosing this gauge
instead of a covariant one is that the propagator looks more
complicated and it diverges when $p^{\mu}$ becomes parallel to
$n^{\mu}$. In this gauge 
\begin{equation*} 
\Delta_{g,{\mu\nu}}(p,n)=\frac{\ri}{p^{2}}\,d_{\mu\nu}(p,n) 
\end{equation*} 
with 
\begin{equation} 
d_{\mu\nu}\lp(p,n\rp)= 
-g_{\mu\nu}+\frac{p_{\mu}n_{\nu}+n_{\mu}p_{\nu}}{p\,\ldot\,n}= 
\sum_{\lambda=1,2}\,\epsilon_{\mu}^{(\lambda)}(p) 
\epsilon_{\nu}^{(\lambda)}(p)^{\ast} 
\,, 
\label{eq:polsum}
\end{equation} 
where $\epsilon_{\mu}^{(\lambda)}(p)$ is the polarization vector of the 
gauge field (photon in QED, gluon in QCD). 

\subsection{Feynman rules for QCD}

Propagators (Feynman's `$+\ri \epsilon$'-prescription is assumed, but
not shown): 

\vspace*{10pt}
gluon propagator: 
$\Delta_{g,\mu\nu}^{ab}\lp(p\rp)= 
\delta^{ab}\,\Delta_{g,\mu\nu}\lp(p\rp) \longrightarrow$

\vspace*{-30pt} \hspace*{240pt}
\includegraphics[width=0.2\textwidth]{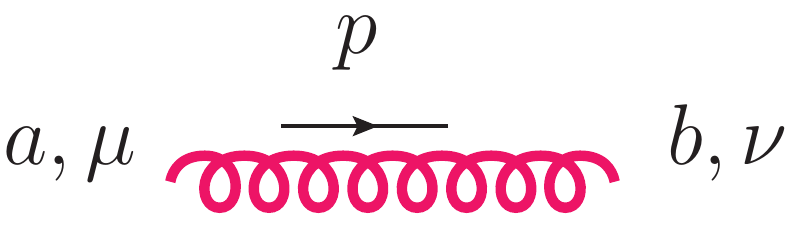}

\vspace*{10pt}
quark propagator: 
$\Delta_{q}^{ij}\,\lp(p\rp)= 
\delta^{ij}\,\ri \frac{\slashed{p}+m}{p^2-m^2}
\longrightarrow$

\vspace*{-30pt} \hspace*{240pt}
\includegraphics[width=0.2\textwidth]{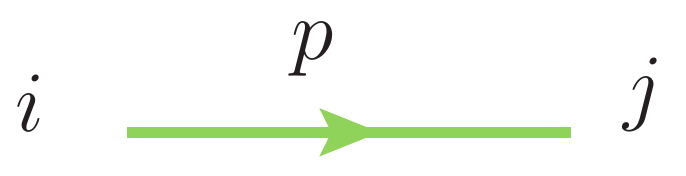}

\vspace*{10pt}
ghost propagator:
$\Delta^{ab}\,\lp(p\rp)= 
\delta^{ab}\,\frac{\ri}{p^{2}} \longrightarrow$

\vspace*{-30pt} \hspace*{240pt}
\includegraphics[width=0.2\textwidth]{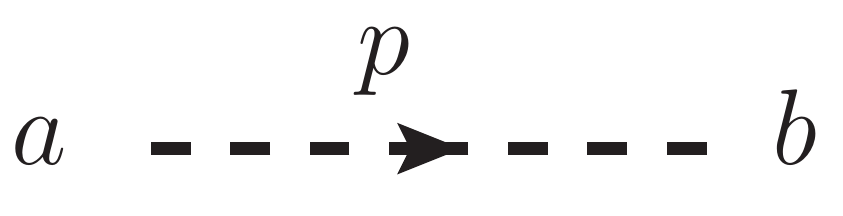}

(not needed in physical gauges)
\bigskip

\noindent
Vertices:

quark-gluon:
$\Gamma_{gq\bar{q}}^{\mu,\,a}= -\ri
g_{S}\,(t^{a})_{ij}\gamma^{\mu}\longrightarrow$

\vspace*{-40pt} \hspace*{180pt}
\includegraphics[width=0.2\textwidth]{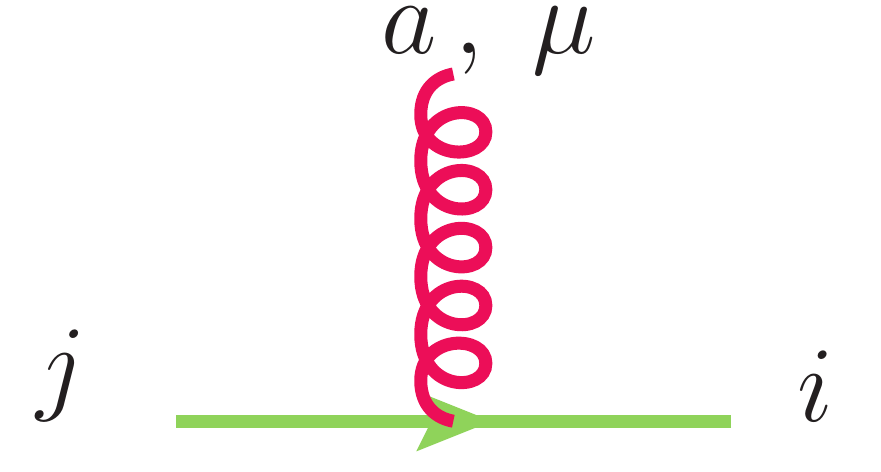}

\vspace*{38pt}
three-gluon:
$\Gamma_{\alpha\beta\gamma}^{abc}\lp(p,q,r\rp)= -\ri
g_{s}\lp(F^{a}\rp)_{bc}\,V_{\alpha\beta\gamma}\lp(p,q,r\rp)\longrightarrow$

\vspace*{-50pt} \hspace*{260pt}
\includegraphics[width=0.15\textwidth]{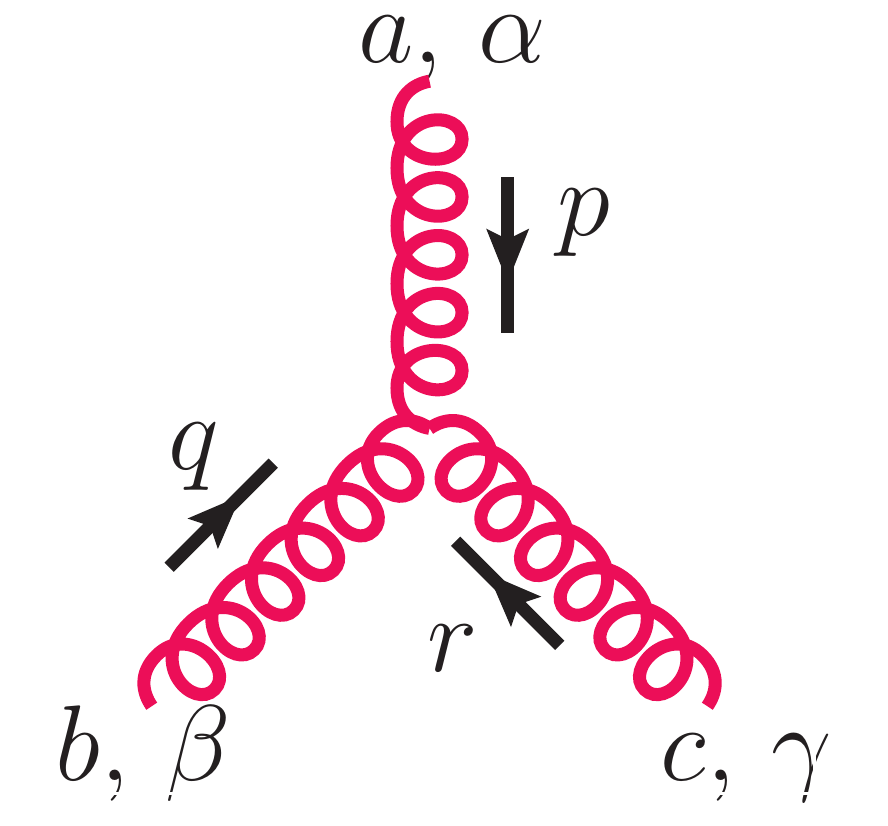}

$V_{\alpha\beta\gamma}\lp(p,q,r\rp)= 
\lp(p-q\rp)_{\gamma}g_{\alpha\beta} 
+\lp(q-r\rp)_{\alpha}g_{\beta\gamma} 
+\lp(r-p\rp)_{\beta}g_{\alpha\gamma}
\,,\quad\,p^\alpha+q^\alpha+r^\alpha=0$

\vspace*{30pt}
four-gluon:
$\Gamma_{\alpha\beta\gamma\delta}^{abcd} = -{\rm i} g_{\rm s}^2\,
\left[
\begin{array}{l}
+f^{xac}\,f^{xbd}\,
(g_{\alpha\beta}g_{\gamma\delta}-g_{\alpha\delta}g_{\beta\gamma})\\
+f^{xad}\,f^{xbc}\,
(g_{\alpha\beta}g_{\gamma\delta}-g_{\alpha\gamma}g_{\beta\delta})\\
+f^{xad}\,f^{xbc}\,
(g_{\alpha\gamma}g_{\beta\delta}-g_{\alpha\delta}g_{\beta\gamma})
\end{array} 
\right]\longrightarrow
$

\vspace*{-56pt} \hspace*{300pt}
\includegraphics[width=0.15\textwidth]{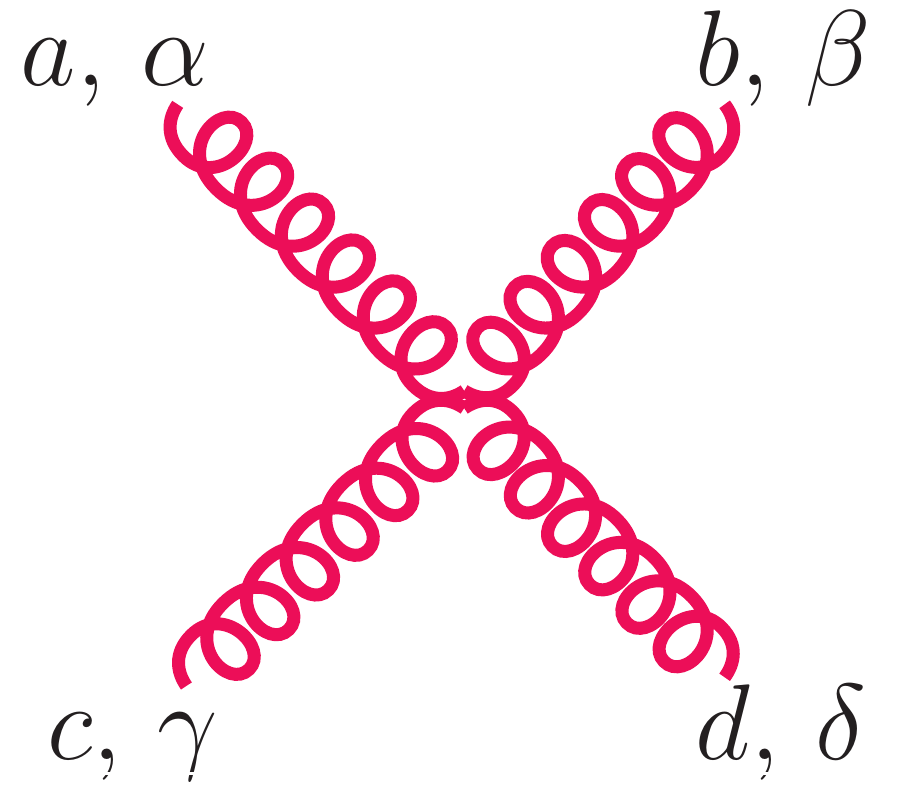}

\vspace*{28pt}
ghost-gluon:
$\Gamma_{g\eta\bar{\eta}}^{\mu,\,a}= 
-\ri g_{S}\,(F^{a})_{ij}\,p^{\mu}\longrightarrow$
\hspace*{100pt}(not needed in physical gauges).

\vspace*{-40pt} \hspace*{200pt}
\includegraphics[width=0.15\textwidth]{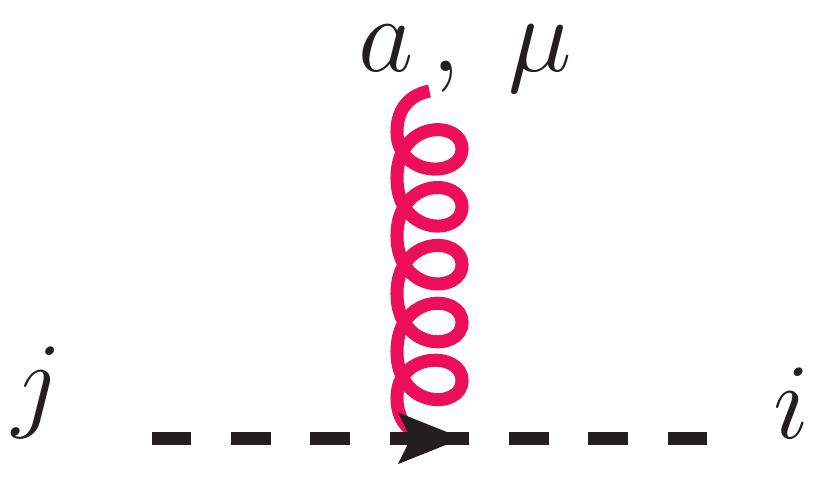}

The four-gluon vertex
differs from the rest of the Feynman rules in the sense that it is not
in a factorized form of a colour and a tensor factor. This is an
inconvenient feature because it prevents the separate summation over
colour and Lorentz indices and complicates automation. We can however
circumvent this problem by introducing a {\em fake field} with propagator

\includegraphics[width=0.15\textwidth]{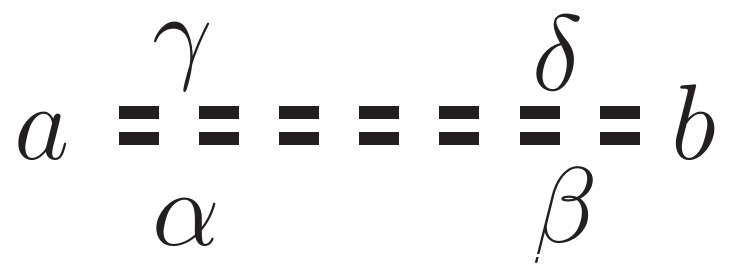}

\vspace*{-30pt} \hspace*{80pt}
$= \frac{{\rm i}}2 \delta^{ab}
(g^{\alpha\beta}g^{\gamma\delta}-g^{\alpha\delta}g^{\beta\gamma})\,,$
that couples only to the gluon with vertex

\vspace*{16pt}
\includegraphics[width=0.15\textwidth]{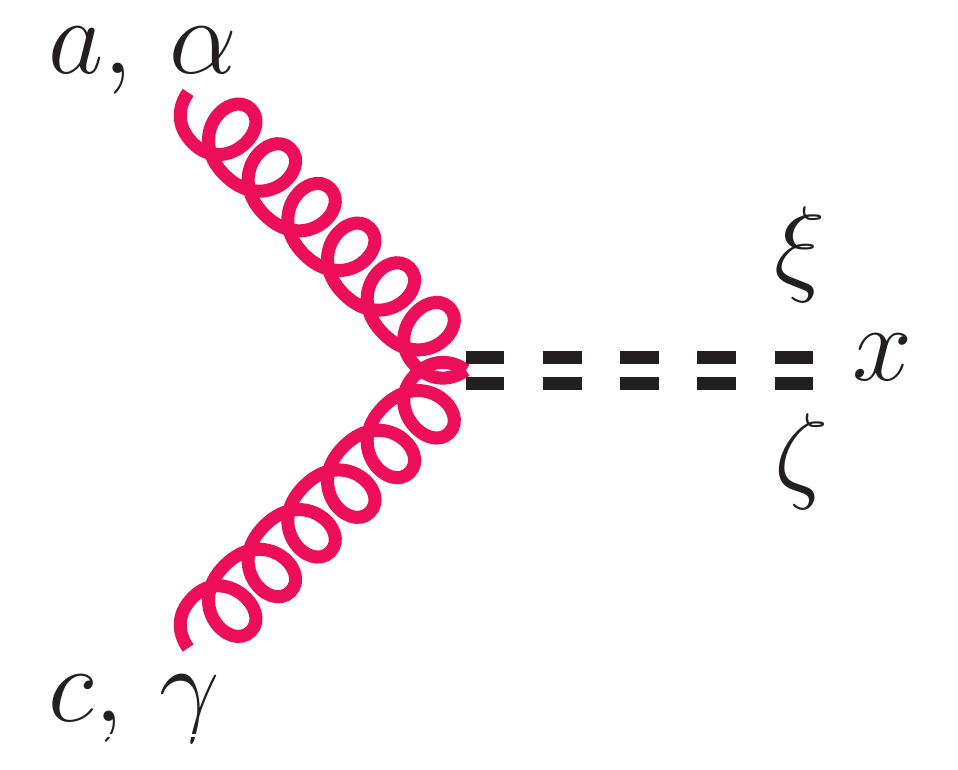}

\vspace*{-46pt} \hspace*{80pt}
$= {\rm i} \sqrt{2} g_{\rm s}\,f^{xac} g^{\alpha\xi}g^{\gamma\zeta}$.

\vspace*{26pt}
\noindent
We can check that a single four-gluon vertex can be written as a sum of
three graphs. This way the summations over colour and Lorentz indices
factorize completely, which helps automation and makes possible for us
to concentrate on the colour algebra independently of the rest of the
Feynman rules.

Finally, we have to supply the following factors for incoming and
outgoing particles:
\begin{displaymath} 
\begin{array}{clccl} 
  \bullet & \textrm{outgoing fermion: }\bar{u}\lp(p\rp) & & \bullet & 
\textrm{outgoing antifermion: }v\lp(p\rp) \\[1mm] 
  \bullet & \textrm{incoming fermion: }u\lp(p\rp) & & \bullet & \textrm{incoming antifermion: }\bar{v}\lp(p\rp) \\ 
  \bullet & \textrm{outgoing photon, or gluon: 
}\epsilon_\mu^{\lp(\lambda\rp)}\lp(p\rp)^{\ast} & & \bullet & 
\textrm{incoming photon, or gluon: }\epsilon_\mu^{\lp(\lambda\rp)}\lp(p\rp)\,. 
\end{array} 
\end{displaymath} 
 
\vspace{-4pt}
\noindent\rule{\textwidth}{1pt}

 
\begin{exe} 
 
Show that the four-gluon vertex can be written as a sum of three graphs,
with the help of the fake field such that in each graph the colour and 
Lorentz indices are factorized:

\vspace*{-15pt}\hspace*{300pt}
\includegraphics[width=0.30\textwidth]{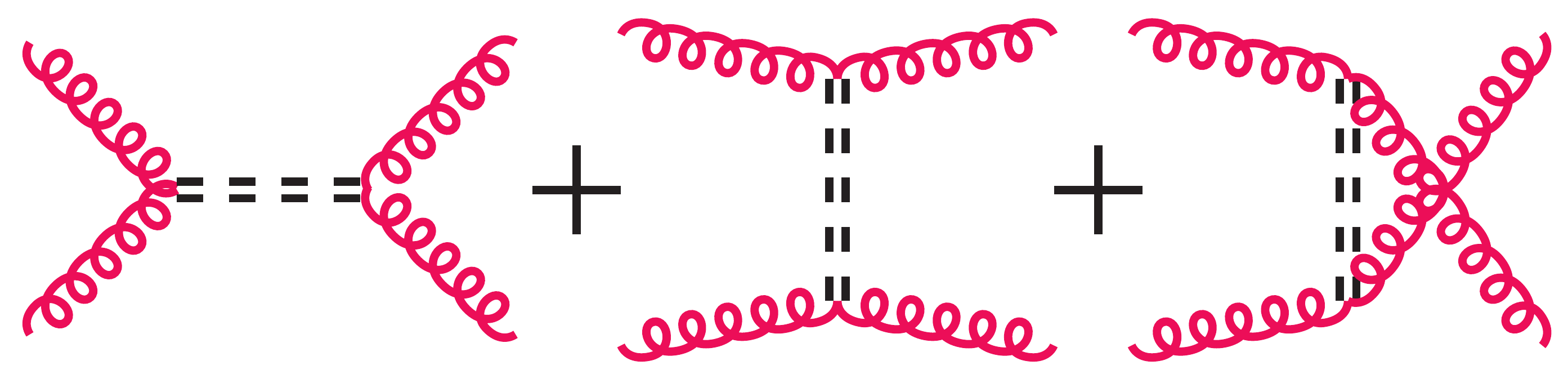}

\end{exe} 
 
\noindent\rule{\textwidth}{1pt}

\subsection{Basics of colour algebra} 
 
Examining the Feynman rules, we find that there are two essential changes
as compared to QED. One is that there is an additional degree of freedom:
colour. The second is that there are new kind of couplings: the self
couplings of the gauge field. We now explore the effect of the first.

In order to see how to treat the colour degrees of freedom, we set to
one all but the colour part of the Feynman rules and try first to
develop an efficient technique to compute the coefficients involving
the colour structure. This is possible because the colour degrees of
freedom factorize from the other degrees of freedom completely. We use
the following graphical representation for the colour charges in the
fundamental representation:
\begin{center} 
\includegraphics[width=0.15\linewidth]{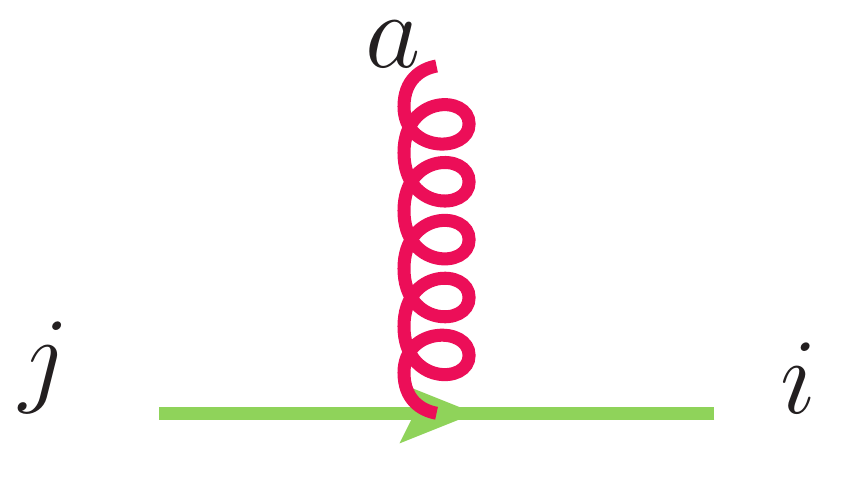}

\vspace*{-40pt} \hspace*{130pt}
$=\lp(t^{a}\rp)_{ij}$
\,. 
\end{center} 
\vspace*{20pt}
The normalization of these matrices is given by Tr$\lp(t^{a}t^{b}\rp)=$ 

\vspace*{-26pt} \hspace*{230pt}
\includegraphics[width=0.3\linewidth]{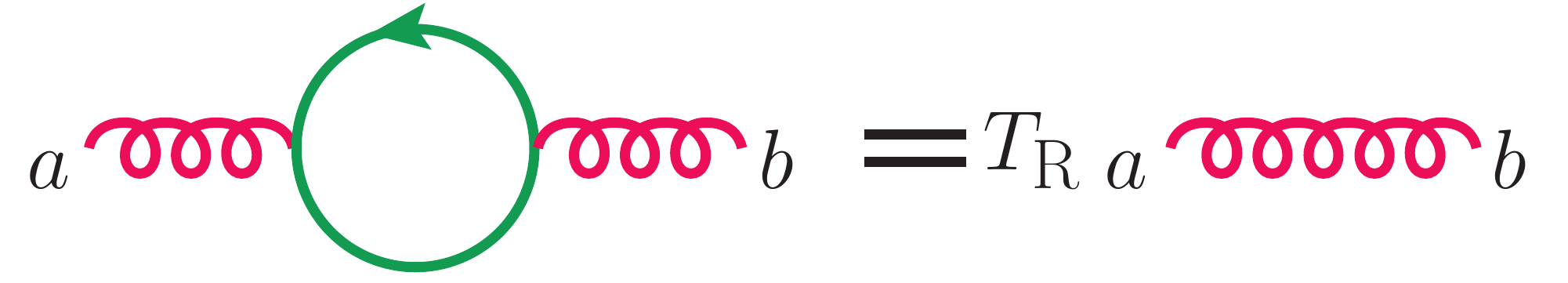}

\vspace*{-26pt} \hspace*{380pt}
{$=T_{\rm R}\,\delta^{ab}$\,.} 
The usual choice is $T_{\rm R}=\frac{1}{2}$, but $T_{\rm R} = 1$ is also used 
often. We shall use both. 
 
In the adjoint representation the colour charge $T^{a}$ is represented by 
the matrix $\lp(F^{a}\rp)_{bc}$ that 
is related to the structure constants by 
\begin{center} 
{$\lp(F^{a}\rp)_{bc}\,=\,\lp(F^{b}\rp)_{ca}\,= 
\,\lp(F^{c}\rp)_{ab}\,=\,-\ri\,f_{abc}=$} 

\vspace*{-40pt} \hspace*{240pt}
\includegraphics[width=0.1\linewidth]{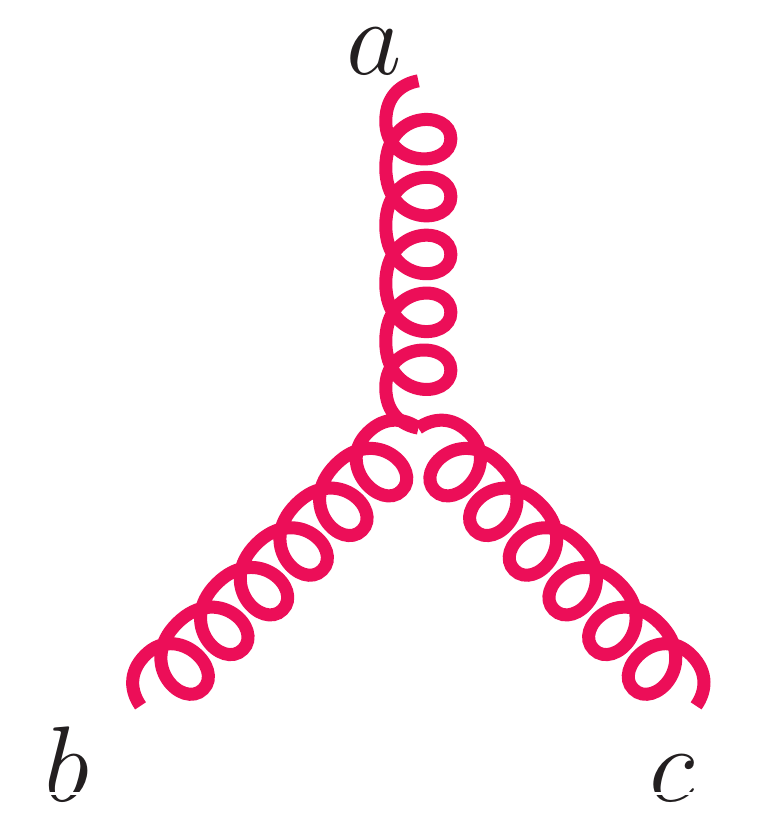} 
\end{center} 
where $F^{a}$ with $a=1,\dots,(N_\rc^2 - 1)$ are 
$\lp(N_{\rc}^{2}-1\rp)\times\,\lp(N_{\rc}^{2}-1\rp)$ matrices 
which again satisfy the commutation relation (\ref{liealg}). The 
graphical notation in the adjoint representation is not unique. For the 
matrix $\lp(F^{a}\rp)_{bc}$ we assume an arrow pointing from index $c$ to 
$b$, opposite to which we read the indices of $\lp(F^{a}\rp)_{bc}$
(similarly as for the matrices $t^a$). On the structure constants the
indices are not distinguished, therefore arrows do not appear. However,
these are completely antisymmetric in their indices, therefore, the
ordering matters. By convention, in the graphical representation, the
ordering of the indices is counter-clockwise.  The representation
matrices are invariant under $SU(N)$ transformations. 
 
The sums $\sum_{a}t^{a}_{ij}t^{a}_{jk}$ and Tr$\lp(F^{a}F^{b}\rp)$ have 
two free indices in the fundamental and adjoint representation, 
respectively. These are invariant under $SU(N)$ transformations, 
therefore, must be proportional to the unit matrix, 
\begin{equation*}
\sum_{j,a}t^{a}_{ij}t^{a}_{jk} = C_{\rm F}\,\delta_{ik}\,, \qquad
{\rm Tr}\lp(F^{a}F^{b}\rp) = C_{\rm A}\,\delta^{ab}\,, 
\label{eq:casimirs}
\end{equation*}
which is depicted graphically as
\begin{center} 
\includegraphics[width=0.4\linewidth]{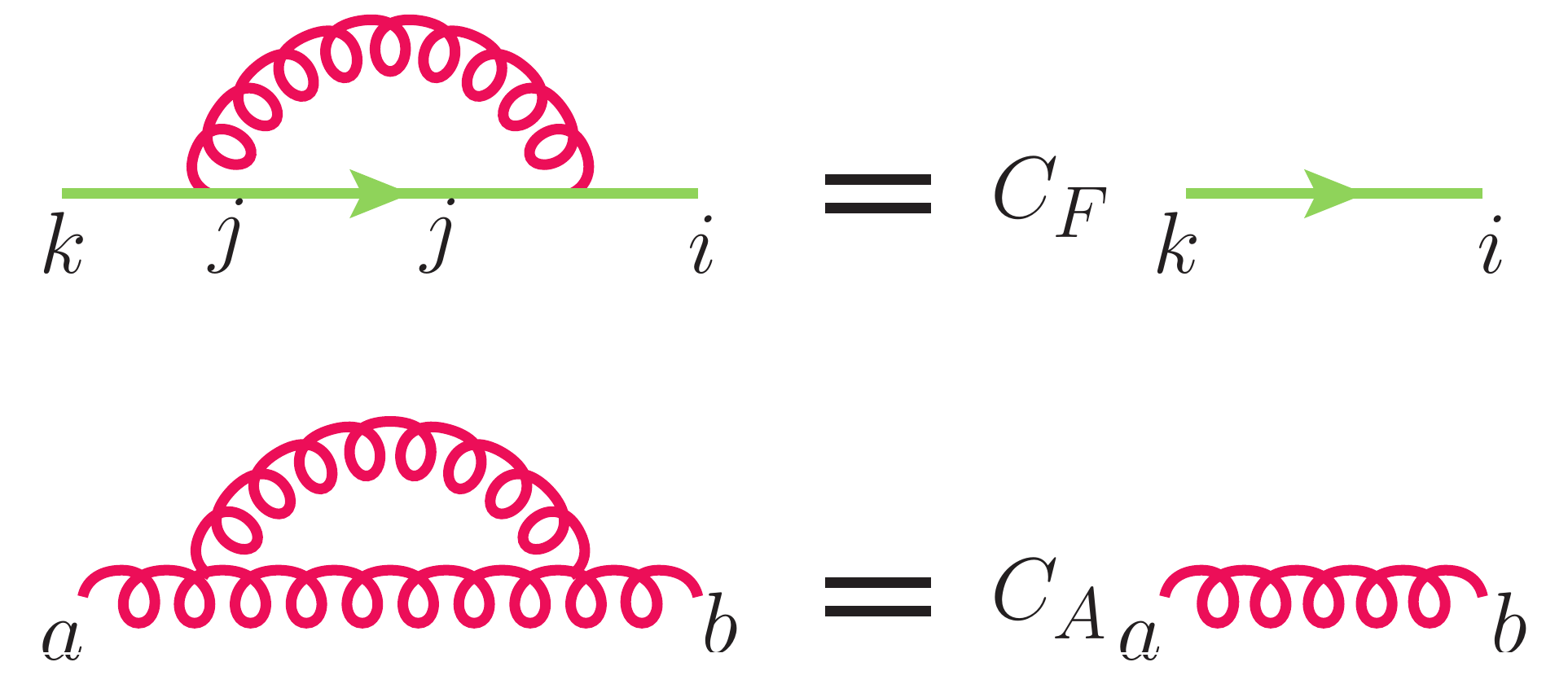} \,.
\end{center} 
Here $C_{\rm F}$ and $C_{\rm A}$ are the eigenvalues of the quadratic 
Casimir operator in the fundamental and adjoint representation,
respectively. In the familiar case of angular momentum operator 
algebra ($SU(2)$), the quadratic Casimir operator is the square of the 
angular momentum with eigenvalues $j(j+1)$. The fundamental 
representation is two dimensional, realized by the (half of the) Pauli 
matrices acting on two-component spinors, when $j=1/2$ and $C_{\rm F} = 
1/2 (1/2 +1) = 3/4$. In the adjoint representation $j=1$ and $C_{\rm A} = 
2$. Below we derive the corresponding values for general $SU(N)$. 
 
The commutation relation (\ref{liealg}) can be represented 
graphically by 
\begin{center}
\includegraphics[width=0.4\linewidth]{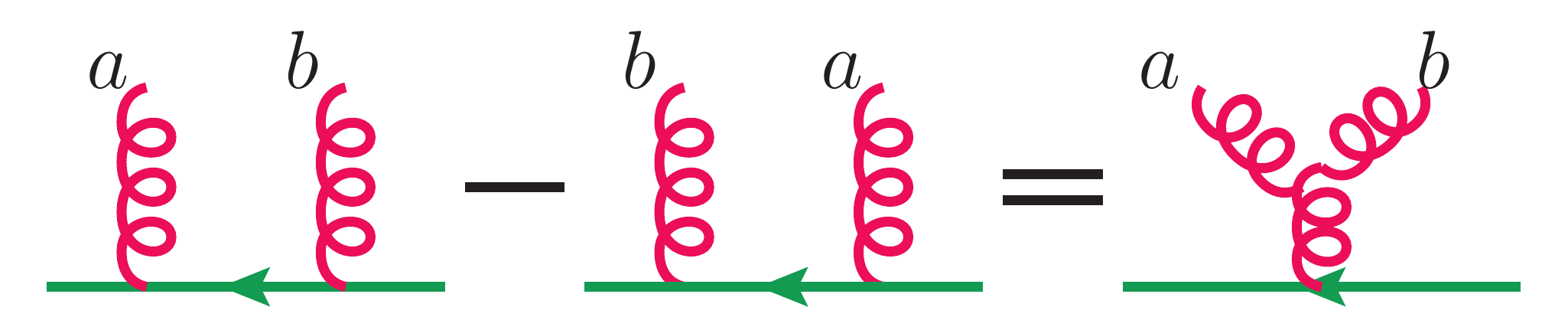}\,.
\end{center}
Multiplying this commutator first with another colour charge operator
with summing over the fermion index and then taking the trace over the
fermion line (\ie multiplying with $\delta_{ik}$) we obtain the
resolution of the three-gluon vertex as traces of products of colour charges:
\begin{center}
\includegraphics[width=0.4\linewidth]{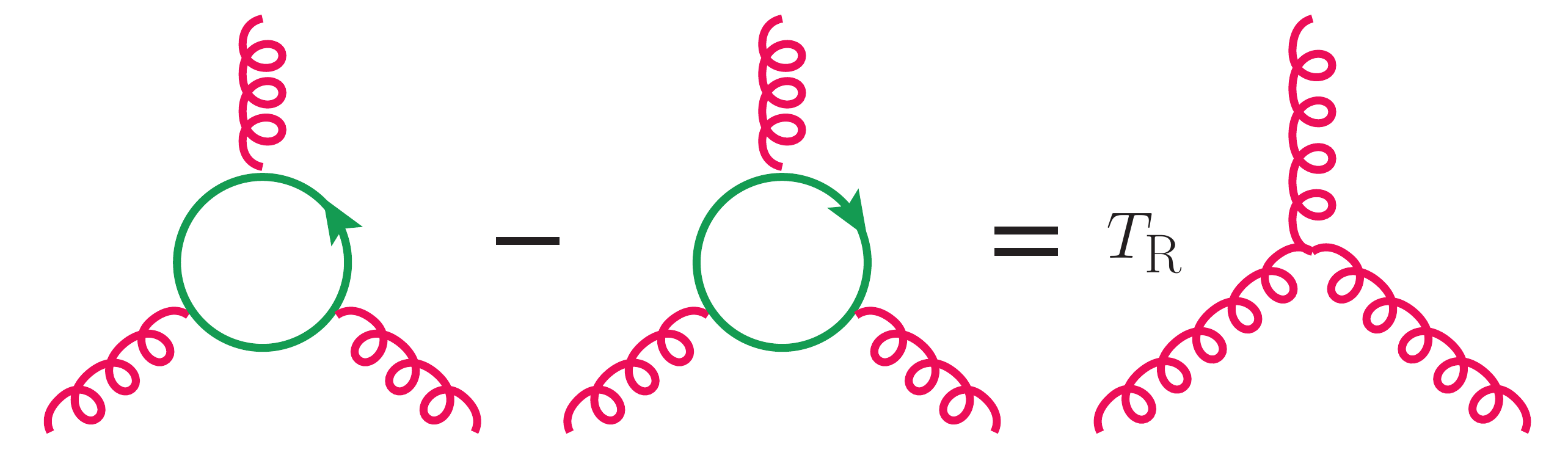}
\\
$= {\rm Tr}(t^a t^b t^c) - {\rm Tr} (t^c t^b t^a) = \ri\,T_{\rm R}
f^{abc}\,.$
\end{center}

We now show some examples of how one can compute the colour 
algebra structure of a QCD amplitude, in particular we will 
also find an explicit value for $C_{F}$ and $C_{A}$. Taking 
the trace of the identity in the fundamental and 
in the adjoint representation we obtain 
\begin{center}
\includegraphics[width=0.08\linewidth]{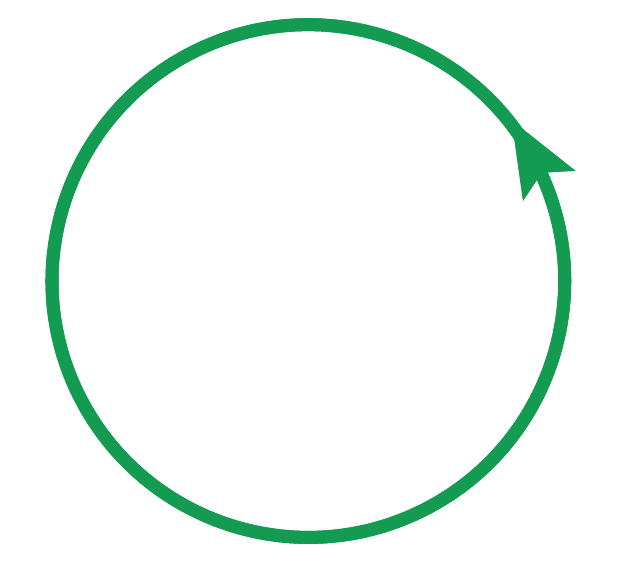}
\hspace*{36pt}
\includegraphics[width=0.08\linewidth]{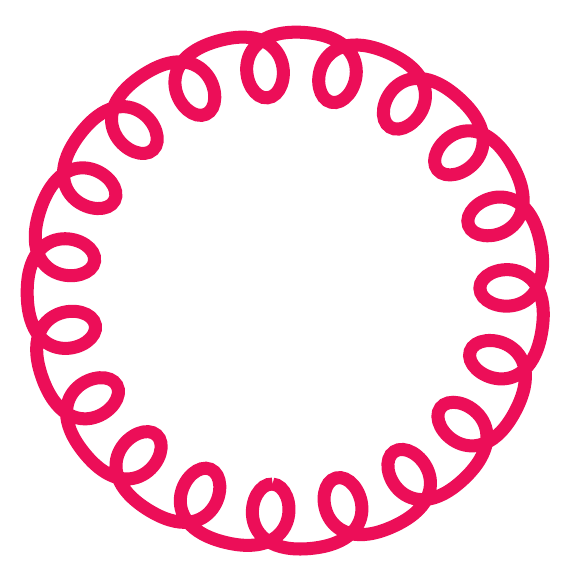}

\vspace*{-32pt} ~\hspace*{100pt}
$=N_{\rc}\,,$ \hspace*{50pt}
$=N_{\rc}^{2}-1$
\,,
\end{center}
\vspace*{10pt}
respectively. Then, using the expressions for the fermion and gluon 
propagator corrections, we immediately find 
\begin{center} 
\includegraphics[width=0.08\linewidth]{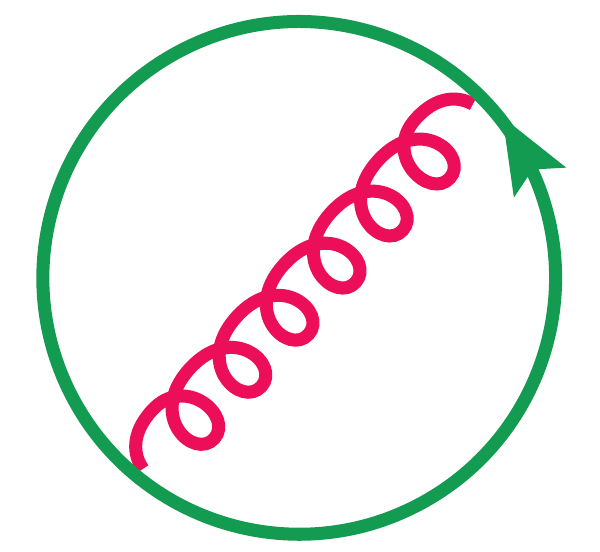} 
\hspace*{66pt}
\includegraphics[width=0.08\linewidth]{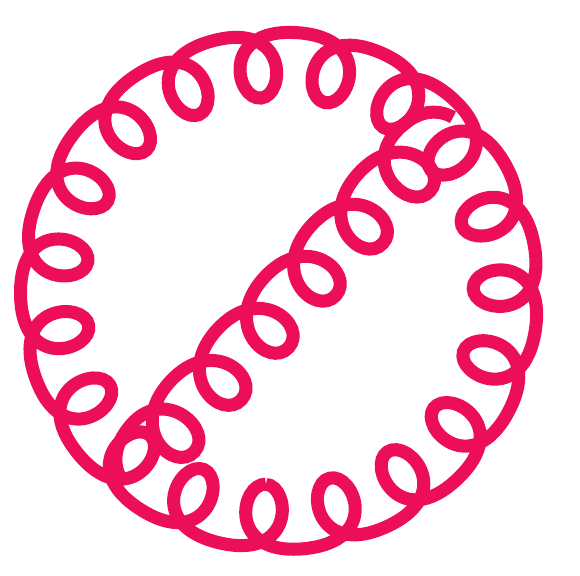} 

\vspace*{-32pt} ~\hspace*{120pt}
$=C_{\rm F}\,N_{\rc}$ \,, \hspace*{60pt}
$=C_{\rm A}\,\lp(N_{\rc}^{2}-1\rp)$\,.
\end{center} 
\vspace*{10pt}
The generators are traceless, 
\begin{center} 
\includegraphics[width=0.12\linewidth]{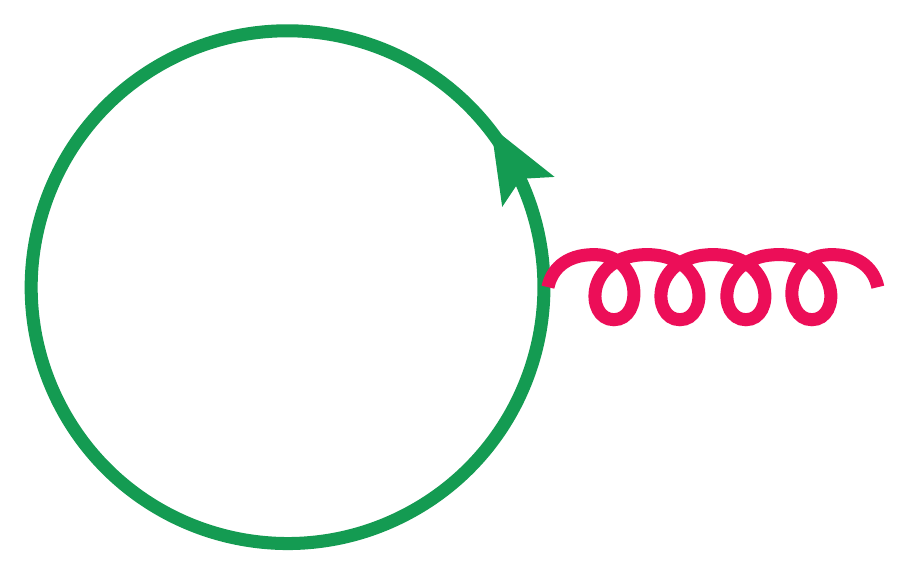} 
\hspace*{80pt}
\includegraphics[width=0.12\linewidth]{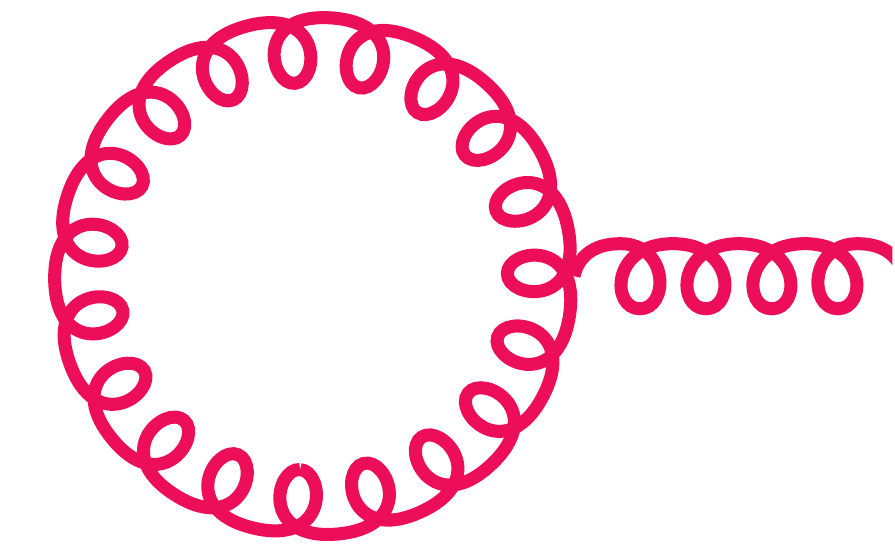} 

\vspace*{-32pt} ~\hspace*{140pt}
$=\textrm{Tr}\lp(t^{a}\rp)\,=\,0$\,,\hspace*{70pt}
$=\textrm{Tr}\lp(F^{a}\rp)\,=\,0$\,. 
\end{center} 

\begin{wrapfigure}{r}{0.4\linewidth}
\includegraphics[width=1.0\linewidth]{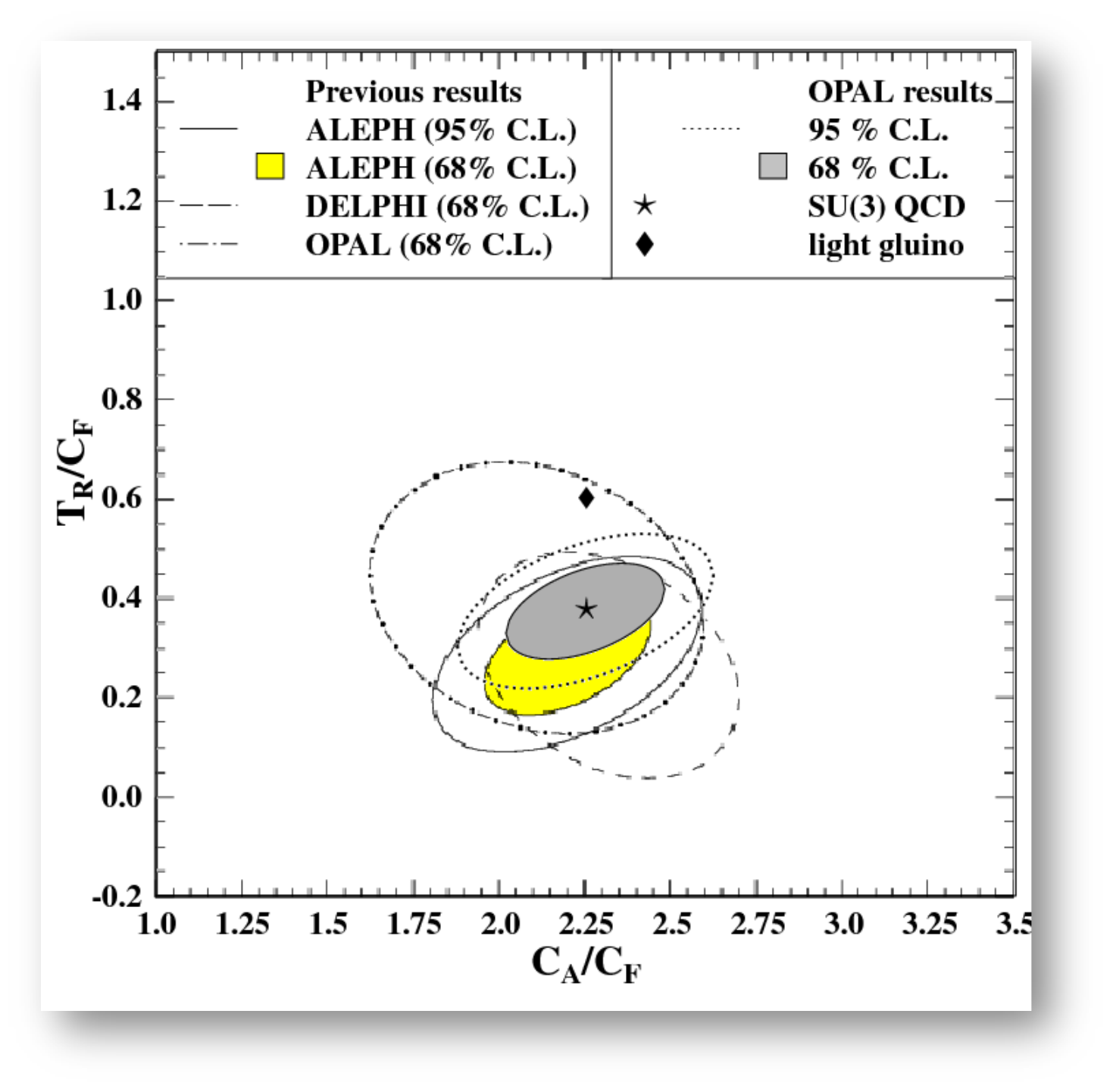}
\caption{Measurement of the colour factors by the LEP collaborations
\cite{Abbiendi:2001qn}
\vspace*{-30pt}
~}
\label{fig:OPAL}
\end{wrapfigure}

\vspace*{10pt}
We can now find the value of $C_{F}$ as follows. On the one 
hand we know that 
\begin{center} 
\includegraphics[width=0.35\linewidth]{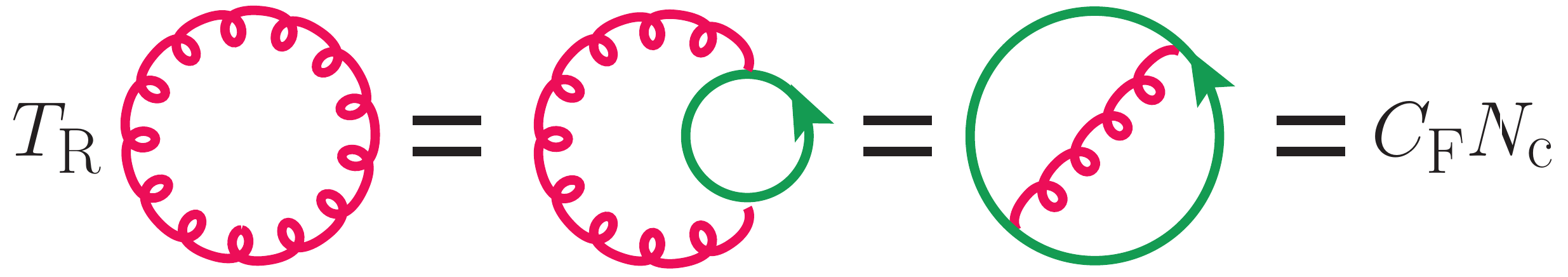}
\end{center} 

\noindent while on the other, the left hand side is also equal to 
$T_{\rm R}\lp(N_{\rc}^{2}-1\rp)$.  Thus 
\begin{equation*}
C_{F}=T_{\rm R}\frac{N_{\rc}^{2}-1}{N_{\rc}}\,. 
\end{equation*}
Analogously one can find 
\begin{equation*}
C_{\rm A}=2\,T_{R}\,N_{\rc}\,. 
\end{equation*}
As the colour factors $C_{\rm F}$ and $C_{\rm A}$ depend on $N_{\rc}$,
their measurement gives information on the number of colours. The
experiments of the Large Electron Positron collider measured the 
values of the colour factors based on fits of theoretical predictions
\cite{Nagy:1997fk} to four-jet angular distributions that are
sensitive to both $C_{\rm F}$ and $C_{\rm A}$. The result of the
simultaneous measurement of the colour factors and the strong coupling
by the OPAL collaboration is shown in \fig{fig:OPAL}
\cite{Abbiendi:2001qn}. The values corresponding to $N_\rc = 3$ are
marked with the star, just in the middle of the confidence-ellipses.  

The expression $\sum_{a}\,t^{a}_{ij}t^{a}_{kl}$ is invariant under 
$SU(N)$ transformations, therefore has to be expressible as a linear 
combination of $\delta_{il}\delta_{kj}$ and $\delta_{ij}\delta_{kl}$ (the 
third combination of Kronecker $\delta$'s is not possible, the direction 
of arrows do not match). The two coefficients can be obtained by making 
contractions with $\delta_{il}\delta_{jk}$ and $\delta_{ij}\delta_{kl}$.
Thus we obtain the Fierz identity, 
\begin{equation*} 
\sum_{a}t^{a}_{ij}t^{a}_{kl} = T_{\rm R} \lp(\delta_{il}\delta_{kj}
-\frac{1}{N_{\rc}} \delta_{ij}\delta_{kl}\rp)
\,,
\end{equation*}
\vspace*{10pt}
or graphically:

\vspace*{-10pt} ~\hspace*{40pt}
\includegraphics[width=0.5\linewidth]{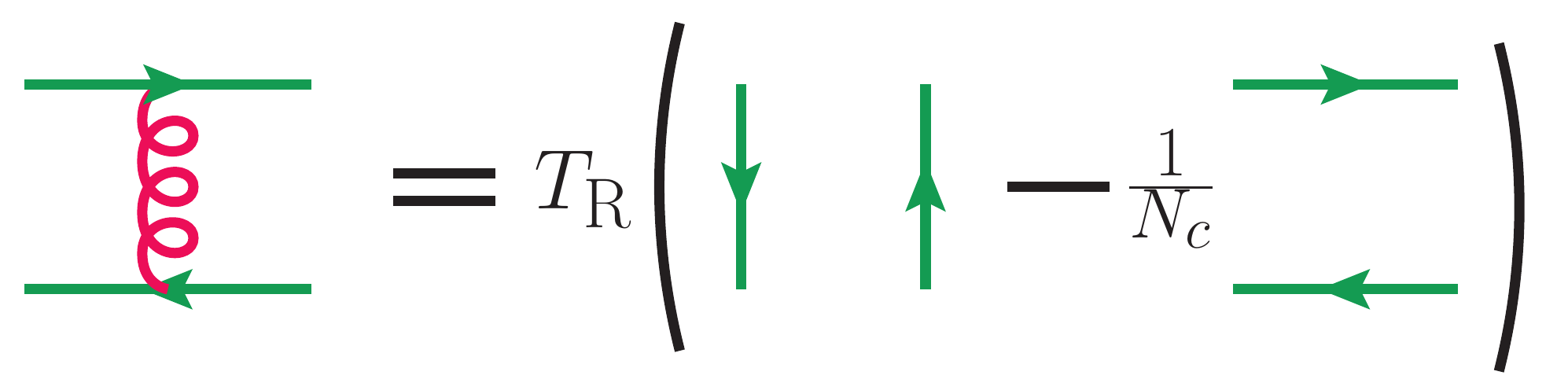} \,.
 
These graphical rules help in computing colour sums easily.
Nevertheless, nowadays computer algebra codes make computation of colour
sums an automated procedure. For instance, you may try 
{\tt In[1]:= Import["http://www.feyncalc.org/install.m"]} in your {\em
Mathematica} session to see one solution.
 
\vspace{-4pt}
\noindent\rule{\textwidth}{1pt}

 
\begin{exe} 
 
Consider the process $q\bar{q}\rightarrow ggg$. Compute the color structures that appear in the squared matrix element. 
 
\end{exe} 
 
 
\begin{exe} 
 
Try using the Fierz identity to obtain 
 
\vspace*{-30pt} ~\hspace*{200pt} 
\includegraphics[width=0.2\linewidth]{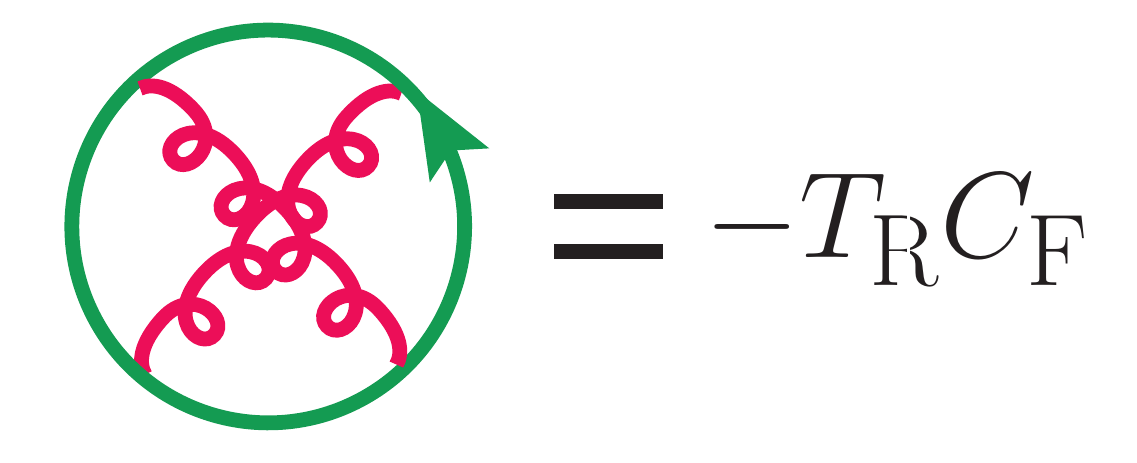}\,. 
 
\end{exe} 
 
 
\begin{exe} 
Determine the color factors $A$,$B$,$C$ in the following equations:
\begin{center} 
\includegraphics[width=0.3\linewidth]{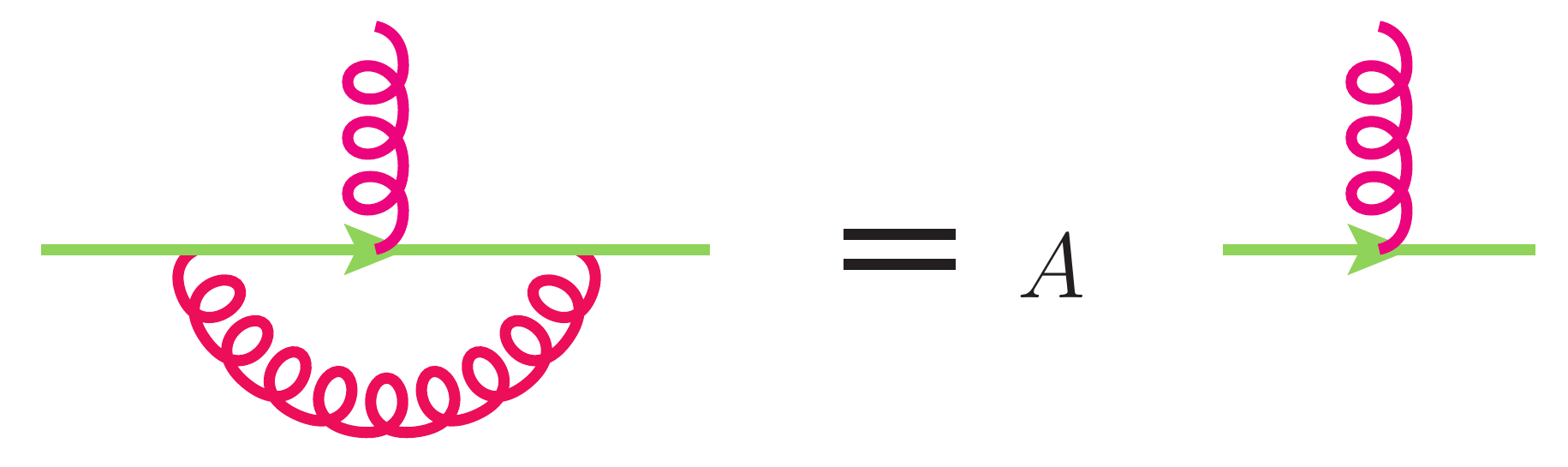}\,, \quad 
\includegraphics[width=0.3\linewidth]{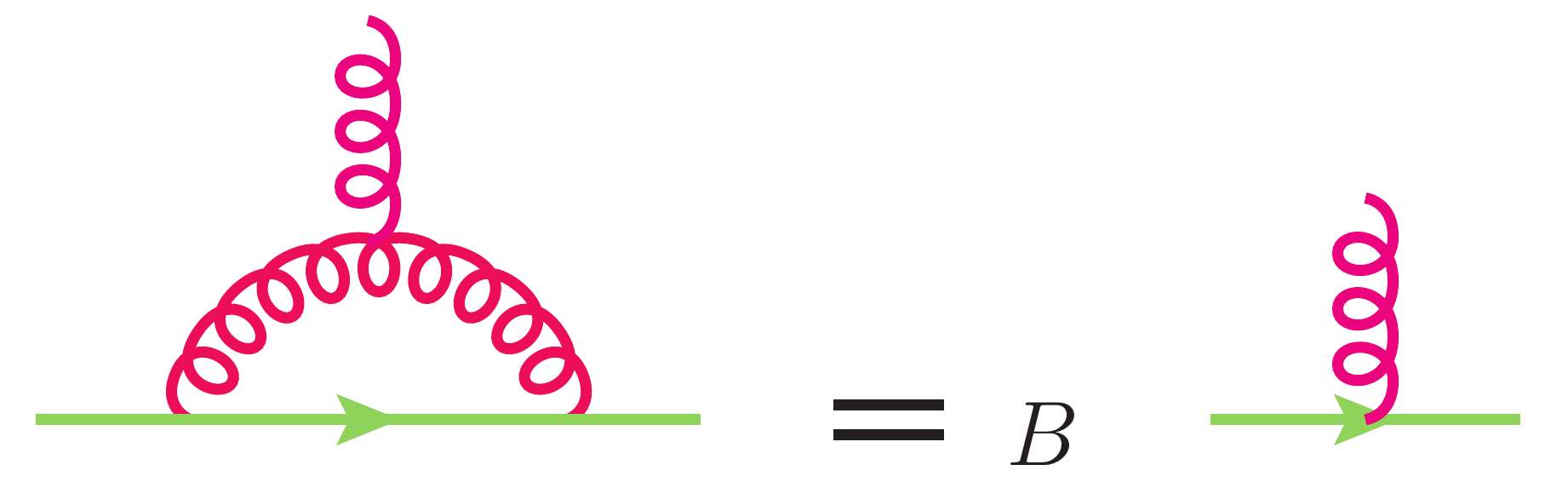}\,, \quad 
\includegraphics[width=0.3\linewidth]{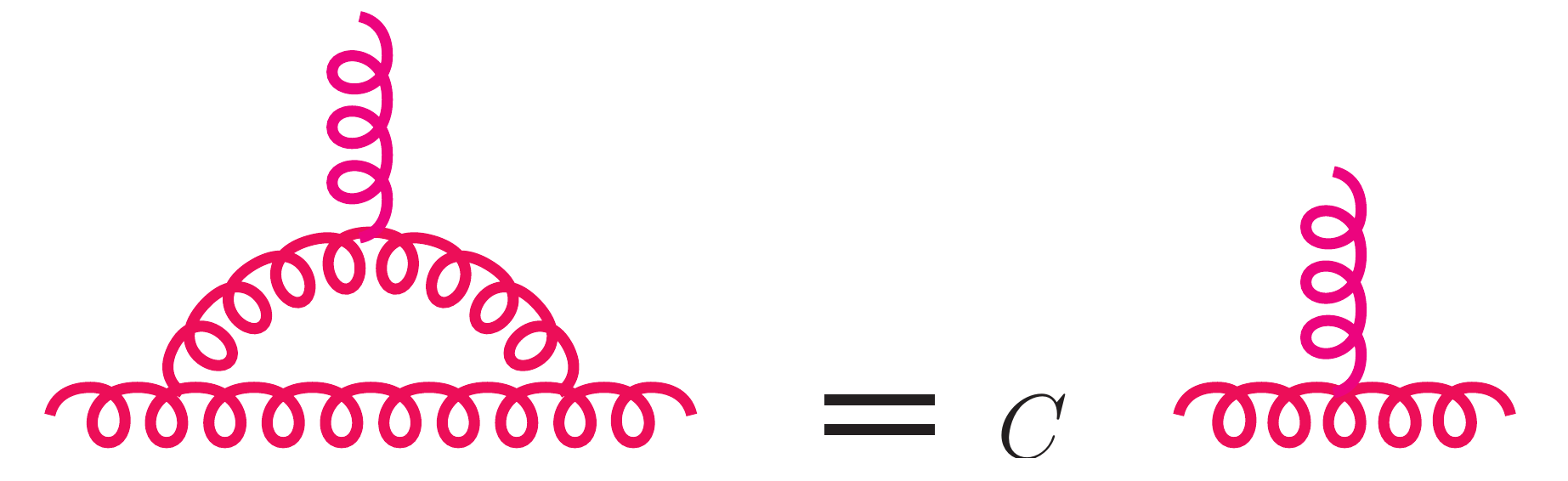}\,.  
\end{center} 
 
\end{exe} 

\noindent\rule{\textwidth}{1pt}

\subsection{Are we done?}

We now have the Feynman rules with colour and the rest factorized, and we
gained some insight how to perform the colour algebra. Thus it seems that
we are in the position to compute the cross section of any process up to
the desired accuracy in perturbation theory (PT), just as we can do in
QED. So it may appear that conceptually we are done. Well, we are going
to see big surprises!

\begin{wrapfigure}{l}{0.54\linewidth}
\includegraphics[width=1.0\linewidth]{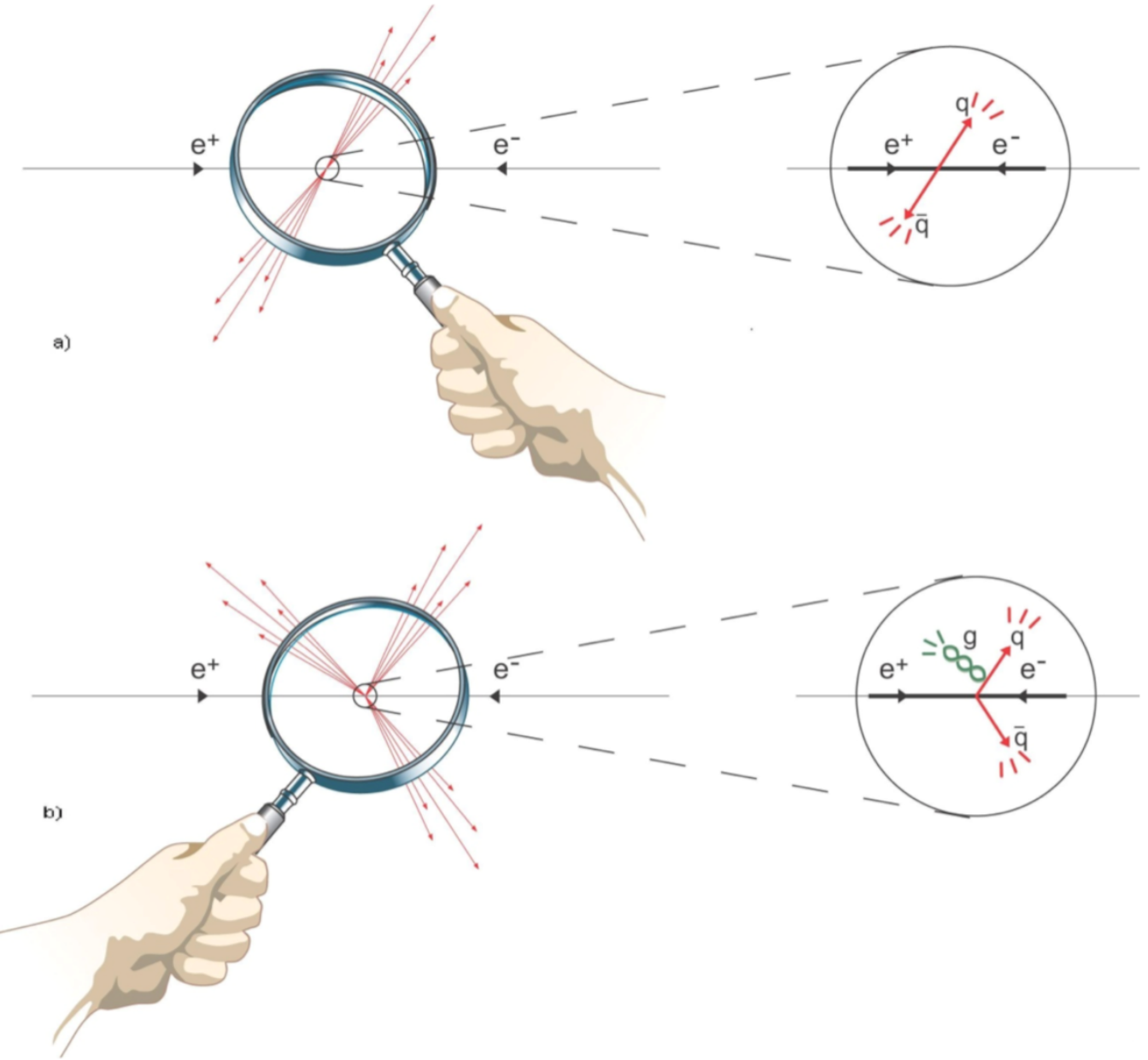}
\caption{Illustration of the approximation of hadronic final states by
partonic events in electron-positron annihilation: the sprays of hadrons
(called jets) are assumed to originate from primary quarks and gluons,
thus approximated by quarks and gluons as shown by the magnification
\vspace*{-30pt}~
}
\label{fig:partonhadron}
\end{wrapfigure}

The first conceptual challenge is due to a phenomenological observation.
In QED, PT is applicable because the elementary excitations of the
quantum fields, the electrons and photons, can be observed as stable,
free particles. Thus asymptotic states are parts of the physical
reality. On the contrary, free quarks and gluons (usually called simply
partons) have never been observed in nature. This experimental fact can
be reformulated saying that the probability of observing a final state
with any {\em fixed number of on-shell partons} is zero.  This negative
result has been turned into the principle of `quark confinement'. Thus
it is questionable whether a QFT of quark and gluon fields can describe
the observed world of particles where in addition to leptons only
hadrons have been found. In fact, a main research project at the 
LEP was to find an answer to this question in a well controlled
quantitative manner.  It turned out that the answer is positive if we
make an assumption that we cannot prove from first principles:
\begin{quote}
{\em The result of a low-order perturbative computation in QCD is an
approximation to sufficiently inclusive hadronic cross section if
(i) the total centre-of-mass energy $Q$ of partons is much larger than
the mass of quarks, $Q >> m_q$, and
(ii) $Q$ is far from hadronic resonances and thresholds.}
\end{quote}
We shall define precisely what `sufficiently inclusive' means later.
Predictions made on the basis of this assumption agree with measurements
(e.g.~made at LEP) within the expected accuracy of the prediction, which
we are to define also later.

Based on this assumption, it makes sense to make predictions with quark
and gluon asymptotic states. However, in QCD the complexity of the
Feynman rules will make higher order computations prohibitive. Indeed,
the largest effort in QCD computations during the past 20 years went
into devising ever more efficient methods to decrease the algebraic
complexity of the computations. This research is driven by the
observation that the QCD Lagrangian is highly symmetric, which has to
be reflected in the final results. Thus the complications somehow appear
mainly because with our rules we artificially introduce complications
at intermediate steps of the computations, which cancels to large extent
in the final formulae.  Learning about the symmetries of QCD is interesting
and useful not only for technical purposes, so let us make an inventory
of those.

\subsection{Symmetries of the classical Lagrangian}

The symmetries can be grouped into two large categories: exact symmetries and
approximate ones. Space-time symmetries are exact. These consist of
invariance against continuous transformations: translations and
Lorentz-transformations (rotations and boosts). In addition
${\cal L}_{\rm cl}$ is invariant under scale transformation:
\[
x^\mu \to \lambda x^\mu
\quad
A_\mu(x) \to \lambda^{-1} A_\mu(\lambda x)
\quad
q(x) \to \lambda^{-3/2} q(\lambda x)
\,,
\]
and conformal transformations, which we do not detail here. The
Lagrangian is also invariant under charge conjugation (C), parity (P)
and time-reversal (T), in agreement with observed properties of strong
interactions (C, P and T violating strong decays are not observed).

We already discussed exact symmetry in colour space: local gauge
invariance. In addition to the classical Lagrangian of \eqn{eq:Lcl}, 
there exists additional gauge invariant dimension-four operator, the
$\Theta$-term:
\begin{equation*}
{\cal L}_\Theta = \frac{\Theta g_s}{32\pi^2}\sum_a
F^a_{\mu\nu}\tilde{F}^{a,\mu\nu}\,,\quad\textrm{with}\quad
\tilde{F}^{a,\mu\nu} =
\frac12\epsilon^{\mu\nu\alpha\beta}F^a_{\alpha\beta}
\,,
\end{equation*}
that violates P and T. As experimentally $\Theta < 10^{-9}$, we set
$\Theta = 0$ in perturbative QCD.

Another interesting feature of ${\cal L}_{\rm cl}$ is that it is almost
supersymmetric. For one massless flavour 
\begin{equation*}
{\cal L}_{\rm cl} = -\frac14 \sum_a F^a_{\mu\nu}F^{a,\mu\nu}
+ \bar{q} \ri \slashed{D} q
\,,
\end{equation*}
which is very similar to the Lagrangian of $N=1$ supersymmetric gauge
theory,
\begin{equation*}
{\cal L}^{\rm SUSY}_{\rm cl} = -\frac14 \sum_a F^a_{\mu\nu}F^{a,\mu\nu}
+ \bar{\lambda} \ri \slashed{D} \lambda
\,.
\end{equation*}
The only difference is that the quark $q$ transforms under the
fundamental, while the gluino $\lambda$ under the adjoint representation
of the gauge group.

An important approximate symmetry of the classical Lagrangian is related
to the quark mass-matrix. Let us introduce the quark flavour triplet 
\[
\psi
 = \left( \begin{array}{c} u\\ d\\ s \end{array} \right)
 = \left( \begin{array}{c} q_1\\ q_2\\ q_3 \end{array} \right)
\,,
\]
with each component being a four-component Dirac spinor, and the combinations
\begin{equation}
P_\pm = \frac12 \big(1\pm \gamma_5\big)
\,.
\label{eq:Ppm}
\end{equation}
The latter are projections:
\[
P_+\,P_- = P_-\,P_+ = 0
\,,\quad
P_\pm^2 = P_\pm
\,,\quad
P_+ + P_- = 1
\,.
\]
It follows from Clifford-algebra that $\gamma_\mu P_\pm = P_\mp
\gamma_\mu$. We define $\psi_\pm = P_\pm \psi$. Using $\gamma_5^2 = 1$,
we find that $\psi_\pm$ are eigenvectors of $\gamma_5$ with $\pm 1$
eigenvalues:
\[
\gamma_5 \psi_\pm = \pm \psi_\pm
\,.
\]
From the definition of the Dirac adjoint, $\overline{\psi} = \psi^\dag
\gamma_0$, we obtain $\overline{\psi_\pm} = \overline{\psi} P_\mp$.
Thus the quark sector of the Lagrangian can be rewritten in terms of the
chiral fields $\psi_\pm$:
\begin{equation*}
\begin{split}
{\cal L}_{\rm cl} &= \overline{\psi}\,\ri\,\gamma_\mu D^\mu\,\psi =
\overline{\psi}(P_++P_-)\,\ri\,\gamma_\mu D^\mu\,(P_++P_-)\psi =
\overline{\psi}P_+\,\ri\,\gamma_\mu D^\mu\,P_-\psi +
\overline{\psi}P_-\,\ri\,\gamma_\mu D^\mu\,P_+\psi =
\\ &=
\overline{\psi}_-\,\ri\,\gamma_\mu D^\mu\,\psi_- +
\overline{\psi}_+\,\ri\,\gamma_\mu D^\mu\,\psi_+ =
{\cal L}_- + {\cal L}_+
\equiv {\cal L}_{\rm L} + {\cal L}_{\rm R}
\,.
\end{split}
\end{equation*}
This decomposition would not work if the gluon field in the covariant
derivative were not Lorentz-vector. In this chiral form the left- and
right-handed fields decouple, so the Lagrangian is invariant under
separate $U(N_{\rm f})$ transformations for the left- and right-handed
fields, \ie under $U_{\rm L}(N_{\rm f})\times U_{\rm R}(N_{\rm f})$,
hence it is called chiral symmetry. Indeed, if $(g_{\rm L}, g_{\rm R})
\in U_{\rm L}(N_{\rm f})\times U_{\rm R}(N_{\rm f})$, then under the
transformation
\[
\psi_{\rm L} \to g_{\rm L} \psi_{\rm L}\,,
\qquad
\overline{\psi}_{\rm L} \to  \overline{\psi}_{\rm L}g^\dag_{\rm L}
\,,
\qquad
g_{\rm R} = 1
\]
${\cal L}_{\rm L}$ remains invariant.  This symmetry is exact if the
quarks are massless. The group elements can be parametrized using
$2 N_{\rm f}^2$ real numbers $\{\alpha, \alpha_a, \beta, \beta_b\}$
($a $, $b=1\,,\dots N_{\rm f}^2 - 1$),
\begin{equation*}
\begin{split}
(g_{\rm L}, g_{\rm R}) &= 
\Bigg(\exp(\ri \alpha) \exp(\ri \beta)
\exp\lp(\ri \sum_a \alpha_a T^a\rp) \exp\lp(\ri \sum_b \beta_b T^b\rp),
\\ &\qquad
\exp(\ri \alpha) \exp(-\ri \beta)
\exp\lp(\ri \sum_a \alpha_a T^a\rp) \exp\lp(-\ri \sum_b \beta_b T^b\rp)\Bigg)
\\ &
\in U_{\rm V}(1)\otimes SU_{\rm L}(N_{\rm f})
\otimes U_{\rm A}(1)\otimes SU_{\rm R}(N_{\rm f})
\,,
\end{split}
\end{equation*}
where the matrices $T^a$ represent the generators of the group
($N_{\rm f}\times N_{\rm f}$ matrices). The transformations
$(\exp\lp(\ri\sum_a\alpha_aT^a\rp)$, $\exp\lp(\ri\sum_a\alpha_aT^a\rp))$,
acting as $\psi \to \exp\lp(\ri \sum_a \alpha_a T^a \mathrm{I}\rp) \psi$,
form a vector subgroup $SU_{\rm V}(N_{\rm f})$. The transformations
$(\exp\lp(\ri\sum_b\beta_bT^b\rp), \exp\lp(-\ri\sum_b\beta_bT^b\rp))$,
acting as $\psi \to \exp\lp(\ri \sum_b \beta_b T^b \gamma_5\rp) \psi$,
however, do not form an axial-vector subgroup because
\[
[T^a\gamma_5, T^b\gamma_5] = \ri \sum_c f^{abc} T^c\mathrm{I}
\qquad
(\gamma_5^2 = \mathrm{I} \ne \gamma_5)\,.
\]

This chiral symmetry is not observed in the hadron spectrum. Therefore,
we assume that vacuum has a non-zero VEV of the light-quark operator,
\[
\left\langle 0| \bar{q} q | 0\right\rangle =
\left\langle 0| \bar{{\rm u}} {\rm u} + \bar{{\rm d}} {\rm d} | 0\right\rangle 
\simeq (250\,{\rm MeV})^3
\,,
\]
a chiral condensate that connect left- and right-handed fields,
\[
\left\langle 0| \bar{q} q | 0\right\rangle =
\left\langle 0| \bar{{\rm q}}_{\rm L} {\rm q}_{\rm R}
+ \bar{{\rm q}}_{\rm R} {\rm q}_{\rm L} | 0\right\rangle 
\,.
\]
The condensate breaks chiral symmetry spontaneously to
$SU_{\rm V}(N_{\rm f})\otimes U_{\rm V}(1)$. This remaining symmetry
explains the existence of good quantum numbers of isospin and baryon
number, as well as the appearance of $N_{\rm f}^2-1 = 8$ massless
mesons, the Goldstone bosons. As non-zero quark masses violate the
chiral symmetry, which is broken spontaneously, the Goldstone bosons
are not exactly massless. Thus we have natural candidates for the
Goldstone bosons: we can identify those with the pseudoscalar meson
octet. In practice, we assume exact chiral symmetry and treat the quark
masses as perturbation. This procedure leads us to chiral perturbation
theory ($\chi$PT) \cite{Gasser:1983yg}, which is capable to predict the
(ratios of) masses of light quarks \cite{Gasser:1982ap,Gasser:1984gg},
scattering properties of pions \cite{Colangelo:2001df} and many more.
Although, $\chi$PT is a non-renormalizable QFT, it can be made
predictive order by order in PT if the measured values of sufficiently
many observables are used to fix the couplings of interaction terms at
the given order.

The QCD Lagrangian was written forty years ago. Since then many attempts
were tried to solve it and mature fields emerged that aim at solving the
theory in a limited range of physical phenomena. For instance, $\chi$PT
is a PT that uses low-energy information (in the MeV range) to explain
the world of hadrons and masses of light quarks. In the same energy
range non-perturbative approaches, notably lattice QCD and sum rules
have been developed for the same purpose. By now it is possible to
explain the light hadron spectrum from first principles using lattice
results \cite{Durr:2008zz}. The main goal at colliders, our focus in
these lectures, is different. We shall prove that PT can give reliable
predictions for scattering processes at high energies, which is the
topic of jet physics.

We have seen that the classical QCD Lagrangian shows many interesting
symmetry properties that can be utilized for (i) easing computations,
(ii) checking results, (iii) hinting on solving QCD. We shall see that
some of these symmetries are violated by quantum corrections, which leads
to important physical consequences. In QCD an important example is
scaling violations. Another example is the axial anomaly which provides
strong constraints on possible QFT's, but it is discussed within the
electroweak theory usually.

\subsection{What is scaling?}
\label{sec:scaling}

Let us consider a dimensionless physical observable $R$ that depends on a
large energy scale $R = R(Q^2)$. Large means that $Q$ is much bigger
than any other dimensionful parameter, for instance, masses of quarks.
Thus we assume that these other dimensionful parameters can be set zero.%
\footnote{We shall study the validity of this assumption in the next
subsection.}
Classically, dim\,$R\,=\,0$ and, since $Q$ is dimensionful, it follows
that $\frac{dR}{dQ}=0$. So $\lim_{Q^{2}\rightarrow\infty}R=$constant,
which is called {\em scaling}.

In these lectures we do not have room for a complete description of
ultraviolet (UV) renormalization of QCD. We simply state that in a
renormalized QFT $R$  depends also on another scale, the
renormalization scale $\mu_\rR$. So
\[
R=\lim_{Q^2\to \infty} 
R\lp(\frac{Q^{2}}{\mu^{2}_{\rR}},\,\as\lp(\mu_\rR^{2}\rp)\rp)\neq 
\textrm{constant},
\]
$R$ need not be a constant. This is called {\em scaling violation}. 
The first term in parenthesis is the only dimensionless combination of
$Q$ and $\mu_\rR$. However, $\mu_\rR$ is arbitrary. If $R$ depended on
$\mu_\rR$, then its value could not be predicted. For simplicity from
now on we drop the subscript ``R'' from $\mu_\rR$. 
As $\mu$ is an arbitrary, un-physical parameter (the classical
Lagrangian did not contain $\mu$), we expect that measurable (physical)
quantities cannot depend on it, which is expressed by the renormalization
group equation (RGE):
\begin{equation*}
0=\mu^{2}\frac{\rm d}{{\rm
d}\mu^{2}}\,R\left(\frac{Q^{2}}{\mu^{2}},\alpha_{\rm
s}\big(\mu^{2}\big)\right)
=\left(\mu^{2}\frac{\partial}{\partial\mu^{2}}
+\mu^{2}\frac{\partial\alpha_{\rm s}}
{\partial\mu^{2}}\frac{\partial}{\partial\alpha_{\rm s}}\right)R
\,.
\end{equation*}
We can simplify this equation a bit by introducing the new variable $t$
and the function $\beta(\alpha_{\rm s})$,
\begin{equation}
t = \ln\frac{Q^2}{\mu^2}
\,,\qquad
\beta(\alpha_{\rm s}) = \mu^2 \frac{\partial \alpha_{\rm s}}
{\partial \mu^2}\bigg|_{\alpha_{\rm s}\:\mathrm{fixed}}
\,.
\label{eq:betafc}
\end{equation}
Then the RGE becomes 
\begin{equation}
\left(-\frac{\partial}{\partial\,t} +\beta(\alpha_{\rm s})
\frac{\partial}{\partial\alpha_{\rm s}}\right)R\big(e^{t},\alpha_{\rm
s}\big)=0
\,.
\label{eq:RGE}
\end{equation}

To present the solution of this partial differential equation, we
introduce the {\em running coupling} $\alpha_{\rm s}(Q^2)$, defined
implicitly by 
\begin{equation}
t=\int_{\alpha_{\rm s}}^{\alpha_{\rm s}\big(Q^{2}\big)}
\!\frac{{\rm d} x}{\beta(x)}\,,\quad \mathrm{with} \quad
\alpha_{\rm s} \equiv \alpha_{\rm s}\big(\mu^{2}\big)
\,,
\label{eq:runningcoupling}
\end{equation}
where $\as \equiv\as\lp(\mu^{2}\rp)$ is an arbitrarily fixed number.
The derivative of \eqn{eq:runningcoupling} with respect to the variable
$t$ gives
\begin{equation*} 
1=\frac{1}{\beta\lp(\as\lp(Q^{2}\rp)\rp)} 
\,\frac{\partial\as\lp(Q^{2}\rp)}{\partial\,t}\,, 
\quad \mbox{which implies}\quad
\beta\lp(\as\lp(Q^{2}\rp)\rp)=\frac{\partial\as\lp(Q^{2}\rp)}{\partial\,t}
\,.
\end{equation*}
The derivative of \eqn{eq:runningcoupling} with respect to $\as$ gives
\begin{equation*} 
0=\frac{1}{\beta\lp(\as\lp(Q^{2}\rp)\rp)} 
\,\frac{\partial\as\lp(Q^{2}\rp)}{\partial\as} 
-\frac{1}{\beta(\as)}\,\frac{\partial\as}{\partial\as}\,, 
\end{equation*}
from which it follows that
\[\frac{\partial\as\lp(Q^{2}\rp)}{\partial\as}= 
\frac{\beta\lp(\as\lp(Q^{2}\rp)\rp)}{\beta(\as)}\,.\]
It is now easy to prove that the value of $R$ for $\mu^2 = Q^2$,
$R\lp(1,\as\lp(Q^{2}\rp)\rp)$ solves
\eqn{eq:RGE}:
\[-\frac{\partial}{\partial\,t}R\lp(1,\as\lp(Q^{2}\rp)\rp)= 
-\frac{\partial\,R}{\partial\as\lp(Q^{2}\rp)} 
\,\frac{\partial\as\lp(Q^{2}\rp)}{\partial\,t}= 
-\beta\lp(\as\lp(Q^{2}\rp)\rp)\,\frac{\partial\,R}{\partial\as\lp(Q^{2}\rp)}\]
and 
\[\beta(\as)\,\frac{\partial}{\partial\as}\,R\lp(1,\as\lp(Q^{2}\rp)\rp)= 
\beta(\as)\,\frac{\partial\as\lp(Q^{2}\rp)}{\partial\as} 
\,\frac{\partial\,R}{\partial\as\lp(Q^{2}\rp)}= 
\beta\lp(\as\lp(Q^{2}\rp)\rp)\,\frac{\partial\,R}{\partial\,\as\lp(Q^{2}\rp)}.\]

It then follows that the scale-dependence  in $R$ enters only through
$\alpha_{\rm s}\big(Q^2\big)$, and we can predict the scale-dependence of
$R$ by solving \eqn{eq:runningcoupling}, or equivalently,
\begin{equation}
\frac{\partial\alpha_{\rm s}\big(Q^{2}\big)}{\partial\,t} =
\beta\left(\alpha_{\rm s}\big(Q^{2}\big)\right)
\,.
\label{eq:runninalpha} 
\end{equation}

So far our analysis was non-perturbative. Assuming that PT is
applicable, which we shall discuss at the end of this subsection, we
may try to solve \eqn{eq:runninalpha} in PT where the $\beta$-function
has the formal expansion:
\begin{equation}
\beta(\alpha_{\rm s}) = -\alpha_{\rm s}\,\sum_{n=0}^{\infty}\beta_{n}
\left(\frac{\alpha_{\rm s}}{4\pi}\right)^{n+1}
\,.
\label{eq:betafunction} 
\end{equation}
The first four coefficients are known from cumbersome computations
\cite{vanRitbergen:1997va}
\begin{equation}
\begin{split}
\beta_{0}&=\frac{11}{3}\,C_{\rA}-\frac{4}{3}\,T_{R}\,n_{f}\,>\,0\,, 
\qquad
\beta_{1}=\frac{34}{3}C_{\rA}^{2}-\frac{20}{3}C_{\rA}T_{R}\,n_{\rm f} 
-4C_{\rF}T_{R}\,n_{\rm f}
\,, \\[2mm]
\beta_{2}&= \frac{2857}{2}- \frac{5033}{18} n_{\rm f}+
\frac{325}{54} n_{\rm f}^2
\,,
\qquad
\beta_{3}= 29243 - 6946.3 n_{\rm f} +405.9 n_{\rm f}^2 +1.5 n_{\rm f}^3
\,.
\label{eq:betan} 
\end{split}
\end{equation}
The first two coefficients in the expansion of the $\beta$ function are
independent of the renormalization scheme. The second two coefficients in
\eqn{eq:betan} are valid in the \msbar\ renormalization scheme.%
\footnote{As we have not gone through the renormalization procedure, we
cannot define precisely what we mean by `renormalization scheme'.
Various schemes differ by finite renormalization of the parameters and
fields in the Lagrangian.}

Another often used convention is
\begin{equation} 
\beta(\as)=-b_{0}\as^{2}\lp[1+\sum_{n=1}^{\infty}b_{n}\,\as^{n}\rp], 
\label{eq:perturbativebeta}
\end{equation}
where $b_{0}=\frac{\displaystyle \beta_{0}}{\displaystyle 4\pi}$ and
$b_{0}b_{1}=\frac{\displaystyle \beta_{1}}{\displaystyle \lp(4\pi\rp)^2}$, thus
$b_{1}=\frac{\displaystyle \beta_{1}}{\displaystyle 4\pi\beta_{0}}$.

If $\as\lp(Q^{2}\rp)$ is small we can truncate the 
series. The solution at leading-order (LO) accuracy is
\begin{eqnarray} 
Q^{2}\,\frac{\partial\as}{\partial\,Q^{2}}= 
\frac{\partial\as}{\partial\,t}=-b_{0}\as^{2} 
\quad
\Rightarrow 
\quad
-\lp[\frac{1}{\as\lp(Q^{2}\rp)}-\frac{1}{\as\lp(\mu^{2}\rp)}\rp]=-b_{0}t 
\quad
\Rightarrow 
\quad
\as\lp(Q^{2}\rp)= 
\frac{\as\lp(\mu^{2}\rp)}{1+b_{0}t\,\as\lp(\mu^{2}\rp)} 
\,, 
\label{eq:alphaLO}
\end{eqnarray}
which gives $\as\lp(Q^{2}\rp)$ as a function of
$\as\lp(\mu^{2}\rp)$ if both are small;
$\as\lp(\mu^{2}\rp)$ is a number to be measured. We observe that:
\begin{equation}
\as\lp(Q^{2}\rp)\stackrel{Q^{2}\rightarrow\,\infty}{\longrightarrow} 
\,\frac{1}{b_{0}t}\,\stackrel{Q^{2}\rightarrow\,\infty}{\longrightarrow}\,0\,.
\label{eq:asymptoticfreedom}
\end{equation}
This behaviour is called {\em asymptotic freedom}. The sign of $b_0$
(positive for QCD) plays a crucial role in establishing whether or not
a theory is asymptotically free. If it is, then the use of PT is
justified: the higher $Q^2$, the smaller the coupling. The coefficient
$b_0$ is easiest to compute in {\em background field gauge}
\cite{Abbott:1980hw} where only three graphs contribute, the quark and
gluon loops:
\begin{equation}
\includegraphics[width=0.4\linewidth]{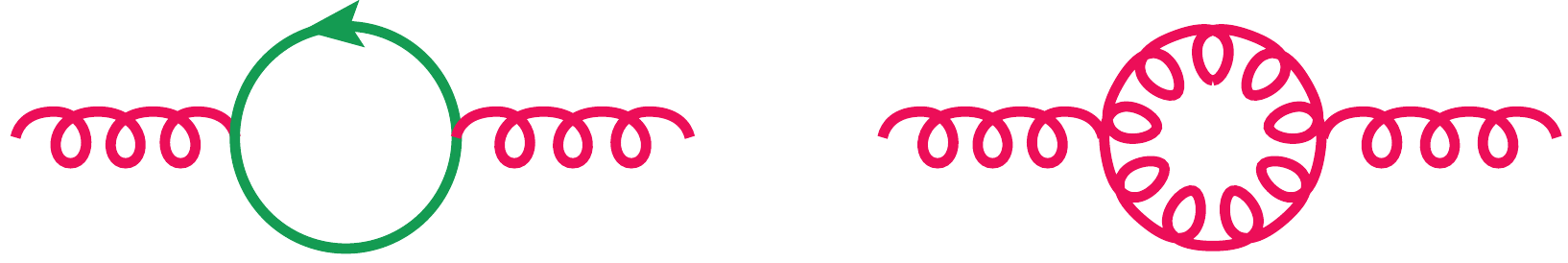} \,,
\label{eq:beta0graphs}
\end{equation}
and a similar ghost loop.
The contribution of the quark loop is negative $-\frac43 T_{\rm R} n_{\rm
f}$, while that of the gluon+ghost loop is positive $\frac{11}{3}C_{\rm A}$.
(We knew the colour factors immediately, only the coefficients have to be
computed!) The net result is positive up to $n_{\rm f} < 17$ in QCD. 
In 2004 D.J.~Gross, H.D.~Politzer and F.~Wilczek were awarded the Nobel
prize for their discovery of asymptotic freedom in QCD
\cite{Gross:1973id,Politzer:1973fx}.

Clearly, it is the gluon self-interaction that makes QCD perfect in 
PT. In QED, in the absence of photon self-interaction, $b_0 < 0$, hence
the coupling increases with energy, but remains perturbative up to the
Planck scale ($\simeq 10^{19}$\,GeV) where we expect that any known
physics breaks down.  

Asymptotic freedom gives rationale to perturbative QCD, but we shall see
that LO accuracy is not enough.  The analysis is also simple at
next-to-leading order (NLO):
\begin{equation*} 
\lp[\as^{2}(1+b_{1}\as)\rp]^{-1}\,\frac{\partial\as}{\partial\,t}=-b_{0}. 
\end{equation*}
$\as\lp(Q^{2}\rp)$ is then given implicitly by the equation
\begin{equation*} 
\frac{1}{\as\lp(Q^{2}\rp)}-\frac{1}{\as\lp(\mu^{2}\rp)} 
+b_{1}\ln\frac{\as\lp(Q^{2}\rp)}{\as\lp(\mu^{2}\rp)} 
-b_{1}\ln\frac{1+b_{1}\as\lp(Q^{2}\rp)}{1+b_{1}\as\lp(\mu^{2}\rp)}=bt\,, 
\end{equation*}
which can be solved numerically.

Using the formula for the sum of the geometric series, $(1+x)^{-1} =
\sum_{j=0}^\infty (-x)^{j}$ and recalling \eqn{eq:alphaLO}, we find that 
the running coupling sums logarithms,
\[
R\big(1, \alpha_{\rm s}\big(Q^{2}\big)\big) = R_0 +
R_1 \alpha_{\rm s}\big(\mu^{2}\big)
\sum_{j=0}^{\infty} \big[-\alpha_{\rm s}\big(\mu^{2}\big)b_0 t\big]^j
\,.
\]
The NLO term $R_2 \alpha_{\rm s}^2$ gives logarithms with one power less
in each term. \\

\begin{wrapfigure}{ht}{0.5\linewidth}
\includegraphics[width=1.0\linewidth]{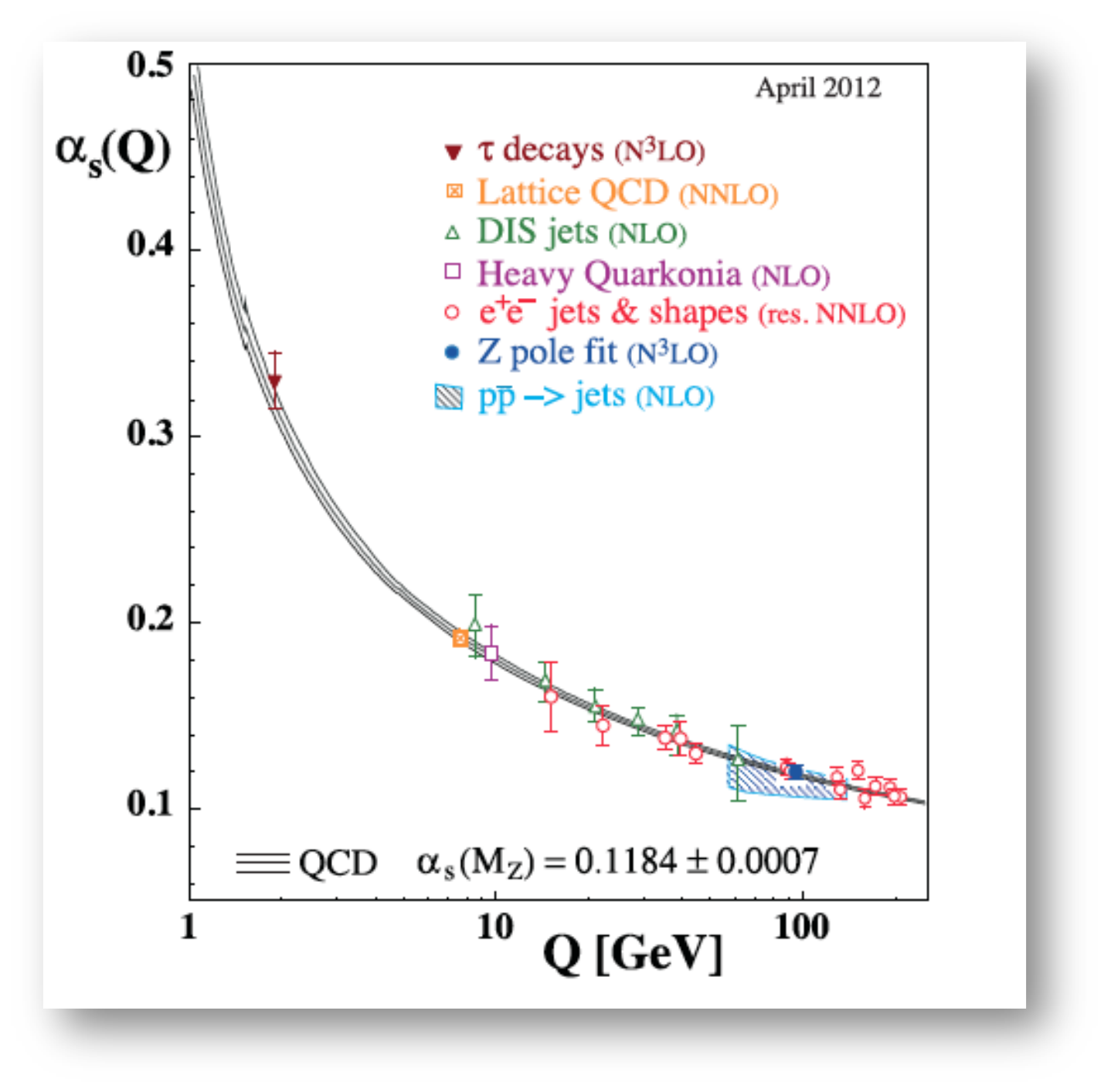}
\caption{Results of measurements of the strong coupling at different
scales. The theoretical prediction with four-loop running, fixed at
$\mu=M_{Z}$ is marked as `QCD'.
\vspace*{-120pt}
}
\label{fig:alphasrun}
\end{wrapfigure}

\subsection{Measuring $\as(\mu^2)$}
\label{sec:measuringalphas}

We know $\as\lp(Q^{2}\rp)$ if $\as\lp(\mu^{2}\rp)$ is known. We
therefore, have to measure $\as$ at some scale $\mu$.  The perturbative
solution of the renormalization group equation (RGE, \eqn{eq:RGE}) is
never unique.  The difference between two solutions at
$\mathcal{O}\lp(\as^{n}\rp)$ is suppressed by $\as$, i.e. at
$\mathcal{O}\lp(\as^{n+1}\rp)$.  Nevertheless, this difference can lead
to significant difference in $\as\lp(Q^{2}\rp)$ if $\mu^{2}$ and
$Q^{2}$ are far from each other, which is important in present day
precision measurements.  Therefore, the scale $\mu$ is chosen to be
$\mu=M_{Z}$ because $M_{Z}=91,2$\,GeV is not far from the scales where
$\as(Q^2)$ is used in current experimental analyses. In
\figs{fig:alphasrun}{fig:alphasMZ} we show the present status of $\as$
measurements from Ref.~\cite{Beringer:1900zz}.

\begin{figure}[ht]
\includegraphics[width=0.4\linewidth]{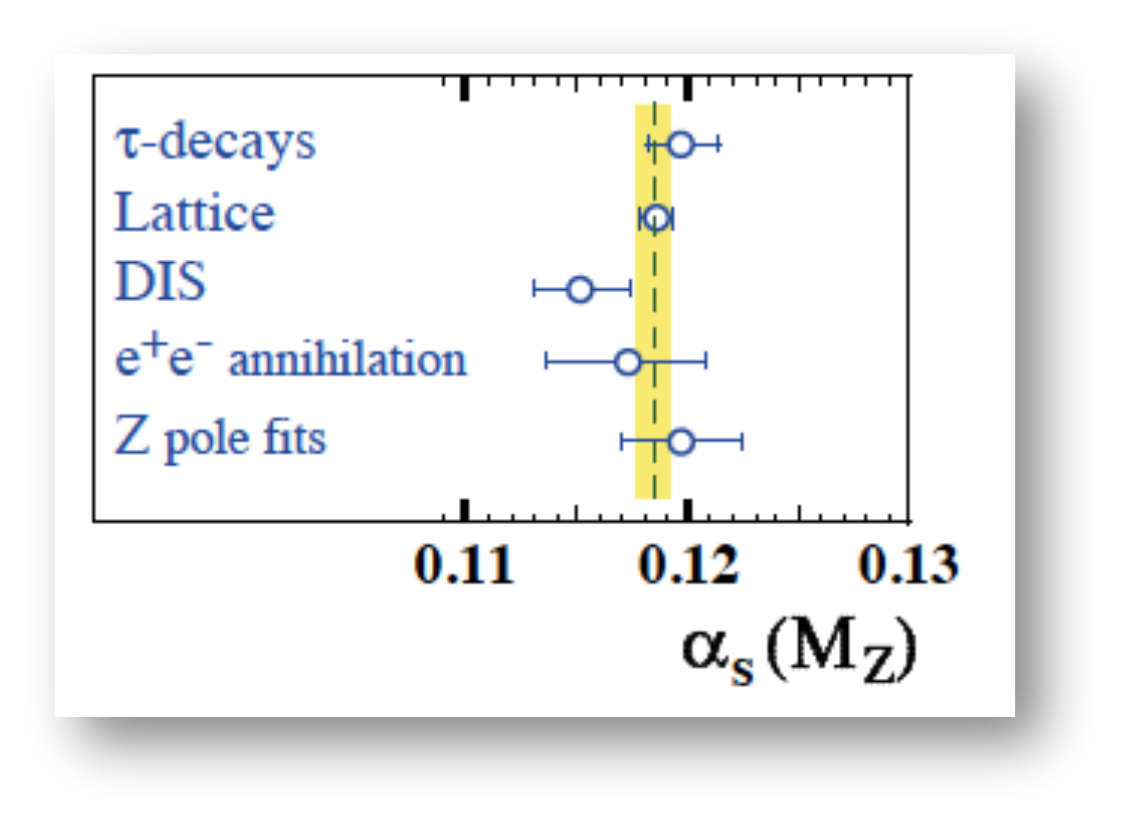}
\vspace*{-10pt}
\begin{flushleft}
\begin{minipage}{0.4\linewidth}
\caption{
Results of different measurements of the strong coupling run to
$\mu=M_{Z}$
}
\label{fig:alphasMZ}
\end{minipage}
\end{flushleft}
\end{figure}

\vspace{-4pt}
\noindent\rule{\textwidth}{1pt}

 
Another approach to solving the RGE is to introduce a refence scale
$\Lambda$ by 
\[
\ln\frac{Q^2}{\Lambda^2} = \int_{\alpha_{\rm s}\big(Q^{2}\big)}^{\infty}
\!\frac{{\rm d} x}{\beta(x)}
\,.
\]
The scale $\Lambda$ indicates where the coupling becomes strong. The
following exercise is to explore the characteristics of this choice.

\begin{exe}

The running of the strong coupling constant is given by \eqn{eq:betafc}. 
The perturbative expansion of the QCD beta function is given by 
\eqn{eq:perturbativebeta} 
with 
$b_0,b_1\geq0$. 
Determine 
(i) the expression for the coupling constant in leading order ($b_0\not = 
0$, $b_1=0$) and the corresponding scale $\Lambda_0$ (see below) 
(ii) the expression for the coupling constant in next-to-leading order 
($b_0\not = 0$, $b_1\not=0$) and the corresponding scale $\Lambda_1$ (see 
below). 
 
Hints:\\ 
\begin{enumerate} 
\item Solve the differential equation for $\alpha(\mu)$; you'll get an 
integration constant. 
\item Express your result in the form 
\begin{equation*} 
\alpha(\mu) =\frac{1}{K\ln (\frac{\mu^2}{\Lambda_0^2})} 
\end{equation*} 
where $K$ is a constant. 
\item Solve the differential equation using $b_1\not=0$ 
\begin{equation*} 
\int\d\alpha \frac{1}{-b_0 \alpha^2-b_1 
\alpha^3}=\frac{b_0+b_1 \alpha\log(\alpha)-b_1 
\alpha\log(b_0+b_1\alpha)}{b_0^2 \alpha}+K 
\end{equation*} 
\item This time the solution cannot be solved for $\alpha$ analytically. 
One can nevertheless find an approximate solution by expanding $\alpha$ in 
$\log \frac{\mu^2}{\Lambda_1^2}$. The constant $K$ is not equal to the 
one in the first part of this exercise. 
\item Cast your equation for $\alpha$ into the form 
\begin{equation*} 
\alpha=\frac{1}{K\ln\frac{\mu^2}{\Lambda_1^2}}\quad\frac{1}{1+c_1\frac{\ln(c_2+b_0\alpha)}{\ln\frac{\mu^2}{\Lambda_1^2}}} 
\end{equation*} 
with a suitable choice of $\Lambda_1$. 
\item Expand the right hand side of your equation in $t=\frac{1}{\ln 
\frac{\mu^2}{\Lambda_1^2}}$ and keep only the first order term. Use the 
expansion 
\begin{equation*} 
\frac{1}{1+C_1\,t\ln(C_3\frac{1}{t}+C_2)}=1+t\,C_1\ln(\frac{1}{t})+O(t)\,. 
\end{equation*} 
\end{enumerate} 
 
\end{exe}
\rule{\textwidth}{1pt}

\subsection{Quark masses and massless QCD}

Quark masses $m_{q}$ are {\em parameters} of $\mathcal{L}_{QCD}$ like the
gauge coupling, which need to be renormalized. In QED the electron mass
is measured in the laboratories at $\mu_{\rm R}^{2}=0$ (classical
limit). We cannot similarly isolate a quark at $\mu_{\rm R}^{2}=0$ (at
low scale quarks are confined). Instead, we can perform a similar RGE
analysis as with $\as$. For simplicity we assume one quark flavour with
mass $m$, which is yet another dimensionful parameter, so the RGE becomes: 
\begin{equation}
\lp[\mu^{2}\frac{\partial}{\partial\mu^{2}}
+ \beta(\as)\frac{\partial}{\partial\as}
- \gamma_{m}(\as)m\frac{\partial}{\partial\,m}\rp]
\,R\lp(\frac{Q^{2}}{\mu^{2}},\as,\frac{m}{Q}\rp)=0\,, 
\label{eq:rengroupeqmass} 
\end{equation}
where $\gamma_{m}$ is called the mass anomalous dimension and the minus
sign before $\gamma_{m}$ is a convention. In PT we can write the mass
anomalous dimension as
\begin{equation*}
\gamma_{m}(\as)\,=\,c_{0}\as\lp(1+c_{1}\as+\mathcal{O}\lp(\as^{2}\rp)\rp), 
\end{equation*}
with known coefficient up to $c_3$. At NLO accuracy we need only
$c_{0}=\frac{1}{\pi}$ and $c_{1}=\frac{303-10\,n_{f}}{72\pi}$.
As $R$ is dimensionless, the dependence on the dimensionful
parameters has to cancel
\begin{equation}
\lp(Q^{2}\frac{\partial}{\partial\,Q^{2}}
+ \mu^{2}\frac{\partial}{\partial\mu^{2}}
+ m^{2}\frac{\partial}{\partial\,m^{2}}\rp)
\,R\lp(\frac{Q^{2}}{\mu^{2}},\as,\frac{m}{Q}\rp)=0\,. 
\label{eq:dimensionles} 
\end{equation}
The difference of \eqns{eq:rengroupeqmass}{eq:dimensionles} gives the
dependence of $R$ on $Q$:
\begin{equation}\label{rengroupeqmasssol} 
\lp[Q^{2}\frac{\partial}{\partial\,Q^{2}}
- \beta\lp(\as\rp)\frac{\partial}{\partial\as}
+ \lp(\frac{1}{2}+\gamma\lp(\as\rp)\rp)m\frac{\partial}{\partial\,m}\rp]
\,R\lp(\frac{Q^{2}}{\mu^{2}},\as,\frac{m}{Q}\rp)=0\,. 
\end{equation}
This equation is solved by introducing the running mass (in
addition to the running coupling) $m\lp(Q^{2}\rp)$ obeying
\begin{equation} 
Q^{2}\frac{\partial\,m}{\partial\,Q^2}=-\gamma_{m}\lp(\as\rp)m\lp(Q^{2}\rp)\,,
\label{eq:runningmass}
\quad\Rightarrow\quad
\ln\frac{m\lp(Q^{2}\rp)}{m\lp(\mu^{2}\rp)}\,=\, 
-\int_{\mu^{2}}^{Q^{2}}\frac{\rd q^{2}}{q^{2}} 
\,\gamma_{m}\lp(\as\lp(q^{2}\rp)\rp)
\,.
\end{equation}
Exponentiating, changing integration variable from $q^2$ to $\alpha_{\rm
s}$ and using the definition of the $\beta$ function, we obtain
\begin{equation}\label{massrunning} 
m\lp(Q^{2}\rp)\,=\,m\lp(\mu^{2}\rp) 
\exp\lp[-\int_{\as\lp(\mu^{2}\rp)}^{\as\lp(Q^{2}\rp)}\! 
\rd\as\frac{\gamma_{m}\lp(\as\rp)}{\beta\lp(\as\rp)}\rp] 
\stackrel{Q^{2}\rightarrow\infty}{\longrightarrow}\,0\,, 
\end{equation}
which means that asymptotically free QCD is a massless theory at
asymptotically large energies. At LO in PT theory the solution of
(\ref{massrunning}) is given by
\begin{equation*} 
-\frac{\gamma_{m}\lp(\as\rp)}{\beta\lp(\as\rp)}= \frac{c_{0}}{b_{0}\as}
\quad\Rightarrow\quad\,
m\lp(Q^{2}\rp)=
\overline{m}\lp[\as\lp(Q^{2}\rp)\rp]^{\frac{c_{0}}{b_{0}}}\,,
\end{equation*} 
where we introduced the abbreviation $\overline{m} = 
{m\lp(\mu^{2}\rp)\lp[\as\lp(\mu^{2}\rp)\rp]^{-\frac{c_{0}}{b_{0}}}}$.
At NLO the solution becomes
\begin{equation*} 
m\lp(Q^{2}\rp)=
\overline{m}\lp[\as\lp(Q^{2}\rp)\rp]^{\frac{c_{0}}{b_{0}}}
\lp(1+\frac{c_{0}}{b_{0}}\lp(c_{1}-b_{1}\rp)
\lp( \as\lp(Q^{2}\rp)-\as\lp(\mu^{2}\rp)\rp)
+\mathcal{O}\lp(\as^{2}\rp)\rp)\,.
\end{equation*} 
In terms of the running coupling and mass,
$R\left(1,\alpha_{\rm s}(Q^2),\frac{m(Q^2)}{Q}\right)$ is a solution of
\eqn{rengroupeqmasssol}, proven similarly as
$R\left(1,\alpha_{\rm s}(Q^2)\right)$ being the solution of \eqn{eq:RGE}.
Expanding around $m(Q^2) =0$, we obtain
\begin{equation} 
R\left(1, \alpha_{\rm s}(Q^2), \frac{m(Q^2)}{Q}\right)
= R\left(\frac{Q^2}{\mu^2}, \alpha_{\rm s}, 0\right) +
\sum_{n=1}^\infty \frac{1}{n!} \left(\frac{m(Q^2)}{Q}\right)^n
R^{(n)} \left(\frac{Q^2}{\mu^2}, \alpha_{\rm s}, 0\right)
\,.
\label{eq:RGEsolution}
\end{equation}  
We see from \eqn{eq:RGEsolution} that derivative terms are suppressed
by factors of $1/Q^n$ at large $Q^2$.  From the dependence of $R$ on
$\frac{m\lp(Q^{2}\rp)}{Q}$ we can conclude that the effect of mass is
suppressed at high $Q^{2}$ by its physical dimension and also by its
anomalous dimension, which justifies the assumption about negligible
quark masses.  The expansion in \eqn{eq:RGEsolution} has a deeper
consequence. The dimensionless observable $R$ may depend on
$\ln \frac{m\lp(Q^{2}\rp)}{Q}$ that can become large when $Q^2$ is large.
If we want to avoid such large logarithms, we should consider physical
observables (that is physically measurable quantities) that have a
finite zero-mass limit.

\section{Predictions in perturbative QCD}

\begin{wrapfigure}{r}{0.6\linewidth}
\vspace*{-15pt}
\includegraphics[width=1.0\linewidth]{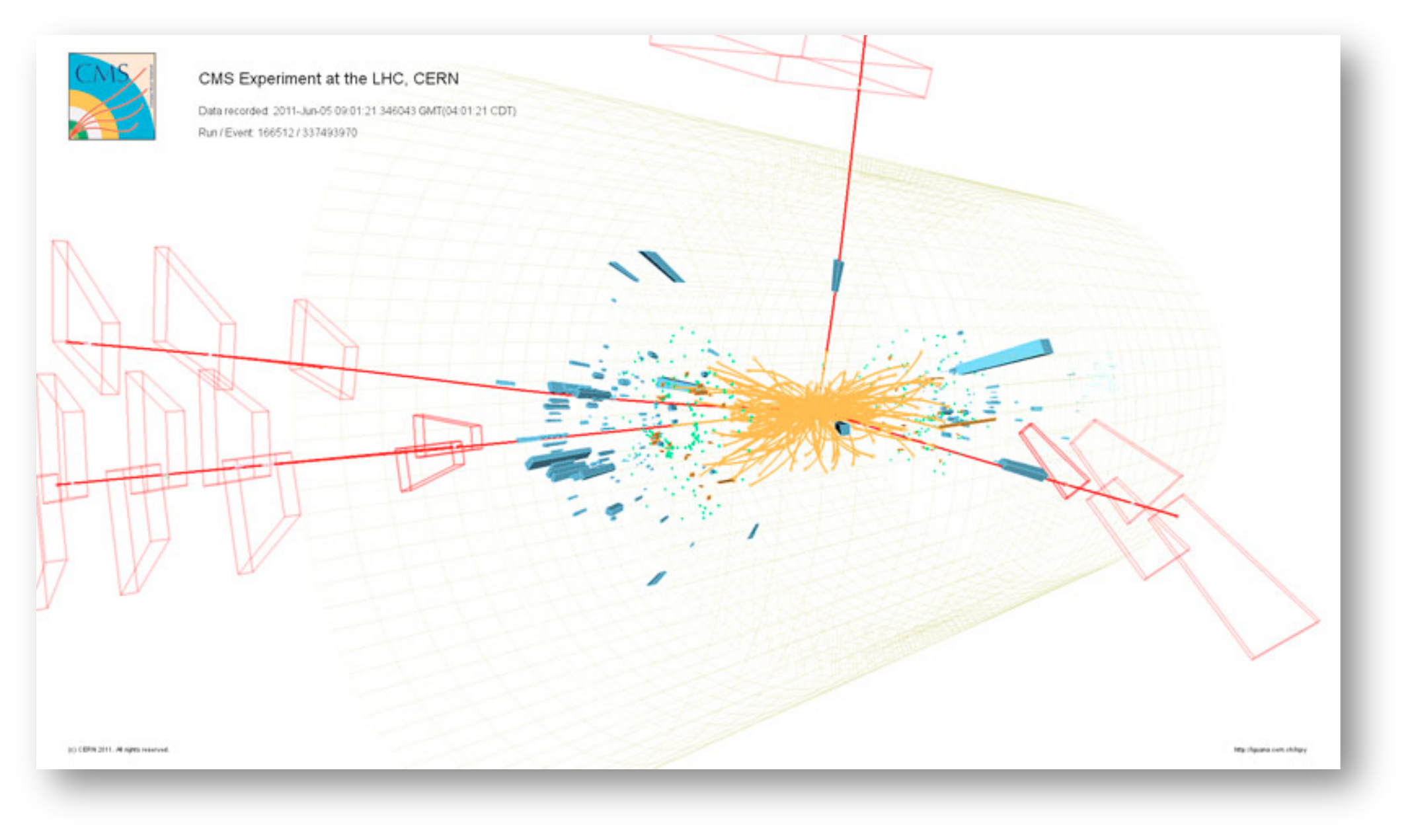}
\vspace*{-20pt}
\caption{An event with four hard muons in the CMS detector
~}
\label{fig:cmsmumumumu}
\end{wrapfigure}
In a typical collider experiment we collect collision events with
something interesting in the final state. For instance, in searching for
the Higgs boson, events with four hard muons such as in
\fig{fig:cmsmumumumu} are interesting. Counting the event rate of such
events we obtain measured cross sections, which compare to theoretical
predictions. Following our assumption about the use of low-order
perturbative predictions in QCD, for such comparisons we need
predictions for cross sections with partons. We start with the simplest
possible case when partons appear only in the final state:
electron-positron annihilation into hadrons (and possibly other
particles).

Let us consider a measurable quantity $O$, that has non-vanishing value for
at least $m$ partons in the final state.  At LO accuracy the basic
formula for the differential cross section in $O$ is
\begin{equation}
\begin{split}
\frac{{\rm d}\sigma}{{\rm d} O} =
\mathcal{N}\,\int\!&{\rm d}\phi_{m}(p_{1},\ldots,p_{m};Q)
\,\frac{1}{S_{\left\{m\right\}}}
\left|\mathcal{M}_{m}(p_{1},\ldots,p_{m})\right|^{2}
\delta\big(O-O_{m}^{(m)}(p_{1},\ldots,p_{m})\big)
\,,
\label{eq:xsection}
\end{split}
\end{equation}
where $\mathcal{N}$ contains non-QCD factors (\eg the flux factor),
${\rm d}\phi_{m}$ is the phase space of $m$ particles, $S_m$ is a
symmetry factor, $\left|\mathcal{M}_{m}(p_{1},\ldots,p_{m})\right|^{2}$
is the squared matrix element (SME), and $O_{m}^{(m)}$ is the value of
$O$ computed from the $m$ final state momenta. The integration is usually
done by Monte Carlo integration and the hard part of the computation is
to obtain the SME. In these lectures we can compute hardly any SME
explicitly. Fortunately, there are freely available computer programs
\cite{Jamin:1991dp,Alwall:2011uj,Pukhov:2004ca,Pukhov:1999gg,Hahn:2000kx}
that can be used to check the formulae. Even more, these programs can
often be used to obtain the cross sections at LO accuracy, too.

We now use \eqn{eq:xsection} to make predictions for the cross section
of electron-positron annihilation into hadrons.  
 
\subsection{$R$ ratio at lowest order}
\label{sec:RLO}

\begin{figure}[h!] 
  \includegraphics[width=1.0\linewidth]{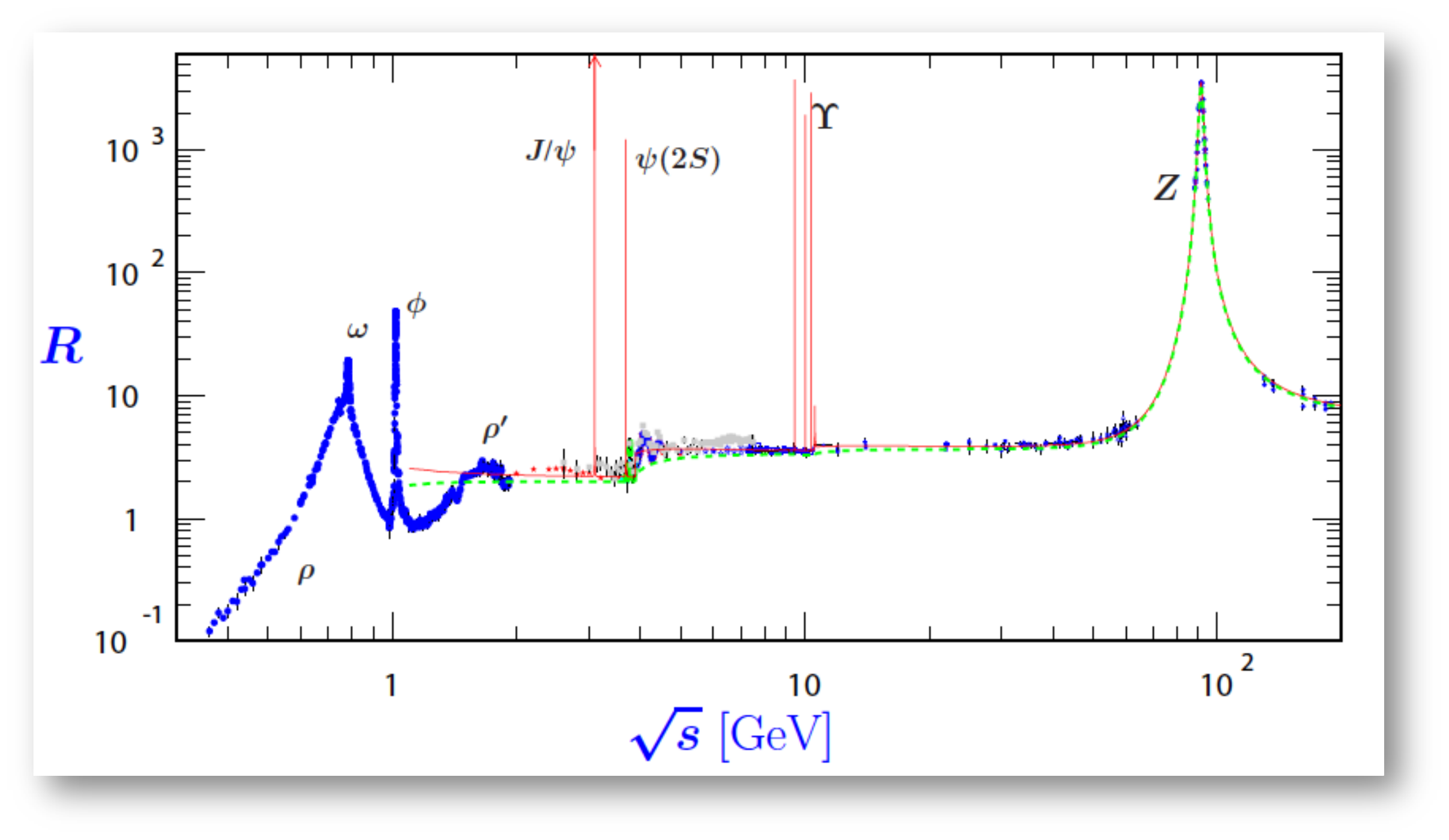}
  \caption{Experimental measurements of the $R$-ratio as a function of
the total centre-of mass energy (taken from Ref.~\cite{Beringer:1900zz}).}
\label{fig:Rratio} 
\end{figure}

The leading-order (LO) perturbative contribution to the cross section 
$\sigma\lp(e^{+}e^{-}\rightarrow\textrm{hadrons}\rp)$ is 
$e^{+}e^{-}\rightarrow\,q\bar{q}$. The calculation is like 
in the case of $e^{+}e^{-}\rightarrow\mu^{+}\mu^{-}$, supplemented with 
colour and fractional electric charge of $q_{j}$. The colour diagram is a 
loop in the fundamental representation which corresponds to a factor 
$N_{\rc}$ as we have seen in the previous chapter. While the 
annihilation into $\mu^{+}\mu^{-}$ contains only one flavour in the 
final state, quarks can have three, four or five flavours depending on 
the centre-of-mass energy.%
\footnote{The sixth flavour, the top is so heavy that it cannot 
contribute at CM energies attained in $e^+e^-$ experiments so far.} 
We have, therefore, to sum over all possible flavours which can appear. 
The ratio of the two cross sections is thus given by 
\begin{equation} 
R\,\equiv\,\frac{\sigma\lp(e^{+}e^{-}\rightarrow\,q\bar{q}\rp)} 
{\sigma\lp(e^{+}e^{-}\rightarrow\,\mu^{+}\mu^{-}\rp)}\,=\, 
\lp(\sum_{q}\,e_{q}^{2}\rp)\,N_{\rc}, 
\label{eq:R} 
\end{equation} 
where $e_{u}=e_{c}=\frac{2}{3}$ and 
$e_{d}=e_{s}=e_{b}=-\frac{1}{3}$. If we 
consider only the up, down, strange and charm quarks 
$\sum_{q}e_{q}^{2}=2\frac{4}{9}+2\frac{1}{9}=\frac{10}{9}$. 
Considering also the bottom quark 
$\sum_{q}e_{q}^{2}=\frac{11}{9}$. This step-wise increasing 
behavior of the $R$-ratio was observed (see \fig{fig:Rratio}), 
providing an experimental confirmation of the existence of 3 families 
of quarks and of the $SU(N_\rc)$ gauge-symmetry of QCD with $N_\rc = 3$. 
 
According to our basic assumption, pQCD cannot give predictions for the
resonances in \fig{fig:Rratio}. However, there is one exception, the
impressive $Z$ peak. The LO prediction uses the cross section for the
$e^+ e^- \to q \bar{q}$ process. A $2\to 2$ process has only a single free
kinematic variable, the scattering angle $\vartheta$. In the full SM
the differential cross section for electron-positron annihilation into
a massless and colourless fermion pair $f\bar{f}$  is obtained from the
square of a single Feynman graph,
$\bigg|$~~~~~~~~~~~~~~~~~~~~$\bigg|^2$ \,, as

\vspace*{-31pt} ~\hspace*{94pt}
\includegraphics[width=0.1\linewidth]{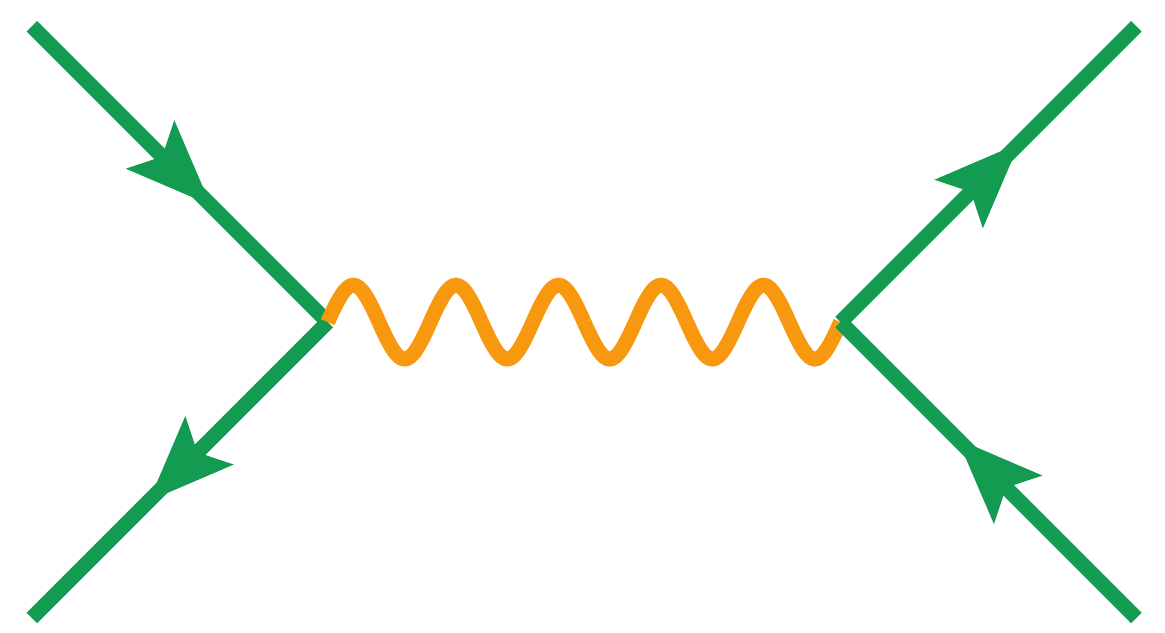}
\begin{equation}
\frac{{\rm d}\sigma}{{\rm d}\cos\vartheta} =
\frac{\pi \alpha^2}{2 s}\bigg\{\big(1+\cos^2\vartheta\big)\bigg[e_f^2
+\big(A_e^2+V_e^2\big)\big(A_f^2+V_f^2\big)
\frac{\kappa^2 s^2}{(s-M_Z^2)^2+\Gamma_Z^2 M_Z^2}
+ \dots\bigg]\bigg\}\,,
\label{eq:epemtoffbar}
\end{equation}
where we neglected terms that vanish at centre-of-mass energy
$\sqrt{s} = M_Z$, or after integration. $e_f$, $A_f$ and $V_f$ denote the
fractional charge, axial-vector and vector electroweak couplings of the
fermions and $\kappa = \sqrt{2} G_{\rm F} M_Z^2/(16 \pi \alpha_{\rm em})
\simeq 0.374$ is a number. Well below the $Z$ peak the $Z$ propagator
becomes negligible and the total cross section is obtained by
integrating over the scattering angle and we find the LO prediction
$\sigma_{\rm LO}(s) = \sigma_0(s) e_f^2$, where 
$\sigma_0(s) = \frac{4 \pi \alpha^2}{3 s}$. On the $Z$ peak the same
integration results in $\sigma_{\rm LO}(M_Z^2) = \sigma_0(M_Z^2)
\bigg[e_f^2 +\big(A_e^2+V_e^2\big)\big(A_f^2+V_f^2\big)
\kappa^2\frac{M_Z^2}{\Gamma_Z^2}\bigg]$. Then we can make prediction for
the hadronic $R$ ratio at LO accuracy by simply counting the contributing
final states and relating their total charge factors to that of the muon
and find
$R_{\rm LO} = 3 \sum_q e_q^2$ away from the $Z$ peak and 
$R_{Z,{\rm LO}} = 3 \sum_q (A_q^2 + V_q^2)/(A_\mu^2 + V_\mu^2)$ on the
$Z$ peak. The factor three is due to the three colours of quarks.
Considering five quark flavours, \ie $m_{\rm b} << s << m_{\rm t}$, we
find $R_{\rm LO} = 11/3$ and $R_{Z,{\rm LO}} = 20.09$. We have seen on
\fig{fig:Rratio} that 11/3 is fairly close to the measured value away from
the $Z$ peak. The measured value of $R_Z$ at LEP is $R_Z = 20.79 \pm
0.04$ \cite{ALEPH:2005ab}. The LO prediction works amazingly well. The
3.5\% difference is mainly due to QCD radiation effects that we call
NLO corrections. Our next goal is to understand the origin of those
corrections.

\vspace{-4pt}
\noindent\rule{\textwidth}{1pt}

 
\begin{exe} 
Derive the result in \eqn{eq:epemtoffbar} (at least below the $Z$ peak,
where you consider only photon intermediate state) and integrate it
over $\vartheta$. 
\end{exe} 
 
\begin{exe} 
Use Mathematica and the Package Tracer.m (or FORM) to 
compute the following traces: 
\begin{eqnarray*} 
&&\tr\left(\slashed{p}_2\gamma^\nu(\slashed{p}_1-\slashed{k}_1)\gamma^\mu\slashed{p}_1\gamma_\mu(\slashed{p}_1-\slashed{k}_1)\gamma_\nu\right)\\ 
&&\tr\left(\gamma^{\mu_1}\gamma^{\mu_2}\gamma^{\mu_3}\gamma^{\mu_4}\gamma^{\mu_5}\gamma^{\mu_6}\gamma^{\mu_7}\gamma^{\mu_8}\gamma^{\mu_9}\gamma^{\mu_{10}} 
\gamma_{\mu_1}\gamma_{\mu_2}\gamma_{\mu_3}\gamma_{\mu_4}\gamma_{\mu_5}\gamma_{\mu_6}\gamma_{\mu_7}\gamma_{\mu_8}\gamma_{\mu_9}\gamma_{\mu_{10}}\right) 
\end{eqnarray*} 
 
\end{exe} 
 
\noindent\rule{\textwidth}{1pt}

\subsection{Ultraviolet renormalization of QCD} 
 
The strong coupling is rather large as compared to the other couplings in 
the SM, and as a result, the QCD radiative corrections are also large. 
Therefore, it is always important to compute at least the NLO accuracy,
but if possible, even higher order corrections.%
\footnote{There is even a more severe reason that we shall discuss later.}
 
The computation of QCD radiative corrections is technically quite 
involved and a good organization of the calculations is very important. 
Thus, first we introduce some notation. The tensor product of the ket 
vectors $\lp|c_{1},\dots,c_{m}\ra\otimes\lp|s_{1},\dots,s_{m}\ra$ 
denotes a basis vector in colour and helicity space, 
$\lp|\mathcal{A}_{m}\lp(p_{1},\dots,p_{m}\rp)\ra$ is a 
state vector of $n = m-2$ final-state particles in colour and 
helicity space. The amplitude for producing $n$ final-state 
particles of colour $\lp(c_{1},\dots,c_{n}\rp)$, spin 
$\lp(s_{1},\dots,s_{n}\rp)$, momentum $\lp(p_1,\dots,p_n\rp)$ is 
\begin{equation} 
\mathcal{A}_{m}^{c_{1}\dots\,c_{m},s_{1}\dots\,s_{m}}\lp(p_{1}\dots\,p_{m}\rp)
\equiv\la 
c_{1}\dots\,c_{m}\rp|\otimes\la 
s_{1}\dots\,s_{m}|\mathcal{A}_{m}\lp(p_{1},\dots,p_{m}\rp)\ra 
\end{equation} 
($m=n+2$), so 
\begin{equation} 
\sum_{\textrm{colour}}\sum_{\textrm{helicity}}\lp|\mathcal{A}_{m}^{{c_{1}\dots\,c_{m},s_{1}\dots\,s_{m}}}
\lp(\{p_i\}\rp)\rp|^{2}=
\la\mathcal{A}_{m}\lp(\{p_i\}\rp)| 
\mathcal{A}_{m}\lp(\{p_i\}\rp)\ra. 
\end{equation} 
The loop expansion in terms of the bare coupling, \ie  the coupling that 
appears in the classical Lagrangian, 
$\gs^{\lp(0\rp)}\,\equiv\,\sqrt{4\pi\as^{\lp(0\rp)}}$ 
is: 
\begin{equation} 
\lp|\mathcal{A}_{m}\ra =
\lp(\frac{\as^{\lp(0\rp)}\,\mu^{2\epsilon}}{4\pi}\rp)^{\frac{q}{2}}
\lp[\lp|\mathcal{A}_{m}^{\lp(0\rp)}\ra
+ \lp(\frac{\as^{\lp(0\rp)}\,\mu^{2\epsilon}}{4\pi}\rp)
\lp|\mathcal{A}_{m}^{\lp(1\rp)}\ra
+ \mathcal{O}\lp((\as^{\lp(0\rp)})\rp)^{2}\rp]\,, 
\end{equation} 
where $q\,\in\,\mathbb{N}$, $\mu$ is the dimensional regularization
scale, introduced to keep $\as^{(0)}$ dimensionless in $d\,=\,4-2\epsilon$
dimensions. The exponent $\frac{q}{2}$ in the prefactor takes account
of the power of $\as$ at LO, the loop-expansion is an expansion in the
strong coupling $\as$. For instance, $q=0$ for $e^+e^- \to q \bar{q}$,
while $q=1$ for $e^+e^- \to q \bar{q}g$. The tree amplitude
$\lp|\mathcal{A}_{m}^{(0)}\ra$ is finite, while the one-loop correction
$\lp|\mathcal{A}_{m}^{\lp(1\rp)}\ra$ is divergent in $d=4$ dimensions,
which is manifest in terms of $1/\epsilon^2$ and $1/\epsilon$ poles if
dimensional reglarization is used. These poles have both ultraviolet
(UV) and infrared (IR) origin.

The UV poles can be removed by multiplicative redefinition of the fields
and parameters in the Lagrangian, systematically order by order in PT.
This is a hard task even at one loop, but presently known up to four loops
\cite{Chetyrkin:2004mf} -- a truly remarkable computation! It turns
out that when computing scattering amplitudes in massless QCD at
one-loop accuracy, the renormalization amounts to the simple substitution 
\begin{equation}\label{alphas1loop} 
\as^{\lp(0\rp)}\mu^{2\epsilon}\,\longrightarrow
\,\as\lp(\mu_{\rR}^{2}\rp)\mu_{\rR}^{2\epsilon}S_{\epsilon}^{-1}
\lp[1-\frac{\as\lp(\mu_{\rR}^{2}\rp)}{4\pi}\,\frac{\beta_{0}}{\epsilon}
+\mathcal{O}\lp(\as^{2}\rp)\rp]\,,
\end{equation} 
with 
$S_{\epsilon}=\frac{\lp(4\pi\rp)^{\epsilon}}{\Gamma\lp(1-\epsilon\rp)}$. 
Note that on the left of this substitution $\mu$ is the dimensional
regularization scale to keep $\as^{\lp(0\rp)}$ dimensionless, while on the
right $\mu_{\rR}$ is the renormalization scale. We discussed in 
\sect{sec:measuringalphas} when we extract $\as$ from measurements, we
have to define $\mu_{\rR}$.  The dimensional regularization scale turns
into the renormalization scale through the substitution (\ref{alphas1loop}).

Why does the substitution (\ref{alphas1loop}) work? Each Feynman graph
consists of vertices with propagators connecting those and external
lines. Moreover, 
\begin{itemize} 
\item[$\bullet$]{each vertex receives a factor $Z_{g}$ (or $Z_{g}^{2}$
for quartic vertex) and factors of $\sqrt{Z_{i}}$, ($i=q,\,A$) for each
field connected to the vertex,}
\item[$\bullet$]{each propagator of field $i$ receives a factor of $Z_{i}^{-1}$,} 
\item[$\bullet$]{each external leg of field $i$ receives a factor of $Z_{i}^{-\frac{1}{2}}$.} 
\end{itemize} 
Thus the renormalization field factors cancel from each 
graph and only the charge renormalization ($Z_{g}$) is 
needed in practice! This can be seen as a consequence of 
the fact that in massless QCD the only free parameter besides the 
gauge-fixing parameter $\lambda$ is $\as$. The scattering amplitudes 
are physical, and any physical quantity has to be independent of 
$\lambda$, so the only remaining parameter, which the amplitudes may 
depend on, is the coupling. The renormalization factor $Z_{g}$ is most 
easily computed in background field gauge, defined by 
\begin{equation} 
\mathcal{L}_{\rm GF}=-\frac{1}{2\lambda}\sum_{a} 
\lp(\partial^{\mu}A_{\mu}^{a} 
+g\,f^{abc}\,\mathcal{A}_{\mu}^{b}A^{\mu\,c}\rp)^{2}, 
\end{equation} 
where $\mathcal{A}_{\mu}^{b}$ is a background field and 
$A_{\mu}^{c}$ describes the quantum fluctuations on this 
background. It can be shown \cite{Abbott:1980hw}
that in this gauge the field and coupling renormalization factors are
related by the Ward identity $Z_{A}^{-\frac{1}{2}}=Z_{g}$, 
and $Z_{A}$ can be computed from loop insertions into the 
propagator shown in (\ref{eq:beta0graphs}).

The simple substitution rule (\ref{alphas1loop}) for the coupling leads to
a simple shift in the amplitude. As
\[
\left[1-\frac{\alpha_{\rm s}\big(\mu_{\rm R}^{2}\big)}{4\pi}
\,\frac{\beta_{0}}{\epsilon}\right]^{\frac{q}2} =
1-{\frac{q}2} \frac{\alpha_{\rm s}\big(\mu_{\rm R}^{2}\big)}{4\pi}
\,\frac{\beta_{0}}{\epsilon} + O(\alpha_{\rm s}^2)
\,
\]
we obtain for the renormalized amplitude $\big|\mathcal{M}_{m}\big\rangle$
\begin{equation}
\begin{split}
\big|\mathcal{M}_{m}\rangle &=
\left(\frac{\as\left(\mu_\rR^{2}\right)\mu_\rR^{2\epsilon}}{4\pi}\,
S_{\epsilon}^{-1}\right)^{\frac{q}{2}}
\!\left(\big|\mathcal{M}_{m}^{\left(0\right)}\rangle
+ \frac{\as\left(\mu_\rR^{2}\right)}{4\pi}\,S_{\epsilon}^{-1}
  \big|\mathcal{M}_{m}^{\left(1\right)}\rangle\right)\quad q\in\mathbb{N},
\\
\big|\mathcal{M}_{m}^{(0)}\big\rangle &=
\big|\mathcal{A}_{m}^{(0)}\big\rangle ,\quad
\big|\mathcal{M}_{m}^{(1)}\big\rangle =
\mu_{\rm R}^{2\epsilon}\big|\mathcal{A}_{m}^{(1)}\big\rangle
- \frac{q}{2}\,\frac{\beta_{0}}{\epsilon}\,S_{\epsilon}
\big|\mathcal{A}_{m}^{(0)}\big\rangle
\,.
\label{eq:Mm}
\end{split}
\end{equation}


The renormalized theory is UV finite, yet
$\lp|\mathcal{M}_{m}^{\lp(1\rp)}\ra$ is still infinite in $d=4$
dimensions, as it is divergent also in the infrared. {\em After} UV
renormalization is achieved we can use dimensional regularization to
regulate the amplitudes in the IR by continuing into $d>4$
($\epsilon<0$). The integrals that are scaleless in $d=4$ have
$\lp(q^{2}\rp)^{-\epsilon}$ mass dimension in $d=4-2\epsilon$
dimensions. Therefore, in the massless limit all integrals can depend
only on momentum invariants raised to a positive fractional power
$\lp(\epsilon\,<\,0\rp)$. We conclude that when all external
invariants vanish, the continued integral must also vanish (``scaleless
integrals vanish in dimensional regularization'').

For {\em IR-safe} observables these IR poles vanish and we can set $d=4$
at the end of the computations,
and we obtain the UV finite, IR regularized SME that can be used to
compute cross sections.

\vspace{-4pt}
\noindent\rule{\textwidth}{1pt}

 
\begin{exe} 
 
Compute the contribution to the beta function from the fermion loop:

\vspace*{-24pt} ~\hspace*{340pt}
\includegraphics[width=0.15\linewidth]{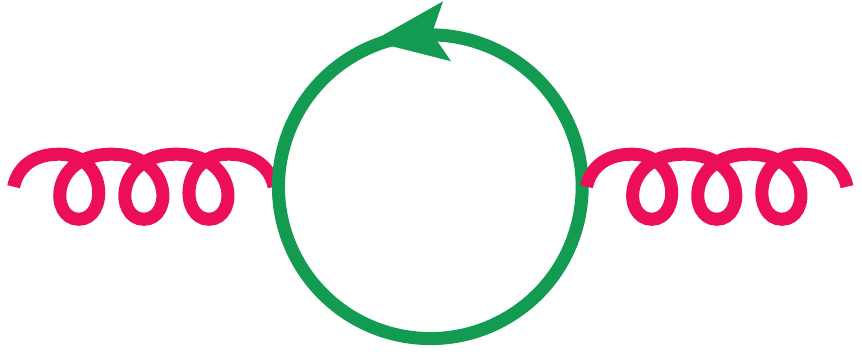}
\begin{enumerate} 
\item Write down carefully the amplitude and compute the trace. 
\item The following types of integrals occur: 
\begin{equation} 
I_2^{\mu}=\int\!\frac{\rd^d\ell}{(2\pi)^d} 
\frac{\ell^\mu}{\ell^2\left(\ell-p\right)^2}
\,,\qquad
I_2^{\mu\nu}=\int\!\frac{\rd^d\ell}{(2\pi)^d} 
\frac{\ell^\mu\ell^\nu}{\ell^2\left(\ell-p\right)^2} 
\end{equation} 
Express these as linear combination of 
\begin{equation} 
I_2(p)=\int\!\frac{\rd^d\ell}{(2\pi)^d}\frac{1}{\ell^2\left(\ell-p\right)^2}\,. 
\end{equation} 
\item Obtain $I_2$ from 
\begin{equation} 
I_{2}(p,m)=\int\!\frac{\rd^{d}l}{\lp(2\pi\rp)^{d}} 
\,\frac{1}{\lp[\lp(l-p\rp)^{2}-m^{2}\rp]l^{2}} 
=\frac{\ri}{\lp(4\pi\rp)^{2-\epsilon}} 
\Gamma\lp(\epsilon\rp)\lp(p^{2}\rp)^{-\epsilon} 
\int_{0}^{1}\!\rd x 
\lp(x\frac{m^{2}}{p^{2}}-x\lp(1-x\rp)-\ri\varepsilon\rp)^{-\epsilon} 
\label{eq:I2scalar} 
\end{equation} 
and find the divergent pieces. 
\end{enumerate} 
The contribution to $\beta_0$ is the coefficient of the $1/\epsilon$ pole
without the coupling factor.
 
\end{exe} 
\noindent\rule{\textwidth}{1pt}

\subsection{$R$ ratio at NLO accuracy} 
\label{sec:RNLO}
 
This is by far the simplest example of computing QCD radiative
corrections. As we saw in \sect{sec:RLO} it requires the total hadronic
cross section that depends only on a single kinematic invariant, the total
centre-of-mass energy $\sqrt{s}$. As a result, the emerging integrals in
this computation can be evaluated exactly. Nevertheless, the complete
computation is still too lengthy, and we shall be able to present the main
step and filling the details is left to the student.

There are two kinds of corrections that contribute at NLO accuracy. One is
the {\em real correction}, with an additional gluon in the final state, so
the SME is computed from Feynman graphs as

$\big\langle\mathcal{M}_{3}^{(0)}\big|\mathcal{M}_{3}^{(0)}\big\rangle=
\bigg|$~~~~~~~~~~~~~~~~~~~~~~~~~~~~~~~~~~~~~~~~$\bigg|^2$ \,, which gives
an $\mathcal{O}(\alpha_{\rm s})$ correction. The other kind of
contribution

\vspace*{-30pt} ~\hspace*{76pt}
\includegraphics[width=0.2\linewidth]{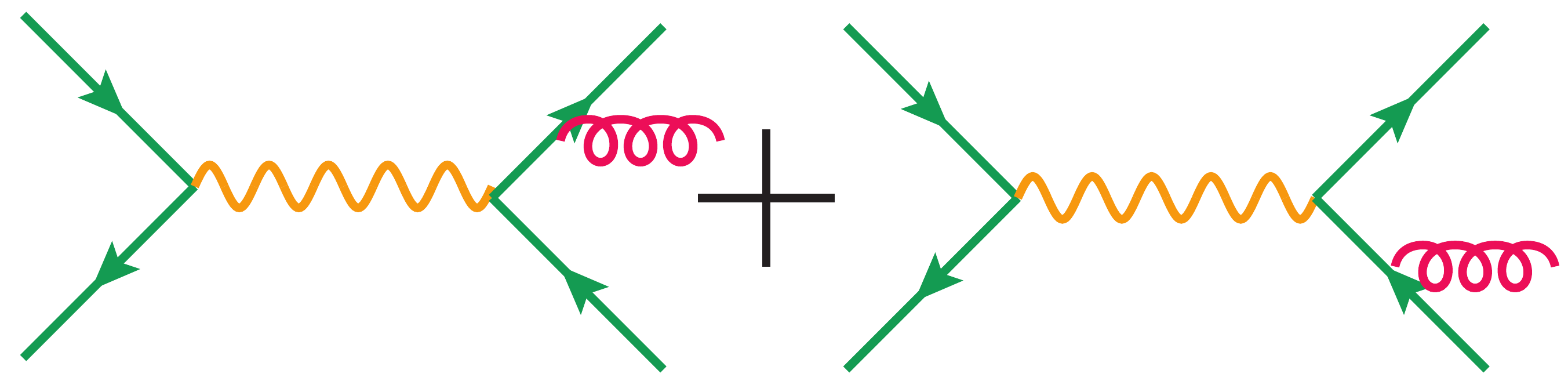}

\noindent is the {\em virtual correction}, with an additional gluon
providing a loop in the final state,

$\big\langle\mathcal{M}_{2}^{(1)}\big|\mathcal{M}_{2}^{(0)}\big\rangle
+\big\langle\mathcal{M}_{2}^{(0)}\big|\mathcal{M}_{2}^{(1)}\big\rangle=
2 \rR\re \bigg\langle$~~~~~~~~~~~~~~~~~~~~\bigg|~~~~~~~~~~~~~~~~~~~~$\bigg\rangle$ \,.

\vspace*{-30pt} ~\hspace*{175pt}
\includegraphics[width=0.1\linewidth]{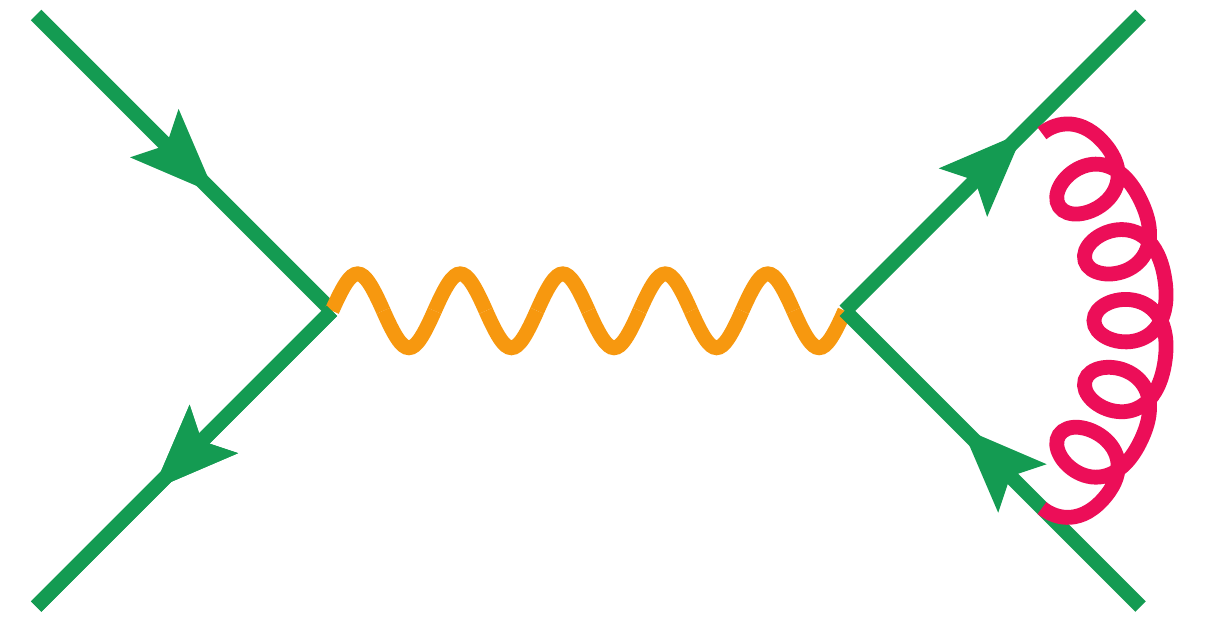}
~~
\includegraphics[width=0.1\linewidth]{figures/Rtree.pdf}

The real correction has three particles in the final state.
The three-particle phase space has five independent variables: two
energies and three angles. As we are looking for the total cross section, we
integrate over the angles and use $y_{ij} = (p_i+p_j)^2/s = 2 p_i\cdot
p_j/s$ scaled two-particle invariants to write both the phase space and
the SME. Momentum conservation implies $1 = (p_1+p_2+p_3)^2/s = y_{12} +
y_{13} + y_{23}$. The complete real contribution to the total cross
section is 
\begin{equation}
\sigma^{\rm R} = \sigma_0 R_0
\int_0^1\!{\rm d}y_{13}
\int_0^1\!{\rm d}y_{23}
\,C_{\rm F}\,\frac{\alpha_{\rm s}}{2\pi}
\left(
\frac{y_{23}}{y_{13}}
+ \frac{y_{13}}{y_{23}}
+ \frac{2y_{12}}{y_{13}y_{23}}
\right)
\Theta(1-y_{13}-y_{23})
\,.
\label{eq:sigmaR}
\end{equation}
\begin{wrapfigure}{l}{0.3\linewidth}
\vspace*{-10pt}
\includegraphics[width=1.0\linewidth]{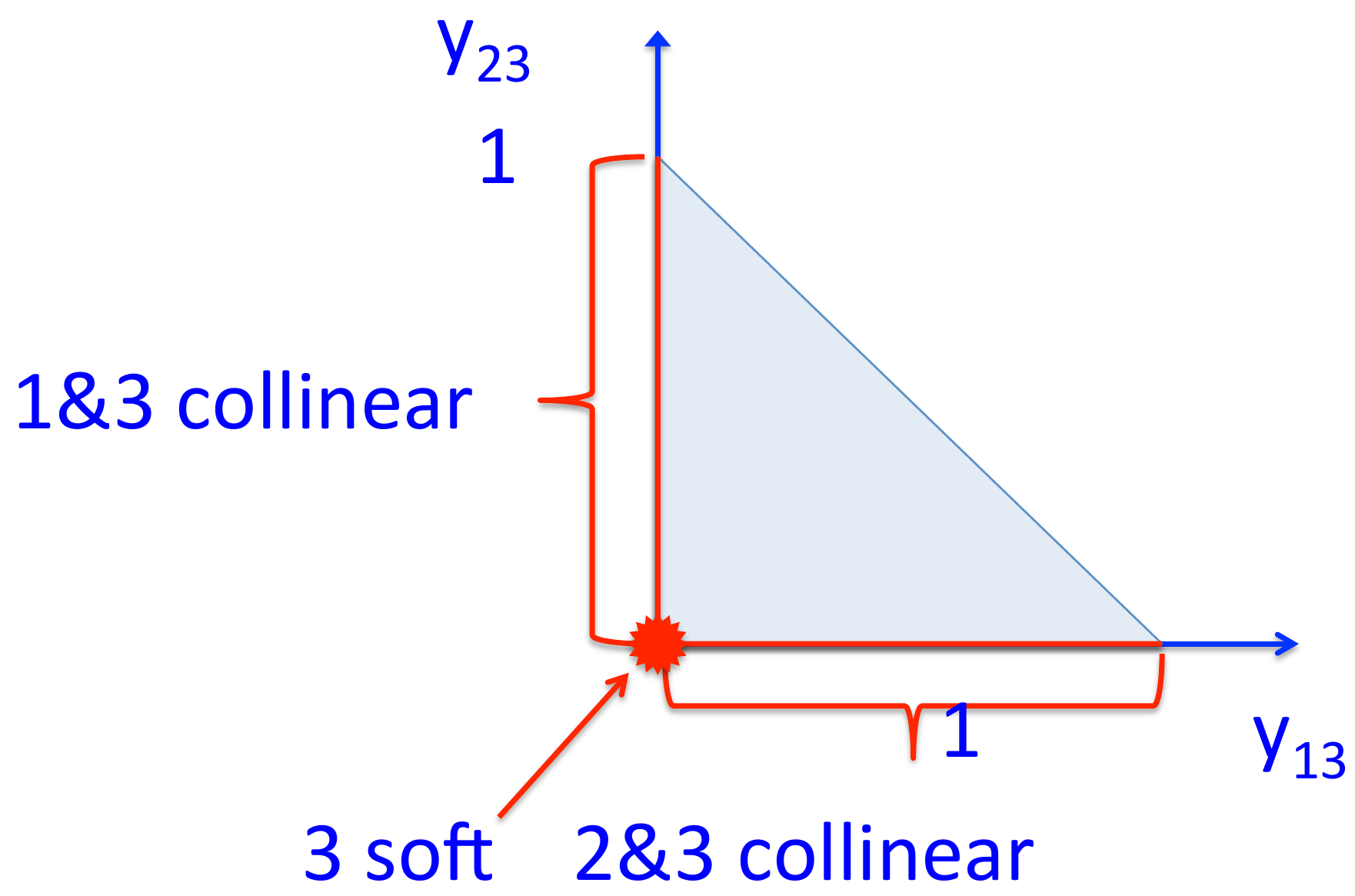}
\vspace*{-15pt}
\caption{Region of integration for real correction
~}
\label{fig:region}
\end{wrapfigure}
This integral is divergent along the boundaries at $y_{13} = 0$, $y_{23}
=0$ as well as in the point $y_{13}y_{23} = 0$, so the singularities
are in the IR parts of the phase space. As 
$y_{i3} s = 2 E_i E_3 (1-\cos \vartheta_{i3})$, the divergence occurs
either when $E_3 \to 0$, which is called {\em soft-gluon} singularity,
or when $\vartheta_{i3} \to 0$, which is called {\em collinear}
singularity (the gluon is collinear to either of the quarks). The
region of integration with the singular places is shown in
\fig{fig:region}.

To make sense of the integral, we use dimensional regularization, which
amounts to the computation of the phase space and the SME in
$d = 4 - 2\epsilon$ dimensions. The result is
\begin{eqnarray}
\sigma^{\rm R}(\epsilon) \aand= \sigma_0 R_0 H(\epsilon)
\int_0^1\!\frac{{\rm d}y_{13}}{y_{13}^\epsilon}
\int_0^1\!\frac{{\rm d}y_{23}}{y_{23}^\epsilon}
\Theta(1-y_{13}-y_{23})
\\ &&\qquad\qquad\times
\,C_{\rm F}\frac{\alpha_{\rm s}}{2\pi}
\left[(1-\epsilon)
\left(
\frac{y_{23}}{y_{13}}
+ \frac{y_{13}}{y_{23}}
\right)
+ \frac{2y_{12}}{y_{13}y_{23}}
-2\epsilon
\right]
\,,
\nonumber
\label{eq:sigmaRdimreg}
\end{eqnarray}
where $H(\epsilon) = 1 + \mathcal{O}(\epsilon)$ (the exact form of this
function will turn out to be irrelevant). The integrals can be evaluated
exactly, but actually the Laurent-expansion around $\epsilon = 0$ is
sufficient,
\begin{equation}
\sigma^{\rm R}(\epsilon) = \sigma_0 R_0 H(\epsilon)
\,C_{\rm F}\,\frac{\alpha_{\rm s}}{2\pi}
\left[ \frac{2}{\epsilon^2} + \frac{3}{\epsilon} + \frac{19}{2}-\pi^2
+ \mathcal{O}(\epsilon)
\right]
\,.
\label{eq:intsigmaRdimreg}
\end{equation}

The computation of the virtual correction is even more cumbersome due to
the loop integral. We present only the result:
\begin{equation}
\sigma^{\rm V}(\epsilon) = \sigma_0 R_0 H(\epsilon)
\,C_{\rm F}\,\frac{\alpha_{\rm s}}{2\pi}
\left[-\frac{2}{\epsilon^2} - \frac{3}{\epsilon} - 8 + \pi^2
+ \mathcal{O}(\epsilon)
\right]
\,,
\label{eq:intsigmaVdimreg}
\end{equation}
We now see that the {\em sum of the real and virtual contribution is
finite in $d=4$}, so for the sum we can set $\epsilon \to 0$ and find the
famous $\alpha_{\rm s}/\pi \simeq 0.037$ correction:
$R = R_0 \big(1+\frac{\alpha_{\rm s}}{\pi}+\mathcal{O}(\alpha_{\rm s}^2)\big)$.
The correction is the same for $R_Z$.

Actually there is a much easier way of computing the radiative corrections
to the total cross section from the imaginary part of the hadronic vacuum
polarization, using the optical theorem ($\sigma \propto \im
f(\gamma\to\gamma))$). The state of the art is $R$ at
$\mathcal{O}(\alpha_{\rm s}^4)$ \cite{Baikov:2012er}. The result of the
computation at next-to-next-to-leading order (NNLO) accuracy,
\begin{equation}
\begin{split}
\frac{R}{R_0} =
1  + c_1 \alpha_{\rm s}(\mu) 
    & + \left[c_2+c_1 b_0 \ln \frac{\mu}{Q^2}\right] \alpha_{\rm s}(\mu)^2
\\ &  + \left[c_3+\left(2 c_2 b_0 + c_1 b_1 + c_1 b_0^2 \ln \frac{\mu^2}{Q^2}\right)
  \ln \frac{\mu^2}{Q^2} \right] \alpha_{\rm s}(\mu)^3
+\mathcal{O}(\alpha_{\rm s}^4)
\label{eq:Rnnnlo}
\end{split}
\end{equation}
satisfies the RGE to order $\alpha_{\rm s}(\mu)^4$. The coefficients
$c_1=1/\pi$, $c_2 = 1.409/\pi^2$, $c_3 = -12.85/\pi^3$ suggest that the
perturbation series is convergent. Our more complicated way of computing
$R$ is instructive for our studies in the next section.

\begin{wrapfigure}{r}{0.5\linewidth}
\vspace*{-10pt}
\includegraphics[width=1.0\linewidth]{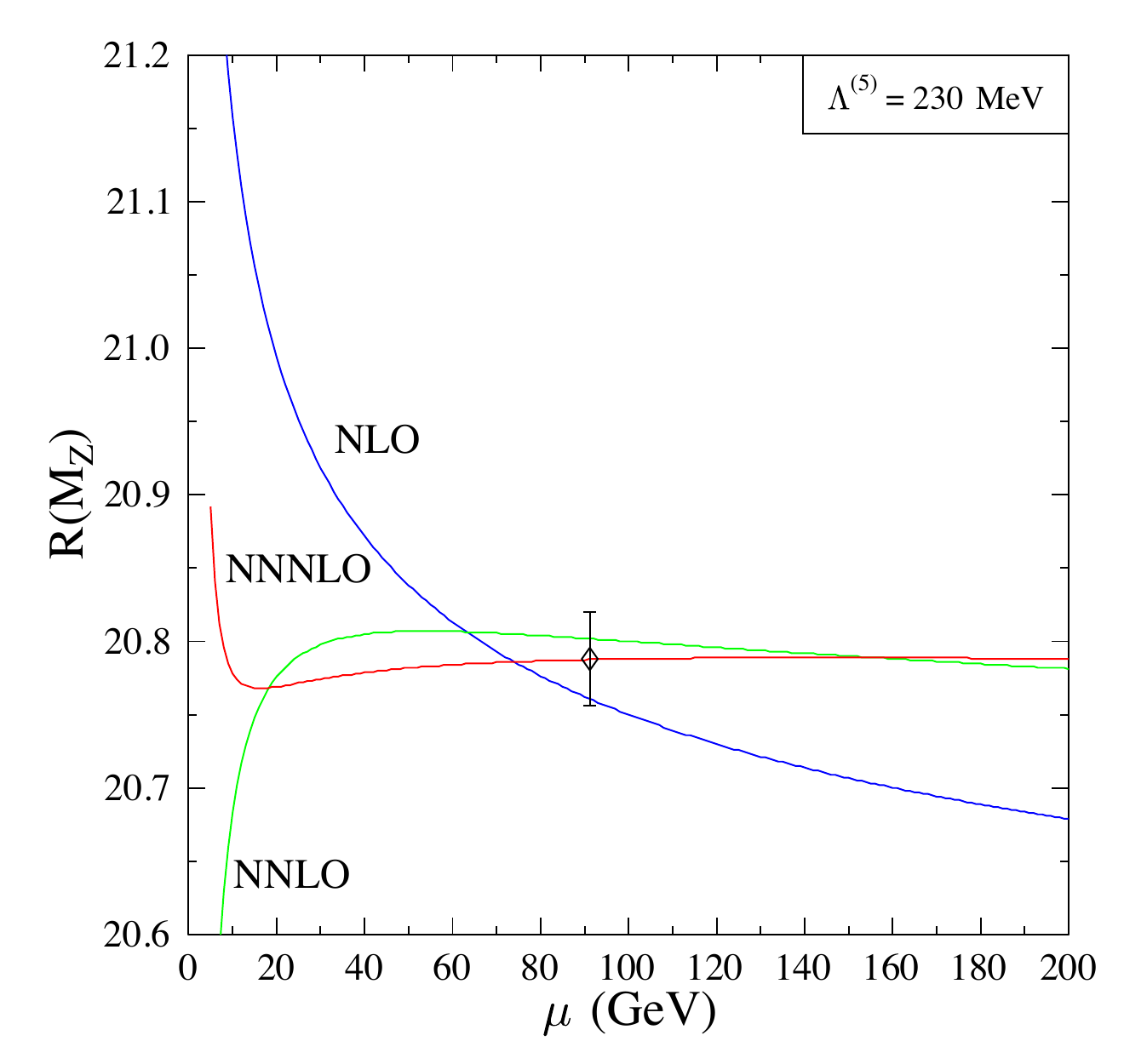}
\vspace*{-15pt}
\caption{Dependence on the renormalization scale of the hadronic ratio on
the $Z$ pole
~}
\label{fig:RZ}
\end{wrapfigure}
The predictions at the first three fixed orders in PT for $R_Z$ are shown
in \fig{fig:RZ}. The $R$ ratio at LO accuracy does not depend on the strong
coupling, hence it is independent of the scale. The figure is meant to show
the general pattern of QCD predictions which, with the exception of the
$R$ ratio, depend on the scale already at LO. The NLO curve shows the
typical feature of LO predictions: it depends on the renormalization
scale in a monotonically decreasing way.  As this scale is unphysical,
in principle, its value can be arbitrary.  Thus the prediction at LO is
in general only an order of magnitude indication of the cross section,
but not a precision result. (In the case of the hadronic ratio the QCD
corrections are actually quite small as compared to many other QCD
cross sections and the precision is actually better than usual.) As a
result, if we want to make reliable predictions in pQCD, the NLO
accuracy (NNLO for $R$) is absolutely necessary unless we have some way
to fix the scale.

However, there is no theorem that tells us the proper scale choice. The
usual practice is to set the scale at a characteristic physical scale of
the process. A reasonable assumption that the strength of the QCD
interaction for a process involving a momentum transfer $Q$ is given by
$\alpha_{\rm s}(Q)$, so $\mu = Q$ is the proper scale choice, to minimize
logarithmic contributions $\ln(\mu/Q)$ in higher-order terms. For
instance, in case of electron-positron annihilation the total
centre-of-mass energy is the usual choice, while for a jet cross
section in proton-proton collisions the transverse momentum of the jet%
\footnote{We discuss jets in the next section.} is used. The
application of this recipe appears clear as long as there is only one
hard scale in the process. In the state of the art computations there
are complex processes with several scales and it is not obvious which one
to choose. For instance, in vector boson hadroproduction in association
with $m$ jets ($m \leq 5$) \cite{Bern:2013gka}, in addition to the
transverse momentum of the vector boson
$E_{V,\perp} = \sqrt{p_{V,x}^2 + p_{V,y}^2}$
there are the transverse momenta of the jets.  In this example, the
choice $\mu=E_{\perp}^V$ was found to result in a badly behaving
perturbation series with corrections driving the $E_\perp$ distribution
of the second hardest jet at NLO accuracy even unphysically negative for
$m=3$ and $E_\perp>475$\,GeV at the LHC. Choosing a dynamical scale,
set event by event, appears a better choice. For instance, half the total
transverse energy of the final-state particles (both QCD partons and
leptons from the decay of the vector boson), $\mu = \hat{H}_{\rm T}/2$,
leads to much milder scale dependence and a similar shape of the
distributions at LO and NLO accuracies.

There are suggestions on making educated guesses for the best scale.
Among those are the principle of fastest apparent convergence (FAC),
that of minimal sensitivity (PMS), or the BLM scale choice
\cite{Grunberg:1980ja,Stevenson:1981vj,Brodsky:1982gc}, beyond the
scope of these lectures. The experience is that in hadron collisions
there is no choice that works well for any process and it is best to
choose a dynamical scale chosen by examining the process.

As there is no unique scale, the standard procedure is to choose a default
scale $\mu_0$, related to the typical momentum transfer in the process,
and to assign a theoretical uncertainty by varying the scale within a
certain range around the default choice $\mu_0$. The usual range is
between half and twice the default choice. However, this is again an
indication only of the scale uncertainties and there is no mathematical
theorem that states this procedure yields the true theoretical
uncertainty due to neglected higher order terms. In order to have a
measure on the effect of neglected higher orders, \ie to understand the
reliability of the assigned theoretical uncertainty one has to compute
the NNLO corrections. The latter are very demanding computations both
technically and numerically and predictions at NNLO accuracy for some
fairly simple processes, with one or two final-state particles in the
prediction at LO,  constitute the state of the art of pQCD.

\vspace{-4pt}
\noindent\rule{\textwidth}{1pt}


\begin{exe}

Show that the $d$-dimensional three-particle phase space
for $q \to p_1+p_2+p_3$ can be expressed in terms of the
Lorentz-invariants $s_{ij} = (p_i+p_j)^2$
\[
\rd \phi_3 =
(2\pi)^{3-2d}\, 2^{-1-d}\, (q^2)^{\frac{2-d}{2}} \,\rd^{d-2}
\Omega \rd^{d-3} \Omega \,\\
\left(s_{12}s_{13}s_{23}\right)^{\frac{d-4}{2}}
\, \rd s_{12} \, \rd s_{13} \, \rd s_{23} \,
\delta\left(q^2-s_{12}-s_{13}-s_{23}\right)\;.
\]
where $\rd^d \Omega$ is the measure of the hypersurface element in $d$
dimensions, $\int\!\rd^{d-3} \Omega = \Omega_d = 2 \pi^{d/2}/\Gamma(d/2)$.
Hints:
\begin{enumerate}
\item
The $d$-dimensional volume measure in spherical coordinates is
recurisvely given by
\[
\rd^{d+1}p=E\,\rd E\,d^\rd\,p_E=E\,\rd\,E\,E^{d-1}\rd^d\Omega\\
\,,\qquad
\rd^d\Omega=(\sin\theta_{1})^{d-1}
\rd\theta_{1}\rd^{d-1}\Omega\,.
\]
\item
Show that
\[\sin^2\theta_1=
\frac{1}{4}\frac{s_{12}\,s_{13}\,s_{23}}{q^{2}\,E_1^{2}\,E_2^{2}}\,,
\]
where $\theta_1$ is the angle between $p_{1}$ and $p_{2}$.
\end{enumerate}

\end{exe}


\begin{exe}

Let $y_{ij}=\frac{s_{ij}}{q^2}$. Using the previous
exercise, compute the real correction to the process
$e^{+}\,e^{-}\,\rightarrow\,q\bar{q}$ given in \eqn{eq:intsigmaRdimreg}.
Hint: Transform  the triangular integration region into the
unit square and evaluate the $B$ (Euler $\beta$) functions.

\end{exe}
\rule{\textwidth}{1pt}

\section{Jet cross sections} 
\label{sec:jetxsec}

In the first two sections we established our theoretical playground to
make predictions for hadronic cross sections. Based on RGE analysis we
showed that PT can only be fully consistent in an asymptotically free QFT,
like QCD. We found that predictions can be made only for those
quantities that remain finite in the limit of vanishing masses of light
quarks. We computed the radiative corrections for such a quantity, the
total hadronic cross section in electron-positron annihilation. We found
that at intermediate steps of the computations there are singular
contributions of two types: of UV and IR origin. The UV singularities can
be removed by renormalization, and the remaining IR ones can be regularized
in dimensional regularization where IR singularities appear as
$1/\epsilon$ poles. When adding all contributions, these poles cancel and
we obtain the finite correction after setting $\epsilon =0$. Our question
in this section is whether there are more exclusive observables than the
totally inclusive one for which this procedure can be applied.

\begin{figure}[ht]
\begin{center}
\includegraphics[width=0.48\textwidth]{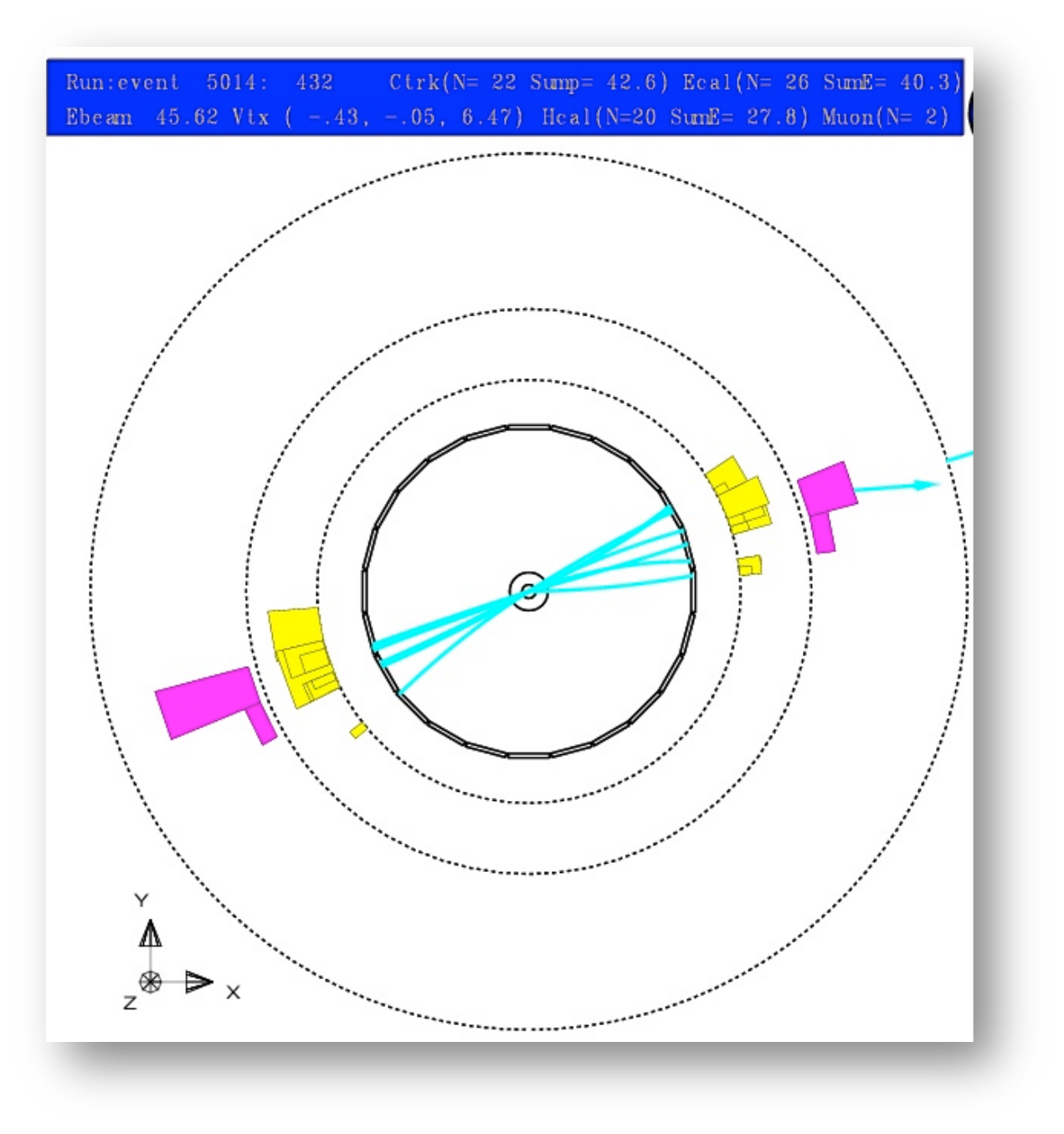}
\includegraphics[width=0.51\textwidth]{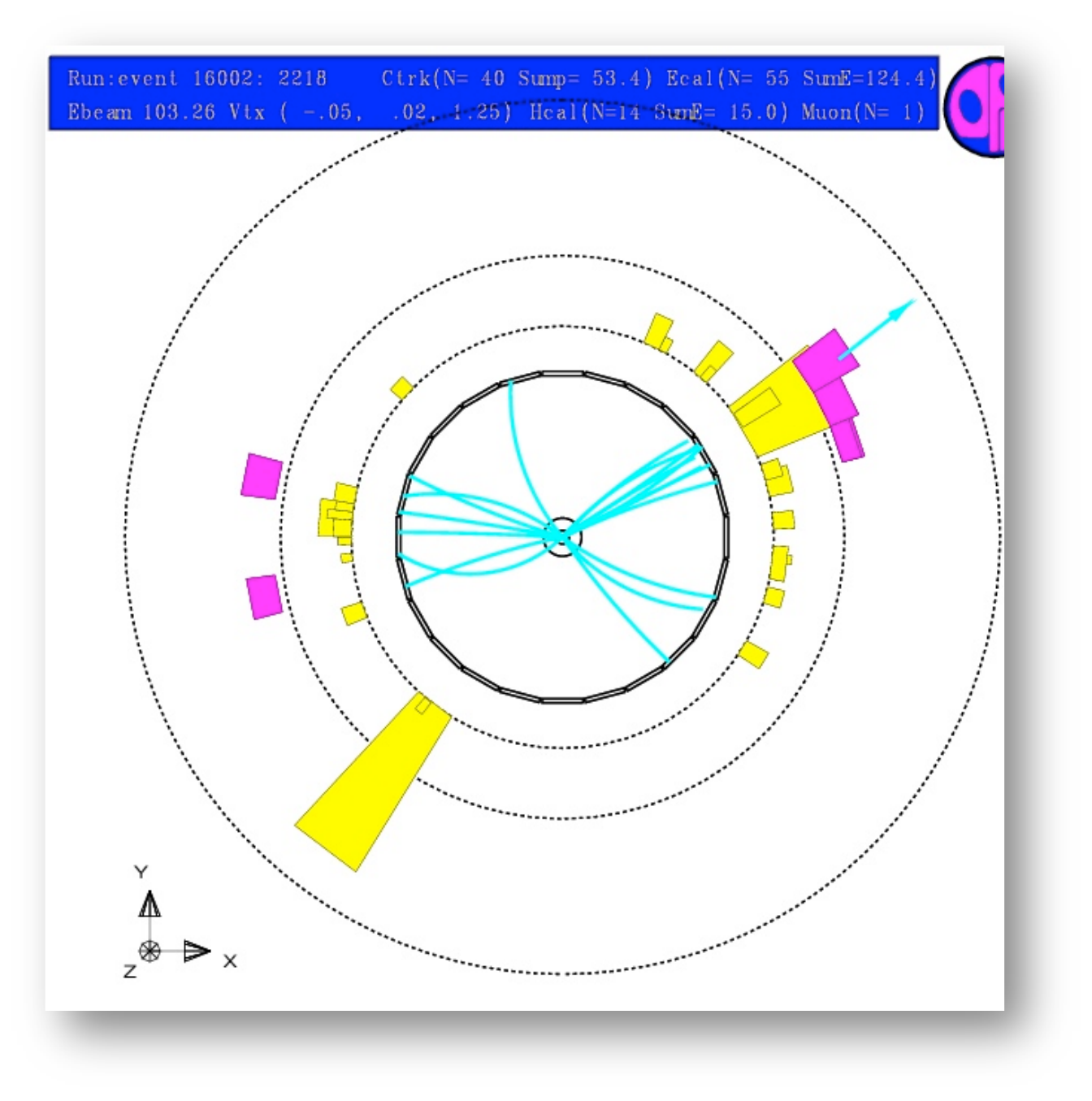}
\caption{Two events observed in the OPAL detector
}
\label{fig:opaljets}
\end{center}
\end{figure}
It is clear from experiments that typical final states have structures.
For instance, \fig{fig:opaljets} shows two events, one with two and the
other with three sprays of hadrons, called {\em hadron jets}. If we count
the relative number of events with two, three, four jets, an interesting
pattern emerges:

\# of events with 2 jets : 
\# of events with 3 jets : 
\# of events with 4 jets
$\simeq \mathcal{O}(\alpha_{\rm s}^0)$ :
$\mathcal{O}(\alpha_{\rm s}^1)$ :
$\mathcal{O}(\alpha_{\rm s}^2)$.

\noindent
Recalling our basic assumption and \fig{fig:partonhadron} we find that
jets reflect the partonic structure of the events.
We now use our pQCD formalism to describe these structures theoretically.
For this purpose, we need a function of the final state momenta
$J_m(\{p_i\})$ that quantifies the structure of the final state in some
ways (we give examples below). This function is called {\em jet function}.

Let us consider again the process $e^+ e^- \to $ hadrons. If we are not
interested in the orientation of the final state events, we can average
over the orientation and find that the SME $|\mathcal{M}_2|^2$ has no
dependence on the parton momenta. Then the two-particle phase space is
$\rd \phi_2 = \rd y_{12} \delta(1-y_{12})$, and contribution of the
process $e^+ e^- \to q\bar{q}$ to the cross section is
\begin{equation}
\sigma_{\rm LO} = |\mathcal{M}_2|^2 \int_0^1\!\rd y_{12} \delta(1-y_{12})
J_2(p_1,p_2)\,,
\end{equation}
which sets our normalization of $|\mathcal{M}_2|^2$. The two kinds of NLO
corrections are
\begin{equation}
\begin{split}
\rd \sigma^{\rm R} &= |\mathcal{M}_2|^2 S_\epsilon 
\frac{{\rm d}y_{13}}{y_{13}^\epsilon}
\frac{{\rm d}y_{23}}{y_{23}^\epsilon}
\,C_{\rm F}\frac{\alpha_{\rm s}}{2\pi}
\left[(1-\epsilon)
\left(
\frac{y_{23}}{y_{13}}
+ \frac{y_{13}}{y_{23}}
\right)
+ \frac{2y_{12}}{y_{13}y_{23}}
-2\epsilon
\right]
J_3(p_1,p_2,p_3)\,,
\\
\rd \sigma^{\rm V} &= |\mathcal{M}_2|^2 S_\epsilon 
\,C_{\rm F}\frac{\alpha_{\rm s}}{2\pi}
\left(\frac{\mu^2}{s}\right)^\epsilon
\left[-\frac{2}{\epsilon^2} - \frac{3}{\epsilon} - 8 + \pi^2
+ \mathcal{O}(\epsilon)
\right]
\rd y_{12} \delta(1-y_{12}) J_2(p_1,p_2)\,.
\label{eq:RandV}
\end{split}
\end{equation}
Contrary to the case of the total cross section, where $J_m = 1$, we
cannot simply perform the integration analytically and combine the
results, neither we can combine the integrands. The general method to deal
with this problem is to regularize both with a properly chosen
subtraction,
\[
\rd \sigma_3^{\rm NLO} = 
\rd \sigma^{\rm R} J_3 - \rd \sigma^{\rm A} J_2 
\,,\quad \mbox{and} \quad
\rd \sigma_2^{\rm NLO} = 
\big(\rd \sigma^{\rm V} + \rd \sigma^{\rm A}\big) J_2 
\,,
\]
such that both terms are separately integrable in $d = 4$ dimensions. This
requires a special property of the jet function $J_n$, called {\rm IR
safety}, expressed analytically as
\begin{equation}
\lim_{y_{13},y_{23} \to 0} J_3 = J_2\,.
\label{eq:IRsafety}
\end{equation}
Qualitatively IR safety means that the jet function is insensitive to
an additional soft particle, or to a collinear splitting in the final
state.

How can we construct such an approximate cross section? For this simple
process we can follow the steps:
\begin{equation}
\begin{split}
 \frac{y_{23}}{y_{13}} + \frac{y_{13}}{y_{23}} + \frac{2y_{12}}{y_{13}y_{23}}
=
\frac{y_{23}}{y_{13}}
+ \frac{1}{y_{13}}&\frac{2y_{12}}{y_{13}+y_{23}} + (1\leftrightarrow 2)
\\ & =
\frac{1}{y_{13}}\bigg[
y_{23}+\bigg(2 \frac{\overbrace{y_{12}+y_{13}+y_{23}}^{=1}}{y_{13}+y_{23}}
-2\bigg)\bigg]+ (1\leftrightarrow 2)
\,.
\label{eq:rewrite}
\end{split}
\end{equation}
Then introduce the new variable
$z_1 \equiv z_{1,2} = \frac{y_{12}}{y_{12}+y_{23}}$, so
that $y_{13} + y_{23} = 1-y_{12} = 1-z_1(1-y_{13})$ and 
\[
\frac{y_{23}}{y_{13}} - \frac{1}{y_{13}} =
\frac{y_{23}(1-y_{13})-y_{12}-y_{23}}{y_{13}(y_{12}+y_{23})}=
\frac{-y_{23}y_{13}-y_{12}}{y_{13}(y_{12}+y_{23})} =
-\frac{y_{23}}{y_{12}+y_{23}}-\frac{z_1}{y_{13}}
\,,
\]
and substitute these into \eqn{eq:rewrite}:
\begin{equation}
\frac{y_{23}}{y_{13}} + \frac{y_{13}}{y_{23}} + \frac{2y_{12}}{y_{13}y_{23}}
=
\bigg[\frac{1}{y_{13}}\left(\frac{2}{1-z_1(1-y_{13})} -1
-z_1\right)
-\frac{y_{23}}{y_{12}+y_{23}} \bigg]
+ (1\leftrightarrow 2)
\,.
\end{equation}
The term $y_{23}/{y_{12}+y_{23}}$ never becomes infinite, 
thus the approximate cross section 
\begin{equation}
\rd \sigma^{\rm A} = |\mathcal{M}_2|^2 S_\epsilon 
\frac{{\rm d}y_{13}}{y_{13}^\epsilon}
\frac{{\rm d}y_{23}}{y_{23}^\epsilon}
\,C_{\rm F} (V_{13,2}+V_{23,1})
\,,\quad\mbox{with} 
\end{equation}
\begin{equation}
\quad
V_{ij,k} = \frac{\alpha_{\rm s}}{2\pi}
\bigg[\frac{1}{y_{ij}}\left(\frac{2}{1-z_{i,k}(1-y_{ij})} -1
-z_{i,k}\right) - \epsilon (1-z_{i,k}) \bigg]
\end{equation}
is a proper subtraction term that regularizes the real contribution in all
of its singular limits in $d$ dimensions. Consequently, the difference 
$\rd \sigma_3^{\rm NLO} = 
\rd \sigma^{\rm R} J_3 - \rd \sigma^{\rm A} J_2$ can be integrated in any
dimensions, in particular, we can set $\epsilon = 0$ and integrate in
$d=4$ numerically.

To obtain $\rd \sigma_2^{\rm NLO}$ we integrate the two terms
separately. For $V_{13,2}$ we change variables in the phase space to
$y_{13}$ and $z_1$, and find
\begin{equation}
\rd \sigma^{\rm A} = |\mathcal{M}_2|^2 S_\epsilon 
\left(\frac{\mu^2}{s}\right)^\epsilon
\int_0^1 \!\rd y_{13} \int_0^1 \!\rd z_1
y_{13}^{-\epsilon}(1-y_{13})^{1-2\epsilon} \,C_{\rm F} V_{13,2}
+ (1\leftrightarrow 2)
\,.
\end{equation}
We shall see that this factorization of the singular terms is universal.
We can now perform the integration over the factorized one-particle phase
space, independently of the jet function, and obtain the {\em integrated
subtraction term} in the form
$\rd \sigma^{\rm A} = |\mathcal{M}_2|^2 \bm{I}(\epsilon)$ with {\em
insertion operator}
\begin{equation}
\bm{I}(\epsilon) = C_{\rm F}\frac{\alpha_{\rm s}}{2\pi}
\frac{1}{\Gamma(1-\epsilon)}
\left(\frac{4\pi\mu^2}{s}\right)^\epsilon
\left[ \frac{2}{\epsilon^2} + \frac{3}{\epsilon} + 10-\frac{\pi^2}3
+ \mathcal{O}(\epsilon)
\right]
\,.
\label{eq:insertionop}
\end{equation}
Comparing this integrated subtraction to \eqn{eq:RandV}, we see that the
sum $\rd \sigma_2^{\rm NLO} = 
\big(\rd \sigma^{\rm V} + \rd \sigma^{\rm A}\big) J_2$ is finite if
$\epsilon = 0$,
\begin{equation}
\sigma_2^{\rm NLO} = |\mathcal{M}_2|^2 C_{\rm F}\frac{\alpha_{\rm s}}{\pi}
\left(1+\frac{\pi^2}{3}\right)
\int_0^1\!\rd y_{12} \delta(1-y_{12}) J_2(p_1,p_2)
+ \mathcal{O}(\epsilon)
\,,
\end{equation}
and so can be integrated in $d=4$ dimensions.

\subsection{Infrared safety} 
\label{sec:IRsafety}

A natural question is if we can construct the approximate cross section
universally, \ie independently of the process and observable. Our
presentation above suggests the affirmative answer. To understand how, we
have to study the origin of the singular behaviour in the SME.  This
singularity arises from propagator factors that diverge\\[5mm]
\hspace*{200pt}
$\propto \frac{1}{(p_{i}+p_{s})^{2}}= \frac{1}{2\,p_{i}\cdot p_{s}}=
\frac{1}{2E_{i}E_{s}(1-\cos\theta)}\simeq \frac{1}{E_{i}E_{s}\theta^{2}}$

\vspace*{-32pt} ~\hspace*{50pt}
\includegraphics[width=0.2\linewidth]{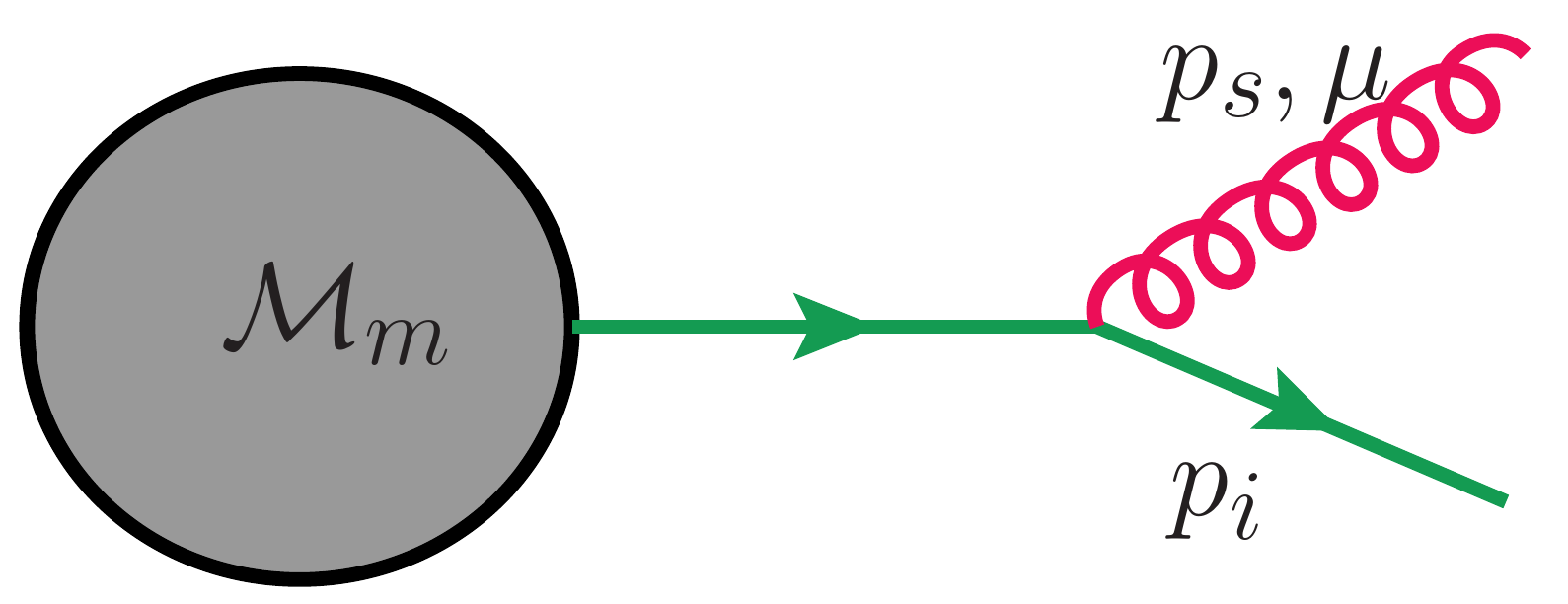}

In the collinear limit, $\theta\,\rightarrow\,0$ and
$\mathcal{M}_{m+1}\simeq \mathcal{M}_m/{\theta}+$ less singular terms
(a factor of $\theta$ appears in the numerator factors).
In the soft limit, $E_{s}\,\rightarrow\,0$ and
$\mathcal{M}_{m+1}\simeq \mathcal{M}_m/{E_s}+$ less singular terms.
The gluon phase space is
\[\frac{{\rm d}^{3}p_{s}}{2E_{s}}=
\frac{1}{2}E_{s}\,{\rm d} E_{s}\,{\rm d}\!\cos\theta\,{\rm d}\phi\simeq
\frac14 E_{s}\,{\rm d} E_{s}\,{\rm d}\theta^{2}\,{\rm d}\phi
\,,
\]
so in the cross section we find logarithmic singularities in both the
soft and the collinear limits: $\frac{\rd E_{s}}{E_{s}}$ or $\frac{\rd
\theta^{2}}{\theta^{2}}$. These are the IR singular limits. In
dimensional regularization the logarithmic singularities appear as poles:
\[\int\!\rd y_{is}\,y_{is}^{-1-\epsilon}\,=\,-\frac{1}{\epsilon}\,.\]
Thus, the singular behaviour arises at kinematically degenerate phase
space configurations, which at the NLO accuracy means that one cannot
distinguish the following configurations: (i) a single hard parton,
(ii) the single parton splitting into two nearly collinear partons,
(iii) the single parton emitting a soft gluon (on-shell gluon with very
small energy).
Then an answer to the question posed at the beginning of \sect{sec:jetxsec}
is given by the Kinoshita-Lee-Nauenberg (KLN) theorem
\cite{Kinoshita:1962ur,Lee:1964is}:
\begin{quote}
\emph{In massless, renormalized field theory in four dimensions,
transition rates are IR safe if summation over kinematically degenerate
initial and final states is carried out.}
\end{quote}
For the $e^{+}e^{-}\rightarrow$ hadrons process, the
initial state is free of IR singularities. Typical IR-safe
quantities are
(i) event shape variables and
(ii) jet cross sections.

\subsection{Event shapes} 
\label{sec:evshapes}

\begin{wrapfigure}{r}{0.4\linewidth}
\includegraphics[width=1.0\linewidth]{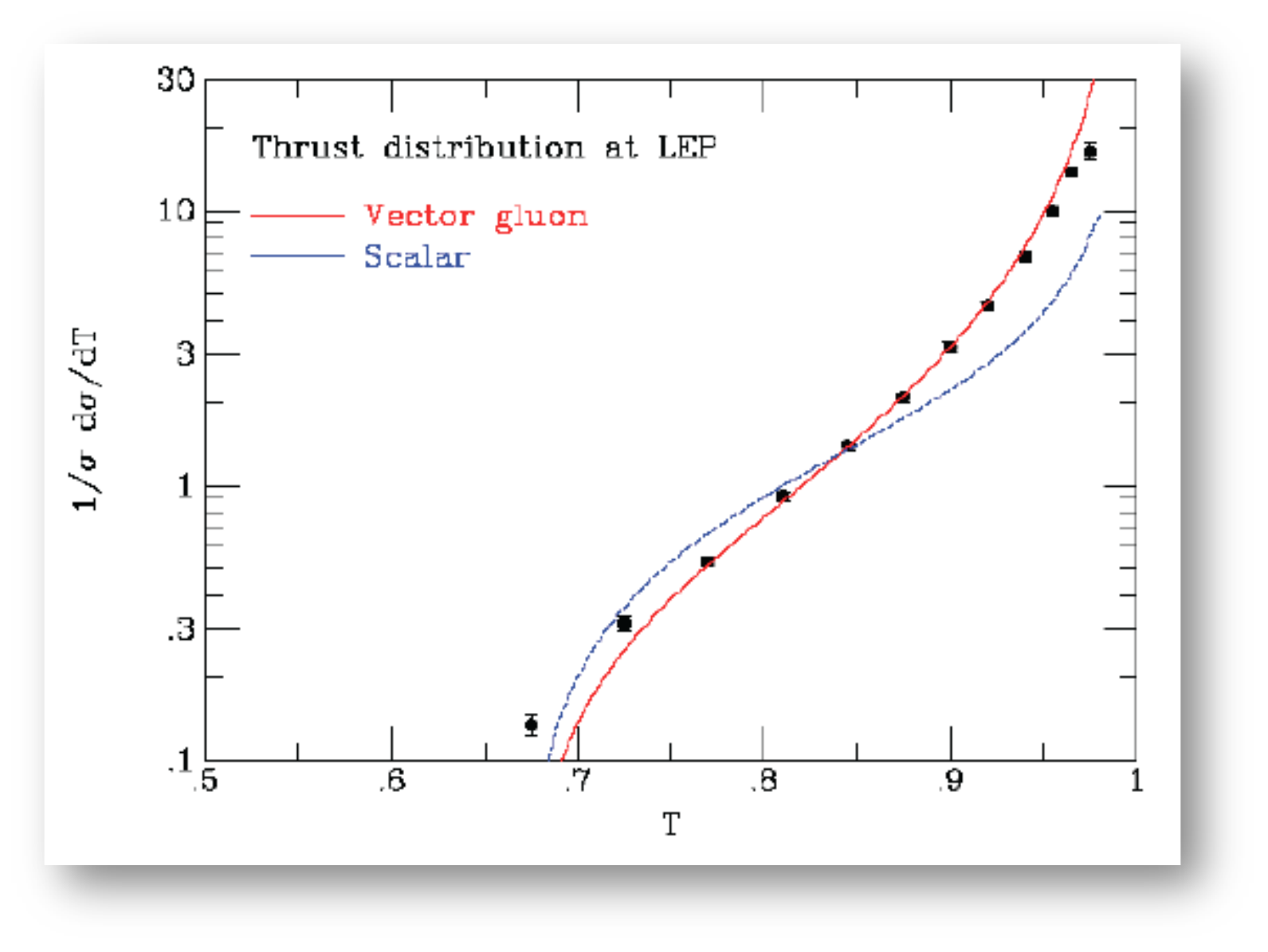}
\vspace*{-20pt}
\caption{Distribution of thrust as measured at LEP compared to pQCD
predictions obtained with vector and scalar gluon
~}
\label{fig:thrust}
\end{wrapfigure}
Thrust, thrust major/minor, C- and D-parameters, oblateness, sphericity,
aplanarity, jet masses, jet-broadening, energy-energy correlation,
differential jet rates are examples of event shape variables. The value
of an event shape does not change if a final-state particle further
splits into two collinear particles, or emits a soft gluon, hence it
is (qualitatively) IR safe.  As as example we consider the thrust $T$,
which is defined by
\begin{equation} 
T=\max_{\vec{n}}\, 
\frac{\sum_{i=1}^{m}\,\lp|\vec{p}_i\,\ldot\,\vec{n}\rp|} 
{\sum_{i=1}^{m}\lp|\vec{p}_{i}\rp|}, 
\label{eq:thrustdef}
\end{equation}
where $\vec{n}$ is a three-vector (the direction of the thrust axis)
such that $T$ is maximal. The particle three-momenta $\vec{p}_{i}$ are
defined in the $e^{+}e^{-}$ centre-of-mass frame. $T$ is an example of the
jet function $J_m$. It is infrared safe because neither $p_j \to 0$, nor
replacing $p_i$ with $z p_i+(1-z) p_i$ change $T$. At LO accuracy it is
possible to perform the phase space integrations and
\begin{equation} 
\frac{1}{\sigma} \frac{{\rm d} \sigma}{{\rm d} T} = C_{\rm F}
\frac{\alpha_{{\rm s}}}{2\pi}
\left[\frac{2 \big(3 T^2-3T+2\big)}{T(1-T)}
\ln\left(\frac{2T-1}{1-T}\right)
- 3(3T-2)\frac{2-T}{1-T}\right]
\,.
\label{eq:thurst}
\end{equation}
We see that the perturbative prediction for the thrust distribution is
singular at $T=1$. In addition to the linear divergence in $1-T$ there is
logarithmic divergence, too. The latter is characteristic to events shape
distributions. In PT at $n$th order logarithms of $1-T$ in the form
$\alpha_{\rm s}^n \ln^m(1/(1-T))\,, m\leq 2n$ appear. These spoil the
convergence of the perturbation series and call for resummation if we
want to make reliable prediction near the edge of the phase space,
for large values of $T$ where the best experimental statistics are
available. Resummations of leading ($m=2n$) and next-to-leading ($m=2n-1$)
logarithms are available for many event shape variables, but the
discussion of this technique is beyond the scope of these lectures.

\vspace{-4pt}
\noindent\rule{\textwidth}{1pt}


\begin{exe}

Verify that $T$ as defined in \eqn{eq:thrustdef} is infrared und collinear safe.
What is the range of values that $T$ can take if
(i) there are only two particles in the final state, or
(ii) $m\rightarrow \infty$ and all $\vec{p}_i$ are distributed spherically?

\end{exe}
\rule{\textwidth}{1pt}

\subsection{Jet algorithms} 
\label{sec:jetalg}

Jets are sprays of energetic, on-shell, nearly collinear hadrons. The
number of jets does not change if a final-state particle further splits
into two collinear particles, or emits a soft gluon, hence it is again
qualitatively IR safe. To quantify the jet-like structure of the final
states jet algorithms have been invented. These have a long history with
rather slow convergence. The reason is that the experimental and
theoretical requirements posed to a jet algorithm are rather different.
Experimentally we need cones that include almost all hadron tracks at
cheap computational price. Theoretically the important requirements are
IR safety, so that PT can be employed to make predictions and
resummability, so that we can make predictions in those region of the
phase space where most of the data appear.

For many years experimenters preferred cone jet algorithms (according to
the `Snowmass accord') \cite{Berger:1992zh}. These start from a cone seed
(centre of the cone) in pseudorapidity ($\eta$) and azimuthal-angle
($\phi$) plane: ($\eta_c,\phi_c$). We define a distance of a hadron
track $i$ from the seed by
$d_{ic} = \sqrt{(\eta_i-\eta^c)^2 + (\phi_i-\phi^c)^2}$. A track
belongs to the cone if $d_{ic} < R$, with a predefined value for $R$
(usually 0.7). It turned out however, that (i) this is an IR unsafe
definition and  (ii) there is a problem how to treat overlapping cones, so
the cone jet definition has been abandoned.

Theoreticians prefer iterative jet algorithms, consisting of the following
steps. (i) First we define a {\em distance between two momenta} (of partons
or hadron tracks) and a {\em rule to combine two momenta}, $p_i$ and $p_j$
into $p_{(ij)}$. (ii) Then we select a {\em value for resolution} $d_{\rm
cut}$ and consider all pairs of momenta. If the minimum of $\{d_{ij}\}$ is
smaller than $d_{\rm cut}$, then we combine the momenta $p_i$ and $p_j$
and start the algorithm again. If the minimum is larger than $d_{\rm
cut}$, then the remaining momenta (after the combinations) are considered
the {\em momenta of the jets}, and the algorithm stops. The drawback of
this algorithm is that it becomes very expensive computationally for many
particles in the final state. This is not an issue in pQCD computations
because according to our basic assumption there are only few partons,
but a major problem for the final states in the detectors where
hundreds of hadrons may appear in a single event.

\begin{figure}[ht]
\begin{center}
\includegraphics[width=0.49\textwidth]{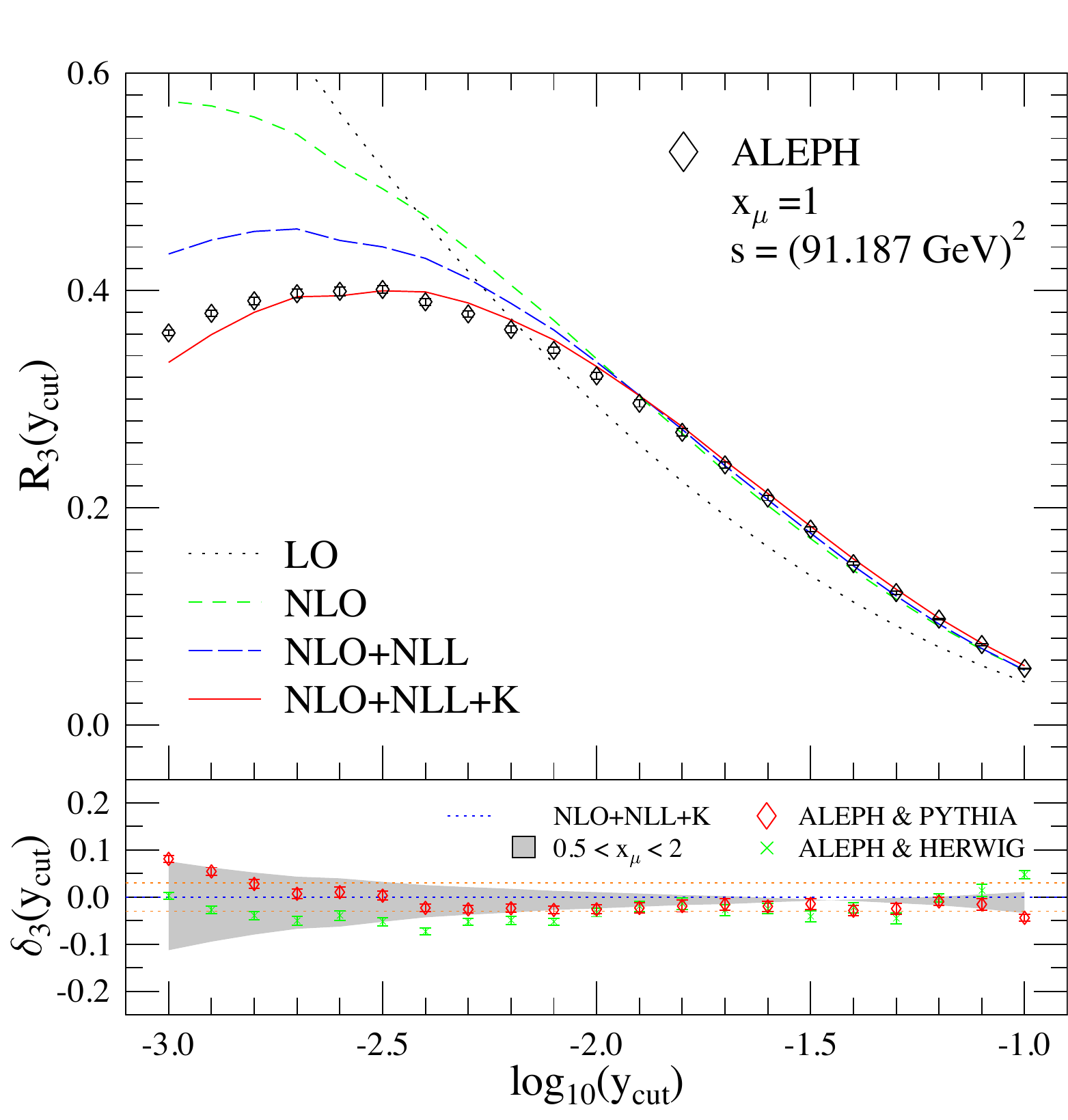}
\includegraphics[width=0.49\textwidth]{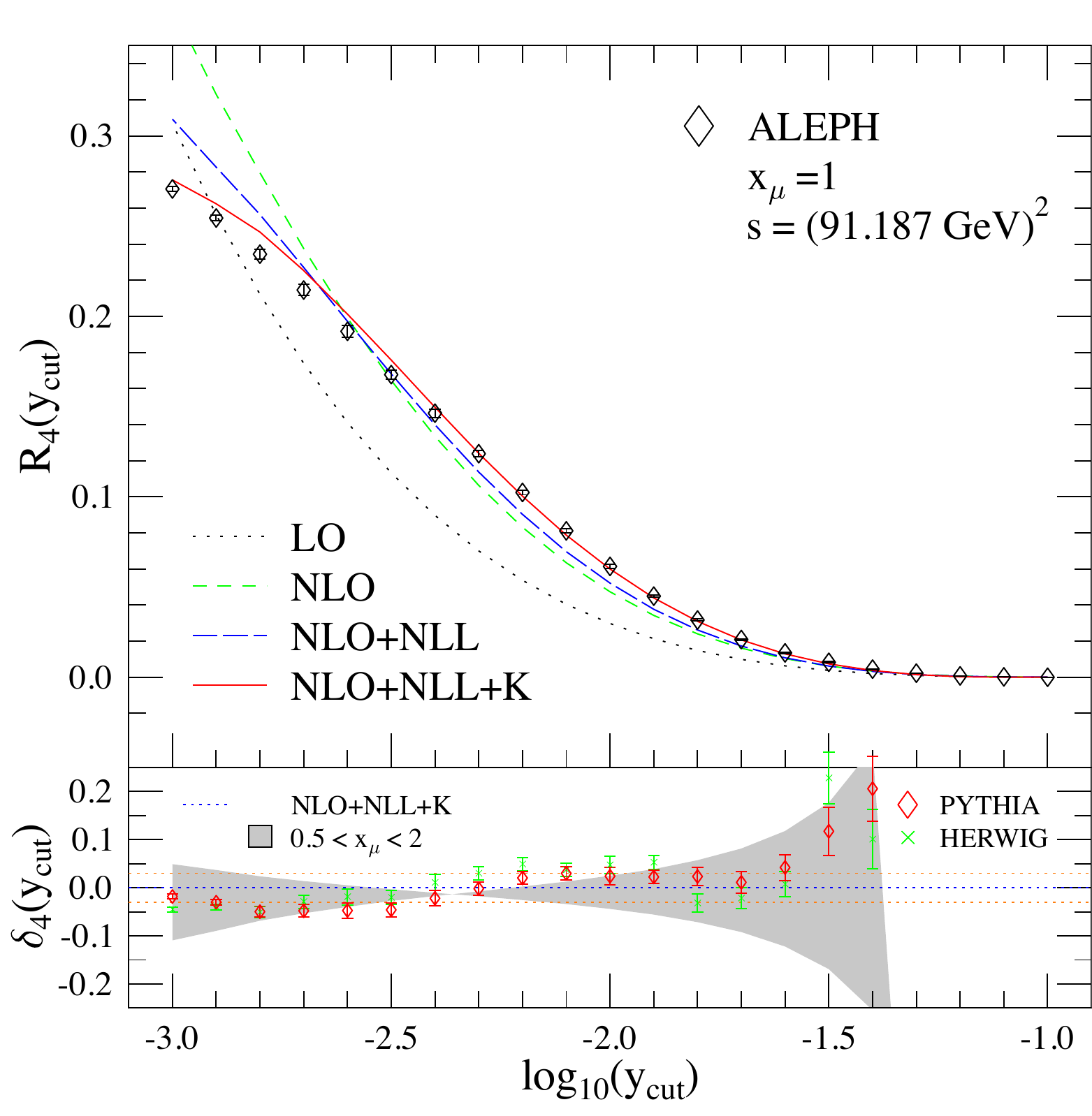}
\caption{Comparison of pQCD predictions to data for three- and four-jet
leptoproduction \cite{Nagy:1998kw}.
The data points include corrections from hadrons to partons based on
Monte Carlo simulations. The '+K' term indicates the inclusion of a 
well-defined subleading (NNLL) term in the resummation formula.
}
\label{fig:34jets}
\end{center}
\end{figure}

At LEP theory won and the Durham (or $k_\perp$) algorithm was used. It was
invented so that resummation of large logarithms could be achieved
\cite{Catani:1991hj}. The distance measure is $d_{ij} = 2
\frac{\min(E_i^2,E_j^2)}{s} R_{ir}$, where $R_{ij} = 1-\cos\theta_{ij}$
and the recombination scheme is simple addition of the four momenta,
$p_{(ij)}^\mu = p_i^\mu + p_j^\mu$.  The resolution parameter $y_{\rm
cut} = d_{\rm cut}/s$ can take values in $[0,1]$. The pQCD prediction contains
logarithmically enhanced terms of the form $\alpha_{\rm s}^n
\ln^m(1/y_{\rm cut})$, at any order, which has to be resummed if we
want to use small value of $y_{\rm cut}$, where we find the bulk of the data
(see Figure~\ref{fig:34jets}). Predictions are
available with leading- ($m = 2n$) and next-to-leading ($2n-1$)
logarithms (LL and NLL) summed up to all orders \cite{Catani:1991hj}.  

Figure~\ref{fig:34jets} shows the fixed order LO and NLO predictions, as
well as predictions where NLO and NLL are matched. The curve at NLO
accuracy gives a good description of the measure data by the ALEPH
collaboration \cite{Barate:1996fi}, but only for $y_{\rm cut} > 0.01$. As
$\alpha_{\rm s} \ln^2(100) = 2.5$, for smaller values of $y_{\rm cut}$
resummation is indispensable. The resummed prediction however, is not
expected to give a good description at large $y_{\rm cut}$ because in the
resummation formula only the collinear approximation of the matrix
element is used. Matching the two predictions gives a remarkably good
description of the data over the whole phase space.

At hadron colliders the $k_\perp$ algorithm needs modifications. First,
instead of energy, the boost-invariant measure of hardness, transverse
momentum is used to define the distance between tracks,
$d_{ij} = \min(p_{\perp,i}^{2},p_{\perp,j}^{2}) \frac{R_{ij}^2}{R^2}$ 
where $R_{ij}^2 = (y_i-y_j)^2 + (\phi_i-\phi_j)^2$ (distance in
rapidity--azimuthal-angle plane), $R$ is a small positive real number,
and we need a distance from the beam $d_{iB} = p_{\perp,i}^{2n}$, too. 
Also, the algorithm needs modification because either $d_{ij}$ or
$d_{iB}$ can be the smallest distance. If a $d_{ij}$ is the smallest
value, then $i$ and $j$ are merged, while if the smallest is a
$d_{iB}$, then momentum $p_i$ becomes a jet momentum and is removed
from the tracks to be clustered. We then call jet candidates with
transverse momentum $p_{\perp,i} > E_R$ resolved jets.  The merging
rule may change as well. In the usual merging we add four-momenta, but
another option is to add transverse momenta,
$p_{\perp,(ij)} = p_{\perp,i} + p_{\perp,j}$, and add rapidities $y$
and azimuthal angles $\phi$ weighted, $y_{(ij)} = (w_i y_i + w_j
y_j)/(w_i + w_j)$ and $\phi_{(ij)}=(w_i \phi_i+w_j \phi_j)/(w_i+w_j)$,
where the weight can be $w_i = p_{\perp,i}$, $p_{\perp,i}^2$,
$E_{\perp,i}$, or $E_{\perp,i}^2$. Such a merging is boost invariant along
the beam. 
The parameter $R$ plays a similar role as $d_{\rm cut}$ in
electron-positron annihilation or the cone radius $R$ in the cone
algorithms: the smaller $R$, the narrower the jet. 

The iterative $k_\perp$-algorithm is infrared safe and resummation of
large logarithmic contributions of the form $\alpha_{\rm s}^n \ln^{2n}$
and $\alpha_{\rm s}^n \ln^{2n-1}$ is possible, which is a clear
advantage from the theoretical point of view.  The logarithms are those
of $1/R$ and/or $Q/E_R$, $Q$ being the hard process scale. By taking
$E_R$ sufficiently large in hadron-hadron collisions, we avoid such
leading contributions from initial-state showering and the underlying
event, so these terms are determined by the time-like showering of
final-state partons (when the virtuality of the parent parton is always
positive). Particles within angular separation $R$ tend to
combine and particles separated by larger distance than $R$ from
all other particles become jets. The algorithm assigns a clustering
sequence to particles within jets, so we can look at jet substructure.

\begin{figure}[ht]
\begin{center}
\includegraphics[width=0.49\textwidth]{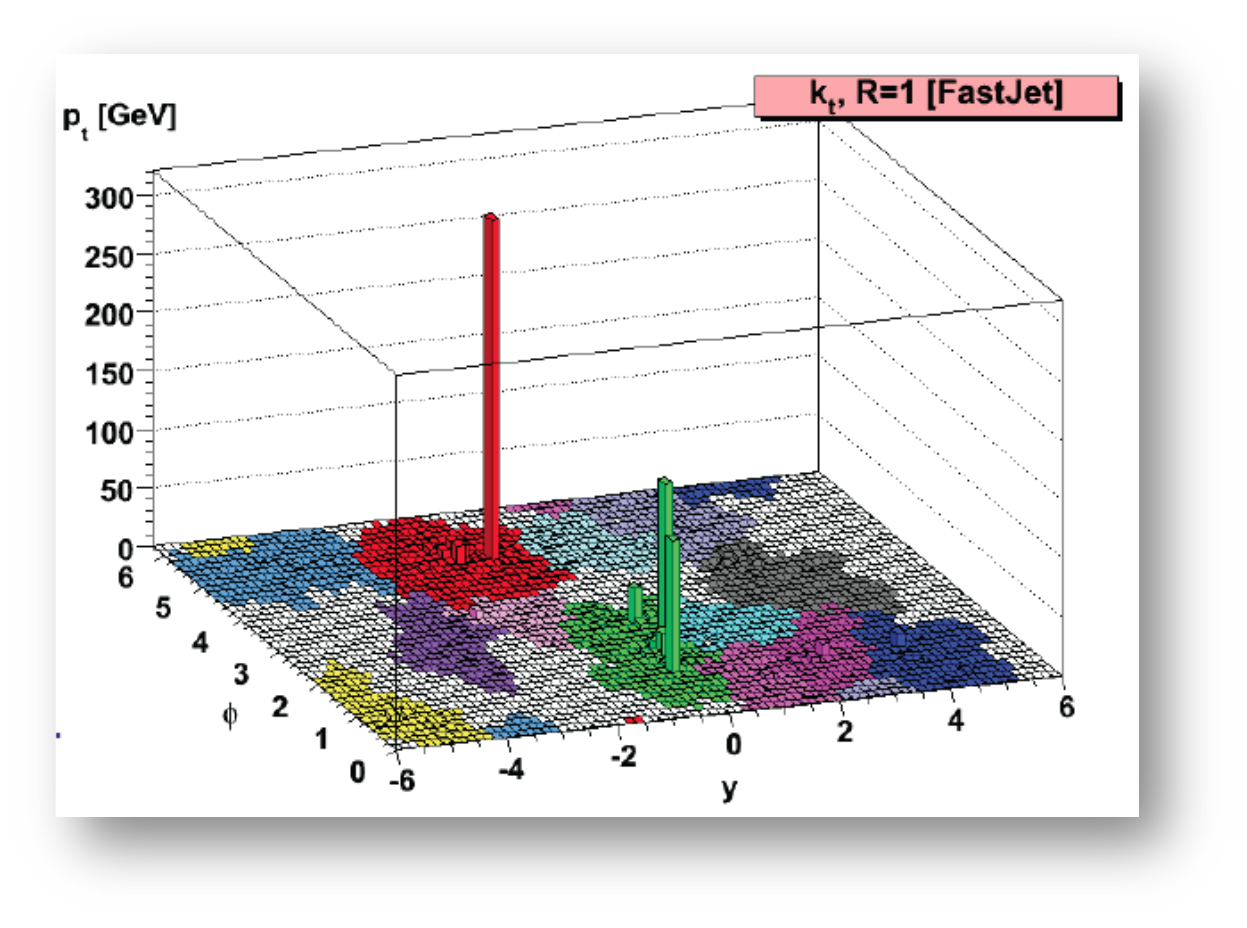}
\includegraphics[width=0.49\textwidth]{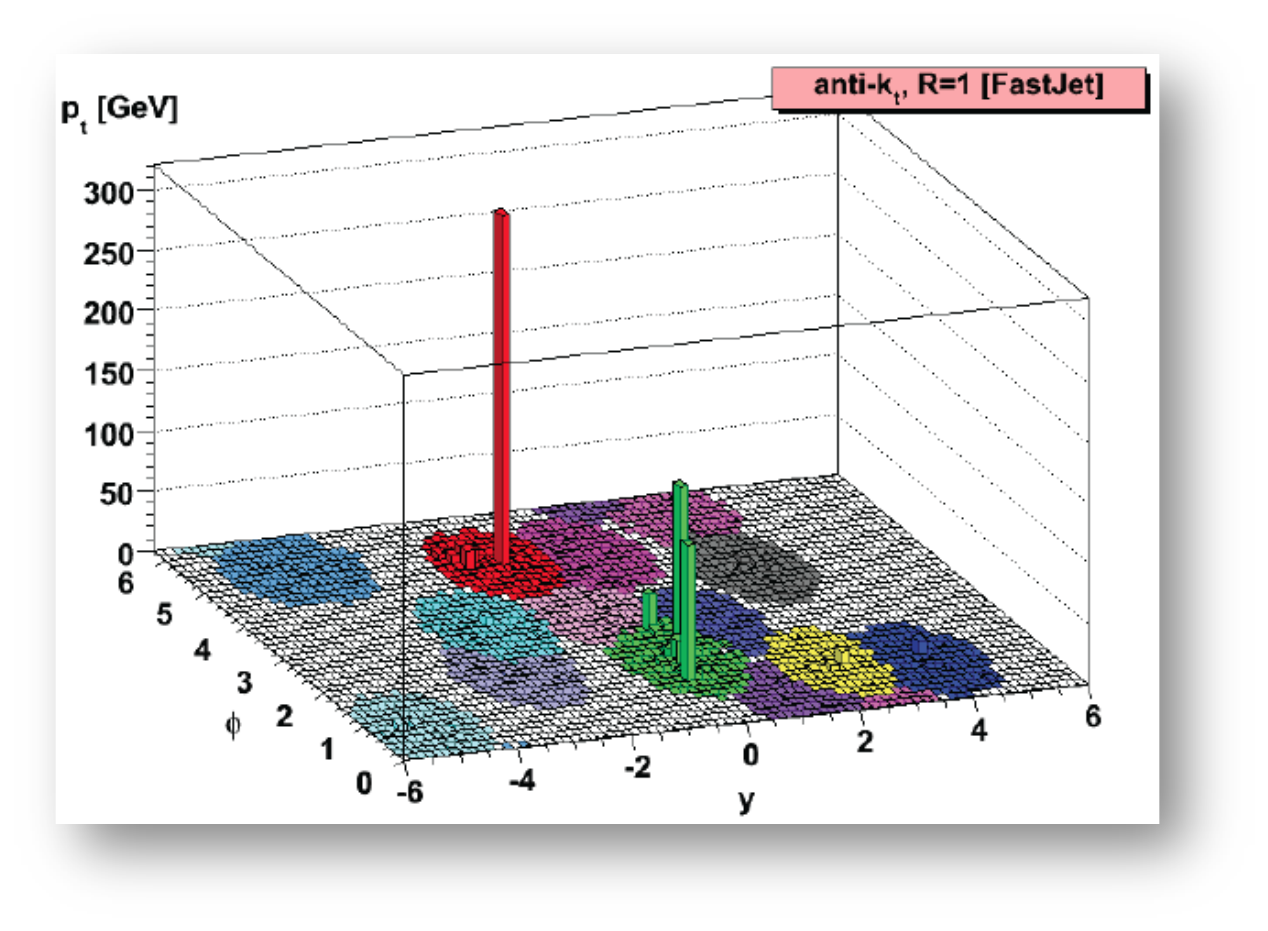}
\caption{Jets in a proton-proton scattering event obtained with the 
(a) $k_\perp$, (b) anti-$k_\perp$ clustering algorithms
}
\label{fig:kTpp}
\end{center}
\end{figure}

Nevertheless, at the TeVatron experiments the $k_\perp$-algorithm did
not become a standard for several reasons.  The jets have irregular,
often weird shapes as seen on \fig{fig:kTpp}(a) because soft particles
tend to cluster first (even arbitrary soft particles can form jets). As
a result there is a non-linear dependence on soft particles, energy
calibration and estimating acceptance corrections are more difficult.
The underlying event correction depends on the area of the jet (in
$\eta-\phi$ plane).  It was also very expensive computationally, so
experimenters had a clear preference of cone algorithms.

\noindent The breakthrough occurred with Refs.~\cite{Cacciari:2005hq,Cacciari:2008gp}
where variants of the $k_\perp$ algorithm and an improved, fast
implementation was introduced. The distance formula was modified to
$d_{ij} = \min(p_{\perp,i}^{2n},p_{\perp,j}^{2n}) \frac{R_{ij}^2}{R^2}$
($n=-1,0,1$). IR-safety is independent of $n$, as well as NLL
resummation of large logarithms.
It was found that with $n=-1$ (called anti-$k_\perp$-alogrihtm) particles
close in angle cluster first, which results in regular cone-like shapes as
seen on \fig{fig:kTpp}(b) without using stable cones. As a result it
became the standard jet algorithm at the LHC experiments. Yet, one should
keep in mind that there is no `perfect' jet algorithm. For instance, the
anti-$k_\perp$ one does not provide useful information on jet
substructure. It is important to remember that {\em in pQCD theoretical
prediction can be made only with IR-safe jet functions, but among those
the goal of the study may help decide which algorithms to use}. 

\section{Towards a general method for computing QCD radiative corrections}
\label{sec:pQCD}

We have seen that (i) in pQCD the computation of radiative corrections at
NLO accuracy is indispensable, (ii) the NLO corrections are of two kinds:
real and virtual, that are separately divergent and contain different
number of particles in the final state, (iii) these singularities cancel
for IR-safe cross sections. To find the finite NLO corrections we have to
develop a method for combining the real and virtual corrections. In order
to be able to automate the NLO computations such a method has to be
general, \ie independent of the measurable quantity and the process. To
devise such a general method, we need to study the origin of the
singularities in a more precise way than we did in the previous section.
We shall find factorization formulae of the SME's that find many important
applications in QCD, and so belong to the most important features of QCD.

\subsection{Factorization of $\lp|\mathcal{M}_{n}\rp|^{2}$ in the soft limit} 
\label{sec:softlimit}

The soft limit is defined by $p_{s}^{\mu}=\lambda\,q^{\mu}$, with    
$\lambda\,\in\,\mathbb{R}^{+}$ and $\lambda\,\rightarrow\,0$ for
$q^{\mu}$ fixed. In this limit the emission of the soft gluon from
(internal) propagators is IR finite. If we consider the emission of a
soft gluon off an external quark we find

\includegraphics[width=0.2\linewidth]{figures/singularity.pdf}

\vspace*{-36pt} ~\hspace*{110pt}
$ \propto
\mathcal{M}_m T_{i}^{s}g_{\rm s}\,\bar{u}(p_{i},s_{i})\gamma^{\mu}
\,\frac{\slashed{p}_{i}+\slashed{p}_{s}}{s_{is}}
\stackrel{p_{s}\to 0}{\simeq}
\mathcal{M}_m T_{i}^{s}g_{\rm s}\,\frac{p_{i}^{\mu}}{p_{i}\cdot p_{s}}
\,\bar{u}(p_{i},s_{i})$. \\[18pt]
In taking the limit, we used the anti-commutation
relation (\ref{eq:Clifford}) to write
$\gamma^{\mu}\slashed{p}_{i}=-\slashed{p}_{i}\gamma^{\mu}+2p_{i}^{\mu}$
and the Dirac equation of the massless bi-spinor, 
$\bar{u}(p_{i})\slashed{p}_{i}=0$. The factor
$\frac{p_{i}^{\mu}}{p_{i}\cdot p_{s}}$ is the ``square root'' of the
eikonal factor $S_{ik}\lp(s\rp)=\frac{2s_{ik}}{s_{is}s_{ks}}$.
In the same limit, we can derive after a bit more algebra the
factorization formula for soft-gluon emission off a gluon line. 
The emission of a soft gluon off an external gluon (in light-cone
gauge) is given by

\includegraphics[width=0.2\linewidth]{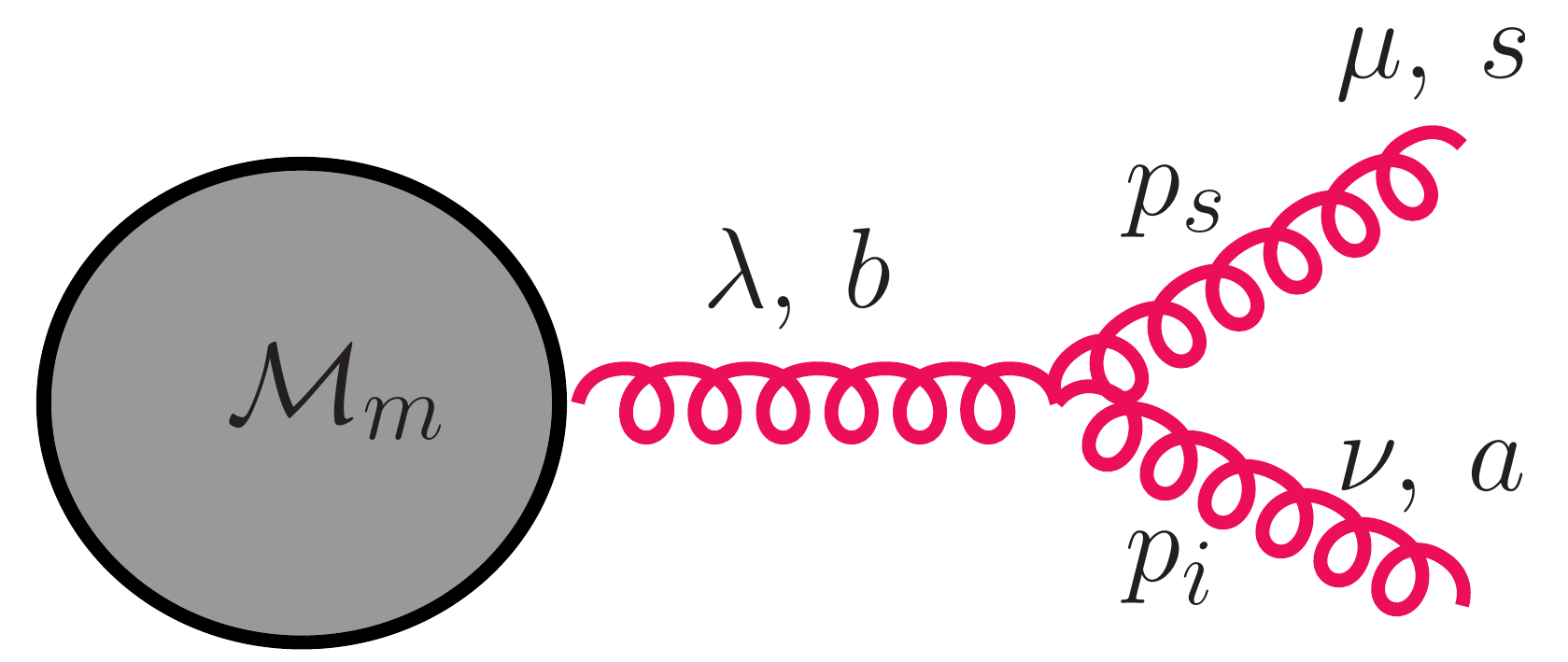}

\vspace*{-36pt} ~\hspace*{110pt}
$\propto\,\mathcal{M}_m \varepsilon^{\mu}\lp(p_{s}, n\rp)\frac{1}{s_{is}} 
\,d^{\lambda\lambda'}\lp(p_{i}+p_{s},n\rp) 
\Gamma_{\nu\mu\lambda'}^{asb}\lp(-p_{i},-p_{s},p_{i}+p_{s}\rp) 
\varepsilon^{\nu}\lp(p_{i},n\rp)$, \\[18pt]
where in the three-gluon vertex
\begin{eqnarray*} 
V_{\nu\mu\lambda}(-p_{i},-p_{s},p_{i}+p_{s})\aand= 
-\lp(p_{i}+2p_{s}\rp)_{\nu}g_{\mu\lambda} 
+\lp(2p_{i}+p_{s}\rp)_{\mu}g_{\nu\lambda} 
-\lp(p_{i}-p_{s}\rp)_{\lambda}g_{\mu\nu} 
\\ && 
= 2p_{i\mu}\,g_{\lambda\nu}+\lp[-\lp(p_{i}+p_{s}\rp)_{\lambda}\,g_{\mu\nu} 
-p_{i\nu}\,g_{\mu\lambda}\rp] 
+\lp[p_{s\mu}\,g_{\nu\lambda} 
+2p_{s\,\lambda}\,g_{\mu\nu}-2p_{s\nu}\,g_{\mu\lambda}\rp] 
\\ && 
\stackrel{p_{s}\rightarrow\,0}{\simeq} 
2p_{i\mu}\,g_{\nu\lambda}-\lp[\lp(p_{i}+p_{s}\rp)_{\lambda}\,g_{\mu\nu} 
+p_{i\nu}\,g_{\mu\lambda}\rp]. 
\end{eqnarray*}
We use
$d^{\lambda\lambda'}\lp(p_{i}+p_{s},n\rp)\lp(p_{i}+p_{s}\rp)_{\lambda} = 
0$ and $\varepsilon^{\nu}\lp(p_{i},n\rp)p_{i\,\nu} = 0$, thus
\begin{equation*} 
\frac{1}{s_{is}}d^{\lambda\lambda'}\lp(p_{i}+p_{s},n\rp) 
\Gamma_{\nu\mu\lambda'}^{asb}\lp(-p_{i},-p_{s},p_{i}+p_{s}\rp) 
\varepsilon^{\nu}\lp(p_{i},n\rp)\stackrel{p_{s}\rightarrow\,0}{\:\simeq\:} 
-T_{b}^{s}\gs\frac{p_{i\mu}}{p_{i}\ldot p_{s}} 
\,\underbrace{\lp[d^{\lambda\lambda'}\lp(p_{i},n\rp) 
g_{\lambda'\nu}\varepsilon^{\nu}\lp(p_{i},n\rp)\rp]}_ 
{-\varepsilon^{\lambda}\lp(p_{i},n\rp)}. 
\end{equation*}

These two results can be unified and formalized by
\begin{equation*} 
\hat{S}_{s}\la\,c_{s}\rp|\left.\mathcal{M}_{m+1}\lp(p_{s},\ldots\rp)\ra= 
\gs\epsilon^{\mu}\lp(p_{s}\rp) 
\,\bJ_{\mu}\lp(s\rp)\lp|\mathcal{M}_{m}\lp(\ldots\rp)\ra, 
\end{equation*}
where $c_{s}$ is the colour index of the soft gluon $s$,
$\hat{S}_{s}$ is an operator which takes the soft limit and
keeps the leading $\frac{1}{\lambda}$ singular term, and the soft gluon
current $\bJ_{\mu}\lp(s\rp)$ is given by
\begin{equation*} 
\bJ_{\mu}\lp(s\rp)=\sum_{k=1}^{m} 
\,\bT_{k}^{s}\,\frac{p_{k\,\mu}}{p_k\,\ldot\,p_{s}}\,. 
\end{equation*}
The soft gluon can be emitted from any of the external
legs, therefore the sum in the previous formula runs over
all external partons. A soft quark leads to an integrable
singularity because the fermion propagator is less singular
than that of the gluon. Colour conservation implies that the current
$\bJ_{\mu}\lp(s\rp)$ is conserved,\\~
\begin{equation*} 
p_{s}^{\mu}\,\bJ_{\mu}\lp(s\rp)\lp|\mathcal{M}_{m}\ra= 
\sum_{k=1}^{m}\bT_{k}^{s}\lp|\mathcal{M}_{m}\ra=
~~~~~~~~~~~~~~~~~~~~~~~~~~~~~~~~~~~~~~~~
=0\,. 
\end{equation*}

\vspace*{-66pt} ~\hspace*{240pt}
\includegraphics[width=0.4\linewidth]{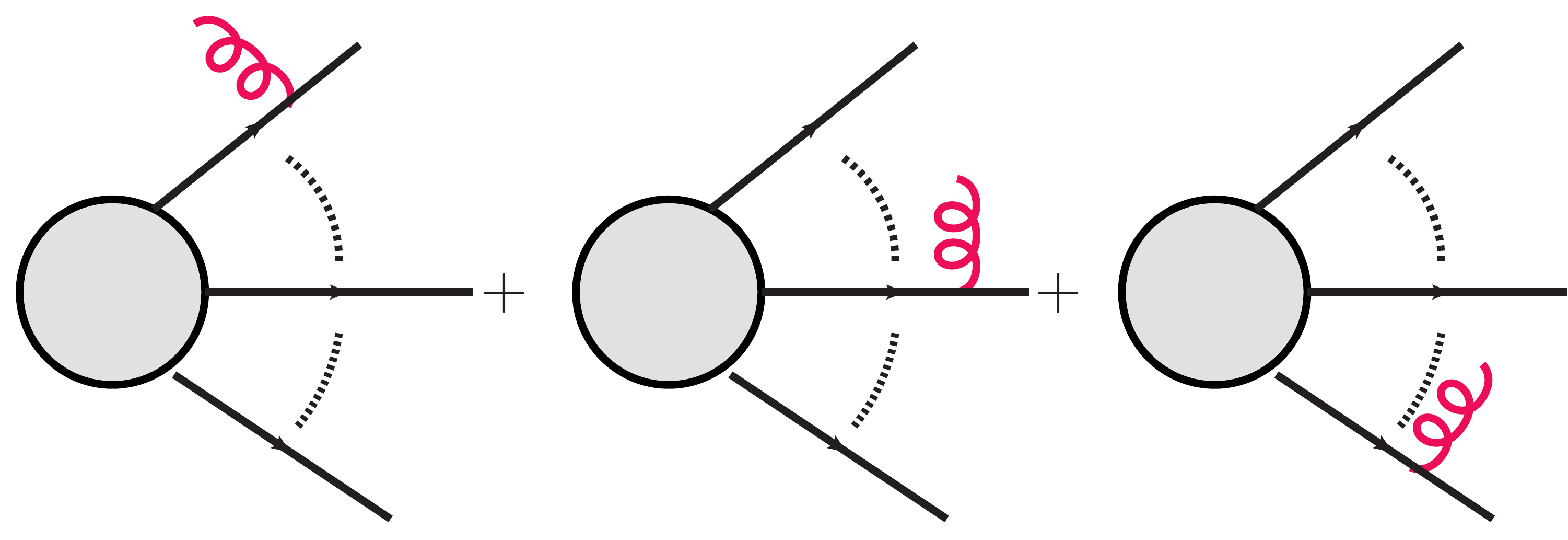}

\noindent Then the soft limit of the SME 
$\la\mathcal{M}_{m}^{\lp(0\rp)}\rp|\left.\mathcal{M}_{m}^{\lp(0\rp)}\ra$
is as follows:
\begin{eqnarray} 
\hat{S}_{s}\lp|\mathcal{M}_{m+1}\lp(p_{s},\dots\rp)\rp|^2\aand= 
4\pi\as\mu^{2\epsilon}\sum_{i=1}^{m}\sum_{k=1}^{m}\,\underbrace{\epsilon_{\mu}\lp(s\rp)\epsilon_{\nu}^{\ast}\lp(s\rp)}_{d_{\mu\nu}\lp(p_{s},n\rp)}\frac{p_{i}^{\mu}p_{k}^{\nu}}{p_{i}\cdot\,p_{s}\,p_{k}\cdot\,p_{s}}\la\mathcal{M}_{m}\rp|\bT_{i}\cdot\bT_{k}\lp|\mathcal{M}_{m}\ra
\label{eq:softfact}
\\ \nonumber
\aand= 
-8\pi\as\mu^{2\epsilon}\sum_{i,k=1}^{m}\frac{1}{2}S_{ik}\lp(s\rp)\lp|\mathcal{M}_{m\,\lp(i,k\rp)}^{\lp(0\rp)}\rp|^{2}\,+\,\textrm{gauge 
terms}  =
~~~~~~~~~~~~~~~~~~~~~~~~~~~~~~~~~~~~
\dots \,. 
\end{eqnarray}

\vspace*{-50pt} ~\hspace*{338pt}
\includegraphics[width=0.2\linewidth]{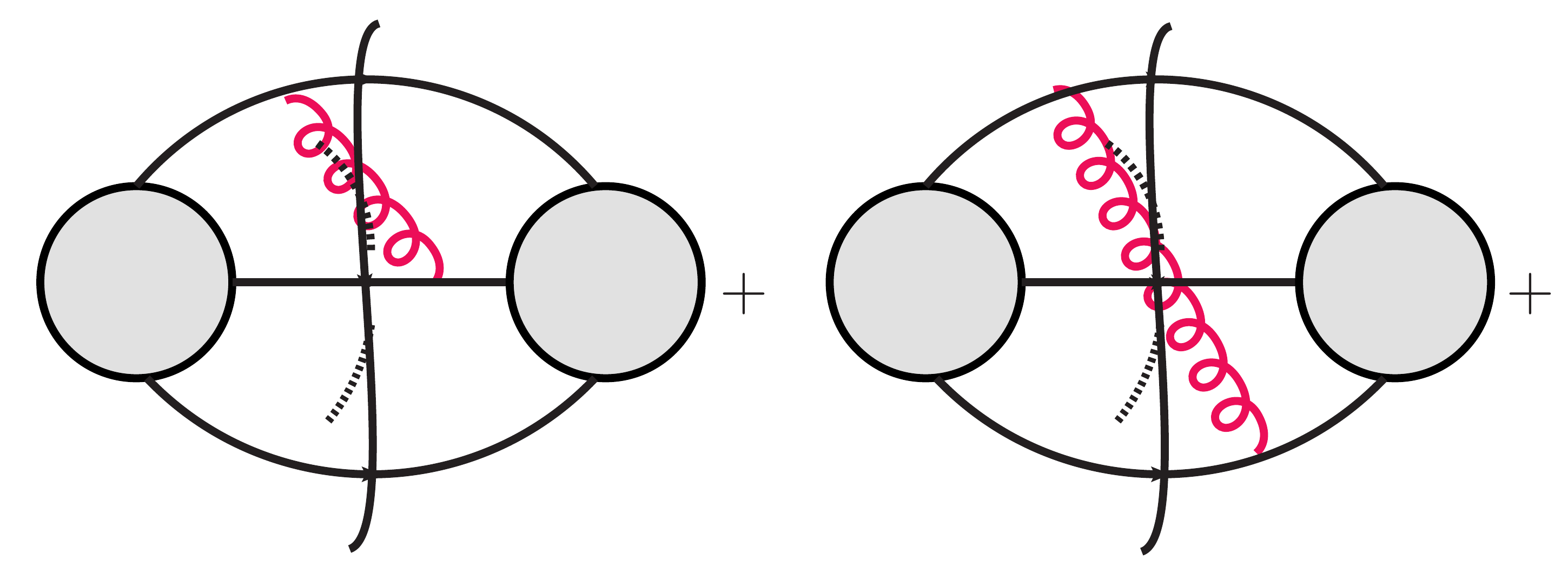}

\noindent
The gauge terms give zero contribution on on-shell matrix elements due to
gauge invariance.

\subsection{Factorization of $\lp|\mathcal{M}_{n}\rp|^{2}$ in the
collinear limit} 
\label{sec:colllimit}

The collinear limit of momenta $p_i$ and $p_r$ is defined by Sudakov
parametrization:
\begin{equation*} 
p_{i}^{\mu}= z_{i}p^{\mu}+k_{i\perp}^{\mu} 
-\frac{k_{i\perp}^{2}}{z_{i}}\,\frac{n^{\mu}}{2\,p\,\ldot\,n}
\,, \qquad
p_{r}^{\mu}= z_{r}p^{\mu}+k_{r\perp}^{\mu} 
-\frac{k_{r\perp}^{2}}{z_{r}}\,\frac{n^{\mu}}{2\,p\,\ldot\,n} 
\end{equation*} 
where $k_{i\perp}^{\mu}+k_{r\perp}^{\mu}\,=\,0$ and 
$z_{i}+z_{r}\,=\,1$. The momentum $p^{\mu}$ is the collinear direction 
and 
\[
p^{2}=p_i^2=p_r^2=n^2=0\,,\qquad
k_{i\perp}\cdot p=k_{r\perp}\cdot n=0
\,,
\]
In the collinear limit 
$k_{i\perp}^{\mu},\,k_{r\perp}^{\mu}\,\rightarrow\,0$ and 
$s_{ir}=-\frac{k_{r\perp}^{2}}{z_{i}z_{r}}$. We now    
state the following theorem
\begin{quote}
\emph{In a physical gauge, the leading collinear singularities are due
to the collinear splitting of an external parton.}
\end{quote}
This means that we need to compute
~~~~~~~~~~~~~~~~~~~~~~~~~~~~~~~~~~~~~~
in the collinear limit. There are three cases:

\vspace*{-35pt} ~\hspace*{128pt}
\includegraphics[width=0.2\linewidth]{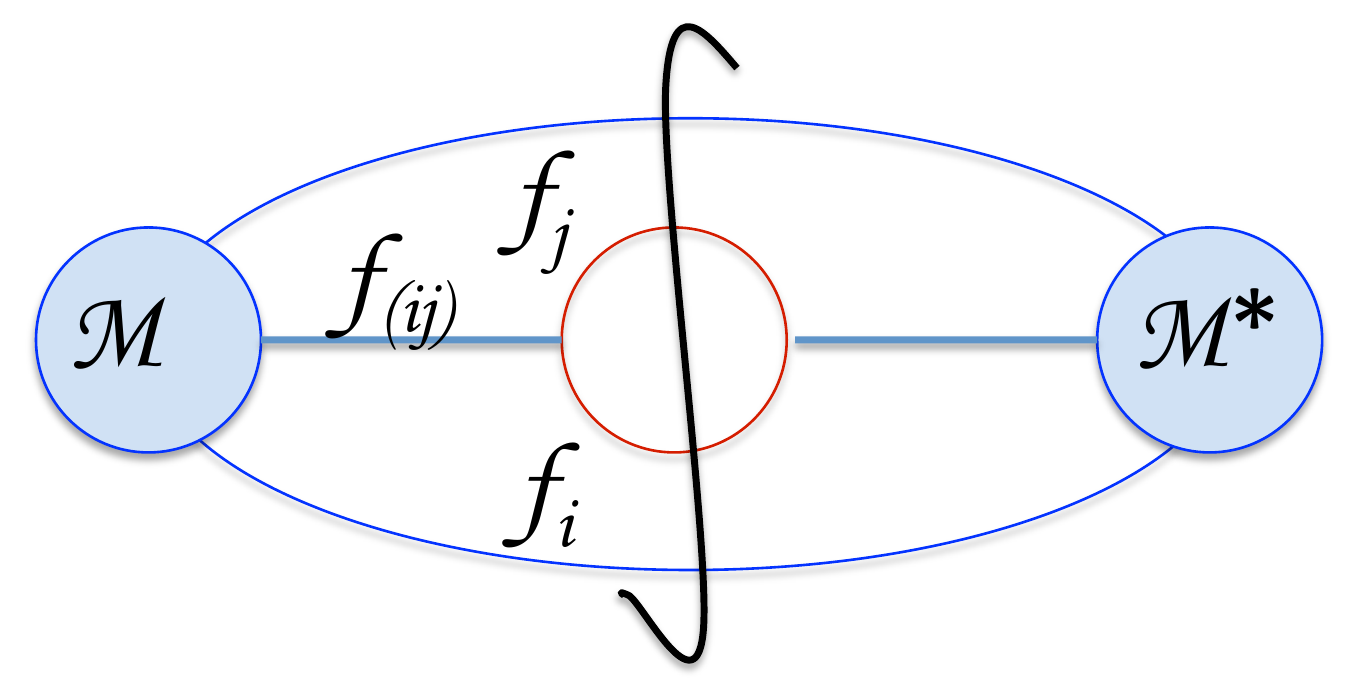}
\begin{center}
\begin{tabular}{ccc}
 $f_{ir}$&$\rightarrow$&$f_{i}\,+\,f_{r}$\\
\hline
 $q$&$\rightarrow$&$q\,+\,g$\\
 $g$&$\rightarrow$&$q\,+\,\bar{q}$\\
 $g$&$\rightarrow$&$g\,+\,g$
\end{tabular}
\end{center}
We compute explicitly the first case and leave the second and the third as
exercise.

For the case of a quark splitting into a quark and a gluon we have
\begin{equation} 
\begin{split} 
&= 
C_{\rF}\gs^{2}\mu^{2\epsilon} 
\,\frac{\slashed{p}_{i}+\slashed{p}_{r}}{s_{ir}} 
\,\gamma^{\mu}\slashed{p}_{i}\gamma^{\nu}\,d_{\mu\nu}\lp(p_{r},n\rp) 
\frac{\slashed{p}_{i}+\slashed{p}_{r}}{s_{ir}}
\\ &
= C_{\rF}\,4\pi\as\mu^{2\epsilon} 
\,\frac{\slashed{p}_{i}+\slashed{p}_{r}}{s_{ir}} 
\lp(-\gamma^{\mu}\slashed{p}_{i}\gamma_{\mu}\, 
+\,\frac{\slashed{p}_{r}\slashed{p}_{i}\slashed{n} 
+\slashed{n}\slashed{p}_{i}\slashed{p}_{r}}{p_{r}\,\ldot\,n}\rp) 
\,\frac{\slashed{p}_{i}+\slashed{p}_{r}}{s_{ir}}. 
\label{eq:Pqggraph} 
\end{split}
\end{equation}

~\hspace*{28pt}
\includegraphics[width=0.1\linewidth]{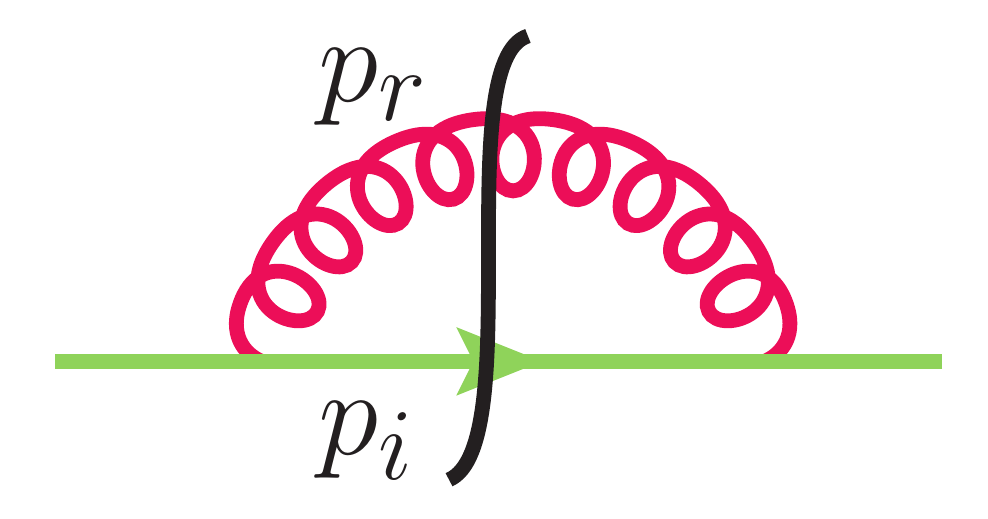}

\noindent
Using
\[ 
-\gamma^{\mu}\slashed{p}_{i}\gamma_{\mu}= \lp(d-2\rp)\slashed{p}_{i}
\,,\qquad 
\slashed{p}_{i}\slashed{p}_{i}=p_{i}^{2}\,\mathbbm{1}
\,,\quad \mathrm{and}\quad
\slashed{p}_{i}\slashed{p}_{r}\slashed{p}_{i}= 
s_{ir}\slashed{p}_{i}-p_{i}^{2}\slashed{p}_{r}=s_{ir}\slashed{p}_{i} 
\,,
\]
we find
\begin{eqnarray*} 
&&\lp(\slashed{p}_{i}+\slashed{p}_{r}\rp)\lp(-\gamma^{\mu}\slashed{p}_{i} 
\gamma_{\mu}\rp)\lp(\slashed{p}_{i}+\slashed{p}_{r}\rp)= 
\lp(d-2\rp)s_{ir}\slashed{p}_{r}\,, 
\\ && 
\slashed{p}_{r}\slashed{p}_{i}\slashed{n}= 
-\slashed{p}_{i}\slashed{p}_{r}\slashed{n}+s_{ir}\slashed{n}\,= 
\,\slashed{p}_{i}\slashed{n}\slashed{p}_{r}-2\slashed{p}_{i}p_{r}\ldot 
n+s_{ir}\slashed{n}= 
=-\slashed{n}\slashed{p}_{i}\slashed{p}_{r}+2p_{i}\ldot n\slashed{p}_{r} 
-2\slashed{p}_{i}p_{r}\ldot n+s_{ir}\slashed{n}\,. 
\end{eqnarray*}
Then
\begin{eqnarray*} 
&&\lp(\slashed{p}_{i}+\slashed{p}_{r}\rp) 
\lp(\slashed{p}_{r}\slashed{p}_{i}\slashed{n} 
+\slashed{n}\slashed{p}_{i}\slashed{p}_{r}\rp) 
\lp(\slashed{p}_{i}+\slashed{p}_{r}\rp)\,= 
\\[1pt]&&\qquad\qquad= 
\,2\lp(\slashed{p}_{i}+\slashed{p}_{r}\rp)\lp(p_{i}\ldot n 
\,\slashed{p}_{r}-p_{r}\ldot n\, \slashed{p}_{i}+p_{i}\ldot p_{r} 
\,\slashed{n}\rp)\lp(\slashed{p}_{i}+\slashed{p}_{r}\rp) 
\\[1pt]&& \qquad\qquad= 
2\lp[p_{i}\ldot n 
\,s_{ir}\slashed{p}_{i}-p_{r}\ldot n\,s_{ir}\slashed{p}_{r}+p_{i}\cdot 
p_{r}\lp(2\lp(p_{i}+p_{r}\rp)\ldot n 
\lp(\slashed{p}_{i}+\slashed{p}_{r}\rp) 
-\lp(p_{i}+p_{r}\rp)^{2}\slashed{n}\rp)\rp] 
\\[1pt]&& \qquad\qquad= 
s_{ir}\lp[4p_{i}\ldot n 
\,\slashed{p}_{i}+2p_{i}\ldot n 
\,\slashed{p}_{r}+2p_{r}\ldot n 
\,\slashed{p}_{i}-s_{ir}\slashed{n}\rp]. 
\end{eqnarray*}
Substituting these results and then the Sudakov parametrization of the
momenta into \eqn{eq:Pqggraph} we obtain
\begin{equation*} 
\begin{split}
&\stackrel{p_{i}\|p_{r}}{\simeq} 
\,\frac{1}{s_{ir}}\,C_{\rF}\,4\pi\as\mu^{2\epsilon} 
\,\lp[2\lp(1-\epsilon\rp)z_{r}+4\frac{z_{i}^{2}}{z_{r}} 
+4z_{i}+\mathcal{O}\lp(k_{\perp}\rp)\rp]\slashed{p} 
\\ &~~= 
\frac{1}{s_{ir}}\,C_{\rF}\,8\pi\as\mu^{2\epsilon} 
\,\lp[2\frac{z_{i}}{z_{r}}+\lp(1-\epsilon\rp)z_{r}\rp]\slashed{p}  
= \frac{1}{s_{ir}}\,C_{\rF}\,8\pi\as\mu^{2\epsilon} 
\left[\frac{1+z_{i}^{2}}{1-z_{i}}-\epsilon\lp(1-z_{i}\rp)\right]\slashed{p} 
\end{split}
\end{equation*}
Similarly to the soft case we can define an operator
$\hat{C}_{ir}$ which takes the collinear limit and keeps the leading
singular ($\mathcal{O}(1/k_\perp^2)$) terms:
\begin{equation} 
\hat{C}_{ir}\lp|\mathcal{M}_{m+1}^{\lp(0\rp)}\rp|^{2}=
8\pi\as\mu^{2\epsilon}\frac{1}{s_{ir}} 
\la\mathcal{M}_{m}^{\lp(0\rp)}\lp(p,\ldots\rp)\rp| 
\hat{P}_{qg}^{\lp(0\rp)}\lp(z_{i},z_{r},k_{\perp};\epsilon\rp) 
\lp|\mathcal{M}_{m}^{\lp(0\rp)}\lp(p,\ldots\rp)\ra\,. 
\label{eq:collfact}
\end{equation}
The kernel $\hat{P}_{qg}$, called Altarelli-Parisi splitting function
for the process $q\rightarrow\,q\,+\,g$, is diagonal in the spin-state
of the parent (splitting) parton:
\begin{equation*} 
\la\,s\rp|\hat{P}_{qg}\lp|s'\ra= 
C_{\rF}\lp[2\frac{z_{i}}{z_{r}}+\lp(1-\epsilon\rp)z_{r}\rp]\delta_{ss'}. 
\end{equation*}
Similar calculations give the splitting kernels for the gluon splitting
processes, which however, contain azimuthal correlations of the parent
parton
\begin{eqnarray} 
&& 
\la\mu\rp|\hat{P}_{q\bar{q}}^{(0)}\lp(z_{i},z_{r},k_{\perp};\epsilon\rp) 
\lp|\nu\ra=T_{R}\lp[-g^{\mu\nu}+4z_{i}z_{r}\, 
\frac{k_{\perp}^{\mu}k_{\perp}^{\nu}}{k_{\perp}^{2}}\rp] 
\\ && 
\la\mu\rp|\hat{P}_{gg}^{(0)}\lp(z_{i},z_{r},k_{\perp};\epsilon\rp)\lp|\nu\ra 
=2C_{\rA}\lp[-g^{\mu\nu}\lp(\frac{z_{i}}{z_{r}} 
+\frac{z_{r}}{z_{i}}\rp)-2\lp(1-\epsilon\rp)z_{i}z_{r} 
\,\frac{k_{\perp}^{\mu}k_{\perp}^{\nu}}{k_{\perp}^{2}}\rp].\quad~ 
\end{eqnarray}
The soft and collinear limits overlap when the soft gluon is also
collinear to its parent parton:
\begin{equation*} 
\hat{C}_{jr}\,\hat{S}_{r}
\,\lp|\mathcal{M}_{m+1}^{\lp(0\rp)}\lp(p_{r},\ldots\rp)\rp|^{2}= 
-8\pi\as\mu^{2\epsilon}\sum_{k\ne j} 
\,\frac{2z_{j}}{s_{jr}\,z_{r}}\lp|\mathcal{M}_{m\lp(j,k\rp)}^{\lp(0\rp)} 
\lp(\ldots\rp)\rp|^{2}\
= 
8\pi\as\mu^{2\epsilon}\,\bT_j^2
\,\frac{2}{s_{jr}}\,\frac{z_{j}}{z_{r}}\lp|\mathcal{M}_{m}^{\lp(0\rp)}\rp|^{2}
\,. 
\end{equation*}

The notation for the splitting kernels in these lectures is different from
the usual notation in the literature. Usually, 
$\hat{P}_{ij}^{(0)}\lp(z,k_{\perp};\epsilon\rp)$ denotes the splitting
kernel for the process $f_i(p) \to f_j(z p) + f_k((1-z) p)$, which does
not lead to confusion for $1 \to 2$ splittings because the momentum
fraction of parton $j$ determines that of parton $k$ as their sum has to
be one. For splittings involving more partons, it is more appropriate
to introduce as many momentum fractions $z_i$ as the number of
offspring partons, with the constraint $\sum_i z_i = 1$, and use the
flavour indices to denote the offspring partons in the order of the
momentum fractions in the argument. For $1\to 2$ splittings this means
the use of $\hat{P}_{ir}^{(0)}\lp(z_{i},z_{r},k_{\perp};\epsilon\rp)$
for the splitting process $f_k(p) \to f_i(z_i p) + f_r(z_r p)$. The
flavour of the parent parton $f_k$ is determined uniquely by the
flavour summation rules, $q+g = q$, $q+\bar{q} = g + g = g$. These flavour
summation rules are unique also for $1\to 3$ splittings.

\vspace{-4pt}
\noindent\rule{\textwidth}{1pt}

 
\begin{exe} 
 
Compute the Altarelli-Parisi-splitting function 
$\hat{P}_{qg}(z)$ for the process $q\rightarrow q g$ from 
the collinear limit of the matrix element for the process 
$e^+e^-\rightarrow q\bar{q}g$: 
\[ 
\lp|{\cal M}\left(e^+e^-\rightarrow 
q\bar{q}g\right)\rp|^{2}\propto 
\left(\left(1-\epsilon\right)\left(\frac{y_{23}}{y_{13}}+\frac{y_{13}}{y_{23}}\right)+2\left( 
\frac{y_{12}}{y_{13}y_{23}}-\epsilon\right)\right) 
\,. 
\] 
 
\end{exe} 
 
 
\begin{exe} 
 
The Altarelli-Parisi splitting function $\hat{P}_{q\bar{q}}\left(z\right)$ for the process $g\rightarrow q\bar{q}$ is defined by the following collinear limit: 
\begin{eqnarray*} 
\langle\mathcal{M}_{n+1}^{\left(0\right)}\left(p_{i},p_r,\ldots\right)\bigl.\bigr|\mathcal{M}_{n+1}^{\left(0\right)}\left(p_{i},p_r,\ldots\right)\rangle\, 
\stackrel{p_i\|p_r}{\simeq} 
\frac{1}{s_{ir}} 8 \pi \alpha_{\rm s} \mu^{2\epsilon}\, 
\langle\mathcal{M}_{n}^{\left(0\right)} 
\left(p,\ldots\right)\bigl|\hat{P}^{(0)}_{q\bar{q}}\left(z,k_\perp\right)\bigr| 
\mathcal{M}_{n}^{\left(0\right)}\left(p,\ldots\right)\rangle\, \aand
\\[2mm] \quad 
=\frac{1}{s_{ir}} 8 \pi \alpha_{\rm s} \mu^{2\epsilon}\, 
\langle\mathcal{M}_{n}^{\left(0\right)}\left(p,\ldots\right)\bigl|\bigr.\mu\rangle 
\langle\mu\left|\hat{P}^{(0)}_{q\bar{q}}\left(z,k_\perp\right)\right|\nu\rangle 
\langle\nu\bigr|\mathcal{M}_{n}^{\left(0\right)}\left(p,\ldots\right)\rangle &&
\\[2mm] \quad 
=\frac{1}{s_{ir}} 8 \pi \alpha_{\rm s} \mu^{2\epsilon}\, 
\langle\mathcal{M}_{n}^{\left(0\right)}\left(p,\ldots\right)\bigl|\bigr.\mu\rangle\,\frac{d_{\mu\rho}}{s_{ir}}\:\Pi^{\rho\sigma}\:\frac{d_{\sigma\nu}}{s_{ir}}\langle\nu\bigr|\mathcal{M}_{n}^{\left(0\right)}\left(p,\ldots\right)\rangle\,. &&
\end{eqnarray*} 
Compute $\langle\mu\left|\hat{P}^{(0)}_{q\bar{q}}\left(z,k_\perp\right)\right|\nu\rangle$ 
in leading order in $k_\perp$. 
Hint: 
In which sense does $\Pi_{\mu\nu}=d_{\mu\rho}\Pi^{\rho\sigma}d_{\sigma\nu}$ hold? 
 
\end{exe} 

\begin{exe} 
Derive the flavour summation rules for $1\to 3$ splittings.
\end{exe} 

\begin{exe} 
Compute the soft limit of \eqn{eq:collfact} and the collinear limit of
\eqn{eq:softfact}.
\end{exe} 

\noindent\rule{\textwidth}{1pt}

\subsection{Regularization of real corrections by subtraction} 
\label{sec:subtractions}

The cross section at NLO accuracy is a sum of two terms, the LO prediction
and the corrections at one order higher in the strong coupling,
\begin{equation*}
\sigma_{\NLO}=\sigma^{\LO}+\sigma^{\NLO}
\,,
\end{equation*}
where $\sigma^{\LO}$ is the integral of the fully differential Born cross
section over the available phase space defined by the jet function, while
$\sigma^{\NLO}$ is the sum of the real and virtual corrections:
\begin{equation*}
\sigma^{\LO}=
\int_{m}\!\rd\sigma^{B}\,J_{m}{\lp(\{p\}_m\rp)}
\,,\qquad
\sigma^{\NLO}=
\int_{m+1}\!\rd\sigma^{R}\,J_{m}{\lp(\{p\}_{m+1}\rp)}
+\int_{m}\!\rd\sigma^{V}\,J_{m}{\lp(\{p\}_m\rp)}\,.
\end{equation*}
Both contributions to $\sigma^{\NLO}$ are divergent in four dimensions,
but their sum is finite for IR-safe jet functions.

The factorization of the squared matrix elements in the soft and collinear
limits allows for a process and observable independent method to regularize
the real corrections in their singular limits. The essence of the method
is to devise an approximate cross section $\rd\sigma^{A}$ that matches the
singular behaviour of the real cross section $\rd\sigma^{R}$ in all
kinematically degenerate regions of the phase space when one parton
becomes soft or two partons become collinear. Then we subtract this
approximate cross section from the real one and the difference can be
integrated in four dimensions. Next, we integrate $\rd\sigma^{A}$ over
the phase space of the unresolved parton and we add it to 
$\rd\sigma^{V}$. The integrated subtraction term cancels the explicit
poles in the virtual correction and the sum can also be integrated in
four dimentions. The key for this procedure is a proper mapping of the
$(m+1)$-parton phase space to the $m$-parton one which respects the
limits, thus the approximate cross section is defined with the
$m$-parton jet function. This way we can rewrite the NLO correction as
a sum of two finite terms, 
\begin{equation}
\sigma^{\NLO}=
\int_{m+1}\lp[\rd\sigma^{R}\,J_{m}{\lp(\{p\}_{m+1}\rp)}
-\rd\sigma^{A}\,J_{m}{\lp(\{\tilde{p}\}_m\rp)}
\rp]_{\epsilon=0}\,
+\,\int_{m}\lp[\rd\sigma^{V}+\int_{1}\!\rd\sigma^{A}\rp]_{\epsilon=0}
\,J_{m}{\lp(\{p\}_m\rp)}
\,.
\label{eq:sigmanlo}
\end{equation}
The definition of the approximate cross section is not unique and the best
choice may depend on further requirements that we do not discuss here. We
also skip the precise definition of the momenta $\tilde{p}^\mu$ which are
obtained by mapping the $(m+1)$-particle phase space onto an $m$-particle
phase space times a one-particle phase space. A widely used general
subtraction scheme that can be used also for processes including
massive partons with smooth massless limits is presented in
Ref.~\cite{Catani:2002hc}, where these definitions are given
explicitly. This method uses the factorization of the SME in the soft
and collinear limits. The challange posed by the overlapping
singularity in the soft-collinear limit is solved by a smooth
interpolation between these singular regions.  

The factorization properties of \eqns{eq:softfact}{eq:collfact} play
other very important roles in pQCD. The numerical implementation of the
SME is in general a process prone to errors. Testing the factorization
in the kinematically degenerate phase space regions serves a good check
of the implementation. The computation is even more difficult for the
virtual corrections. Similar factorization holds for those. The
factorized form of the SME can be used in resumming logarithmically
enhanced terms at all orders, or in devising a parton shower algorithm
for modelling events (see \sect{sec:events}). The splitting kernels
that appear in the collinear factorization have a role in the evolution
equations of the parton distribution functions (see \sect{sec:DGLAP}).

The state of the art in making precision predictions assaults on the one
hand the full automation of computations at NLO, and on the other the
realm of next-to-next-to-leading order (NNLO) corrections. The automation
of computing jet cross sections at NLO accuracy has been accomplished and
several programs are available with the aim to facilitate automated
solutions for computing jet cross sections at NLO accuracy:
\begin{itemize}
\item aMC@NLO (\underline{http://amcatnlo.web.cern.ch})
\item BlackHat/Sherpa (\underline{https://blackhat.hepforge.org})
\item FeynArts/FormCalc/LoopTools (\underline{http://www.feynarts.de})
\item GoSam (\underline{https://gosam.hepforge.org}) 
\item HELAC-NLO (\underline{http://helac-phegas.web.cern.ch})
\item MadGolem (\underline{http://www.thphys.uni-heidelberg.de/~lopez/madgolem-corner.html}).
\end{itemize}

\begin{figure}[ht]
\begin{center}
\includegraphics[width=0.42\textwidth]{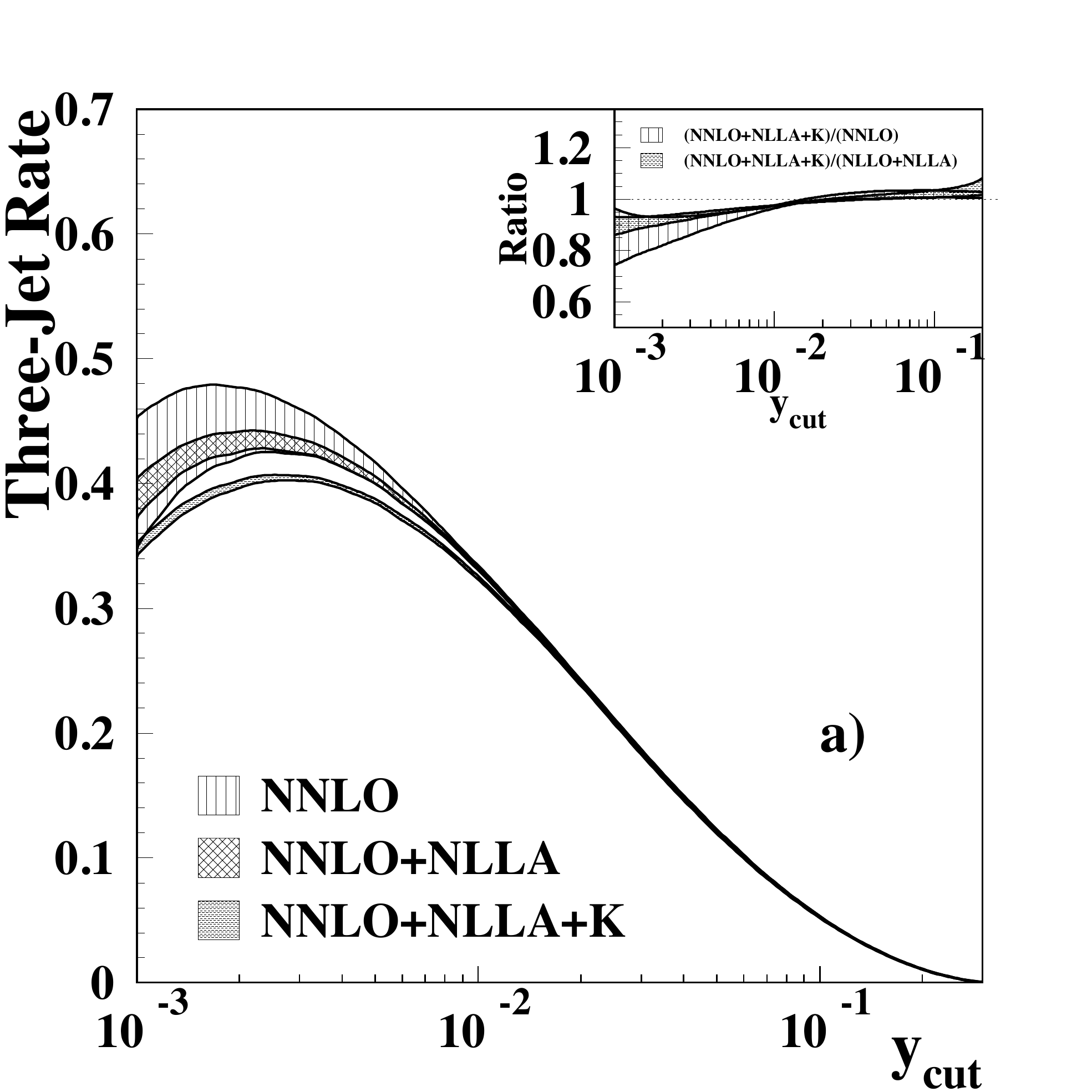}
\includegraphics[width=0.42\textwidth]{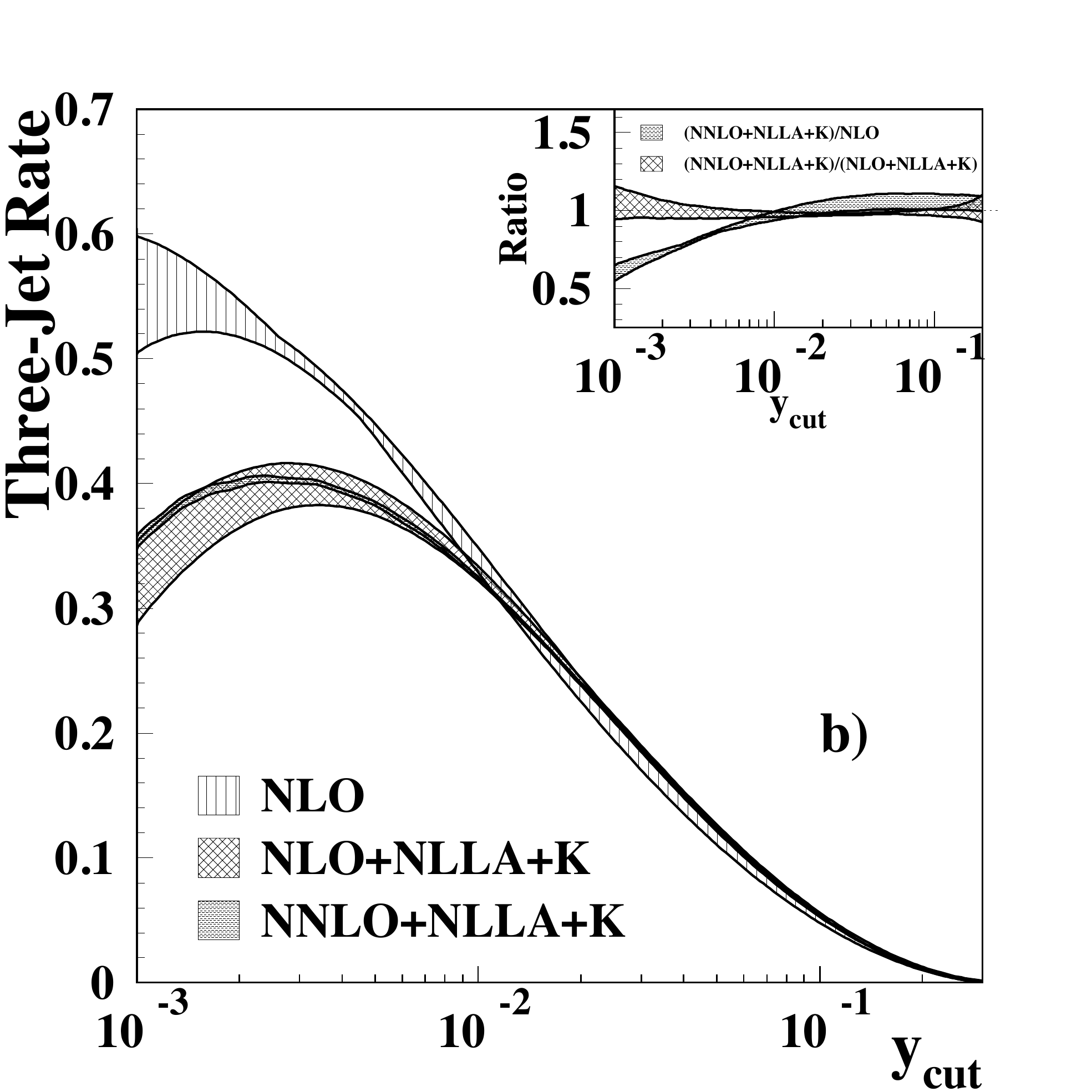}
\caption{QCD predictions for the three-jet rate in electron-positron
annihilation \cite{Schieck:2012mp}
}
\label{fig:3jetnnlo}
\end{center}
\end{figure}

In the NNLO case the IR singularity structure is much more involved
than in the case of NLO computations due to complicated
overlapping singly- and doubly-unresolved configurations. Several
subtraction methods have been proposed for the regularization of the IR
divergences and there is intense research to find a general one that
can be automated. To provide an impression about the importance of
NNLO corrections, we present QCD predictions at various accuracies for
the three-jet rate computed with $\alpha_{\rm s} =0.118$ and at a
centre-of-mass energy of $\sqrt{s}=35$\,GeV in \fig{fig:3jetnnlo}.
Figure \ref{fig:3jetnnlo}(a) shows comparison of prediction at NLO with
that at matched NLO and resummed next-to-leading logarithmic (denoted
by NLLA in the figure) accuracy, while \fig{fig:3jetnnlo}(b) presents
comparison of prediction at NNLO with that at matched NNLO and resummed
NLL accuracy.  The inserts in both cases show the ratio between the
matched and the unmatched predictions.  For all calculations the
uncertainty band reflects the uncertainty due to the variation of the
renormalization scale around the default scale $\mu = \sqrt{s}$ by
factors of 2 in both directions.

\section{Deeply inelastic lepton-proton scattering}
\label{sec:DIS}

\begin{figure}[ht]
\begin{center}
\includegraphics[width=0.49\textwidth]{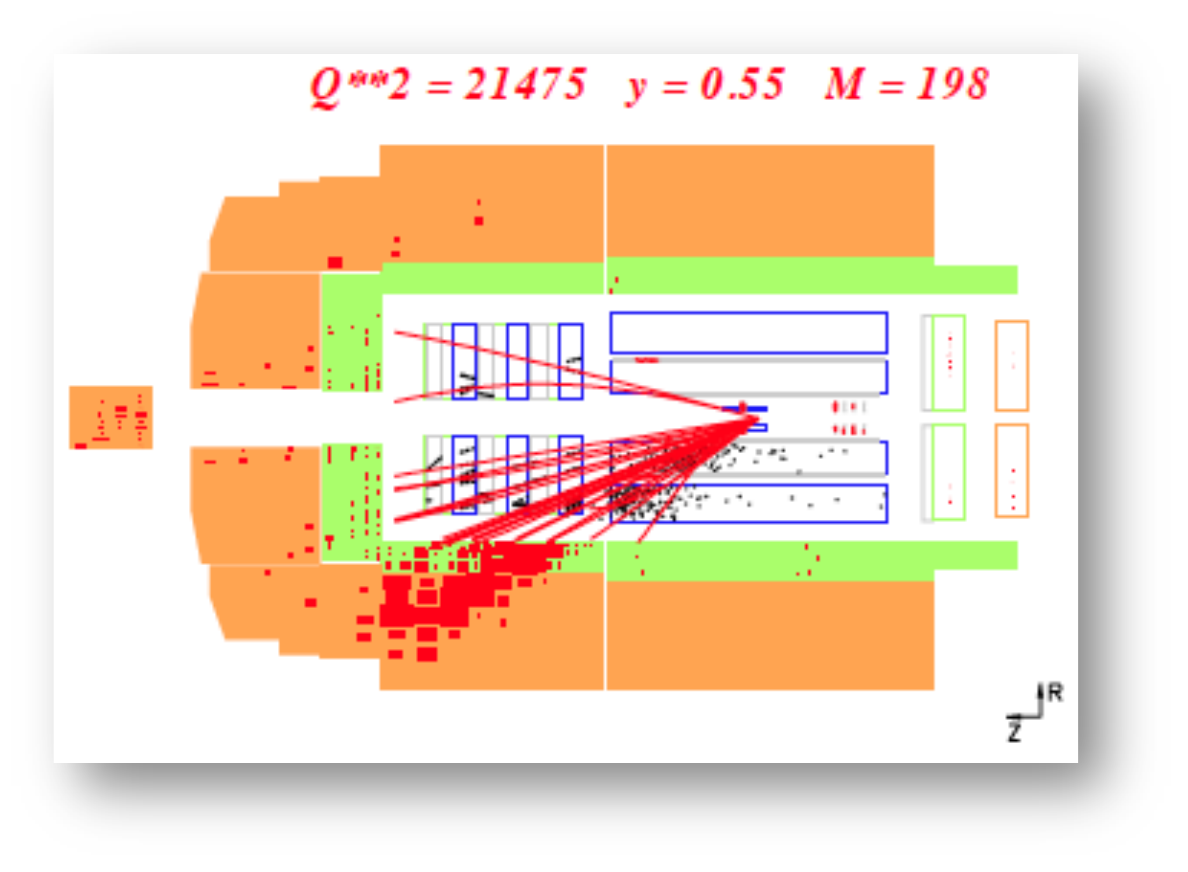}
\includegraphics[width=0.49\textwidth]{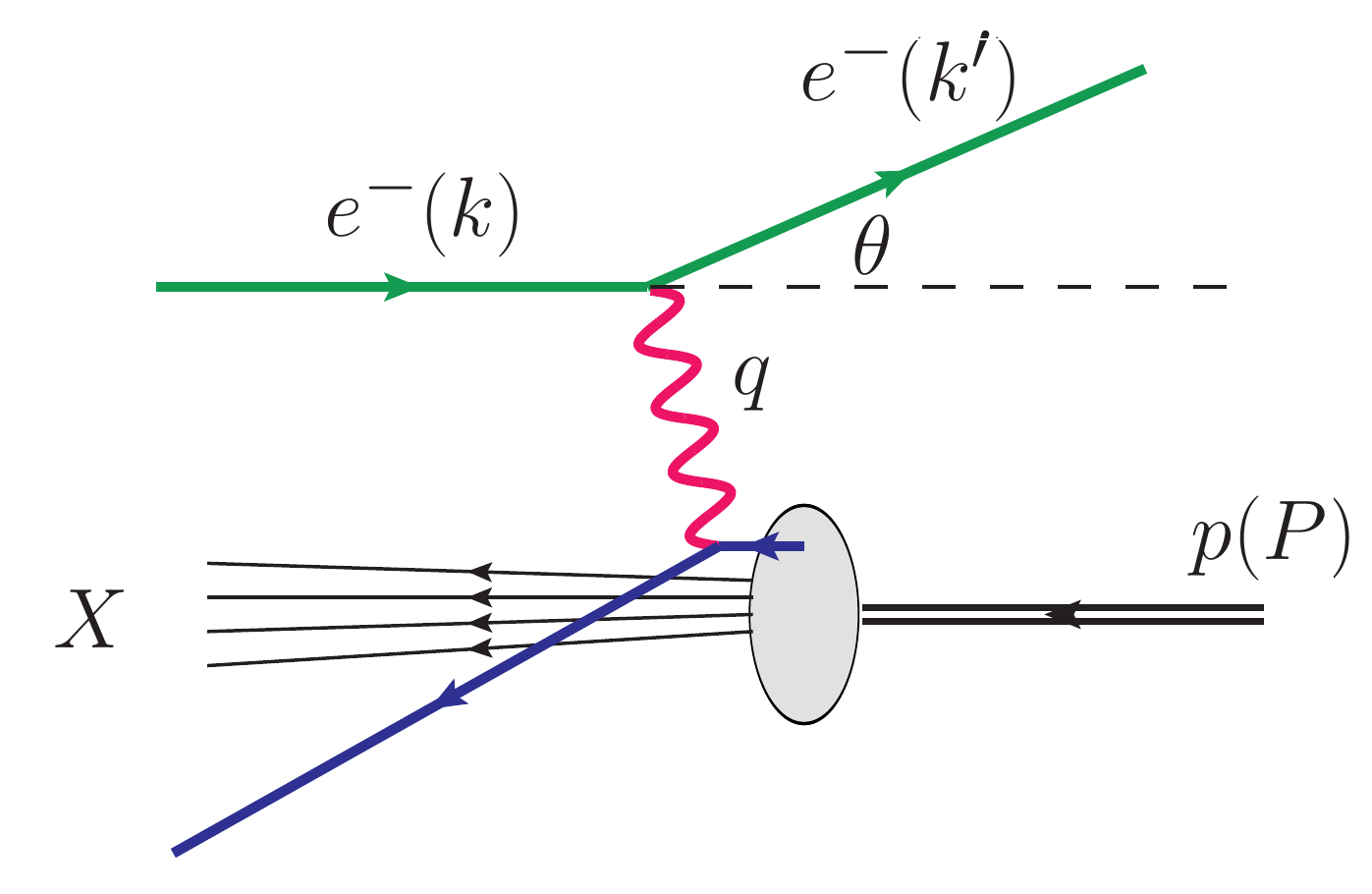}
\caption{Deeply inelastic lepton-proton scattering (a) in the H1
detector and (b) parton model interpretation of such an event
}
\label{fig:DIS}
\end{center}
\end{figure}

Perturbative QCD stems from the parton model that was developed to
understand deeply inelastic lepton-hadron scattering (DIS). The purpose
of those experiments was to study the structure of the proton by
measuring the kinematics of the scattered lepton. In \fig{fig:DIS}(a)
we show a real event in the H1 experiment at the HERA collider. The
value of $Q^2$, which is the modulus squared of the momentum transfer
between the lepton and the proton is 21475\,GeV$^2 >> 1$\,GeV$^2$,
signifying that the scattering is well in the deeply inelastic region.
The parton model interpretation of the event is shown in \fig{fig:DIS}(b):
the lepton is scattered by an angle $\theta$ due to the exchange of a
virtual photon with one of the constituents of the proton (a parton).
The measurement is inclusive from the point of view of hadrons ($X$ means
any number of hadrons that are not observed separately), thus the process
can be described in pQCD.\\

\vspace*{-10pt}
The DIS kinematics is described by the following varibales
\begin{equation*} 
\begin{split} 
\mbox{centre-of-mass energy}^2 &= s = (P+k)^2\,,\\
\mbox{{\color[rgb]{1.000000,0.000000,0.000000} momentum transfer}}
&={\color[rgb]{1.000000,0.000000,0.000000} q^\mu} 
= k^\mu- k'^\mu\,,\\
{\color[rgb]{1.000000,0.000000,0.000000}|\mbox{ momentum transfer}|^2}
&= {\color[rgb]{1.000000,0.000000,0.000000} Q^2} = -q^2 = 2 MExy \,,\\
\mbox{{\color[rgb]{1.000000,0.000000,0.000000} scaling variable}}
&= {\color[rgb]{1.000000,0.000000,0.000000} x} 
= Q^2/(2 P\cdot q)\,,\\
\mbox{energy loss}
&= \nu = (P\cdot q)/M = E-E'\,,\\
\mbox{{\color[rgb]{1.000000,0.000000,0.000000} relative energy loss}}
&= {\color[rgb]{1.000000,0.000000,0.000000} y} 
= (P\cdot q)/(P\cdot k) = 1-E'/E\,,\\
\mbox{recoil mass}^2 &= W^2 = (P+q)^2 = M^2 + \frac{1-x}{x} Q^2
\,,
\end{split} 
\end{equation*} 
where we set the more important ones for these lectures in red. 

\subsection{Parametrization of the target structure} 
\label{sec:structure}

The cross section for $e(k) + p(P) \to e(k') + X$ reads
\begin{equation}
{\rm d}\sigma = \sum_X \frac{1}{4 M E}\int\!{\rm d}\phi\,
\frac14\sum_{{\rm spin}}|{\cal M}|^2
\,.
\label{eq:dsigmaDIS}
\end{equation}
We factorize the phase space and the SME into two parts, one for the
lepton and one for the hadrons:
\begin{equation*}
{\rm d}\phi = \frac{{\rm d}^3 k'}{(2\pi)^3 2 E'}\,{\rm d}\phi_X
\,,\qquad
\frac14\sum_{{\rm spin}}|{\cal M}|^2 =
\frac{e^4}{Q^4}L^{\mu\nu}H_{\mu\nu}
\,.
\end{equation*}
Then the hadron part of the cross section is the dimensionless Lorentz
tensor $W_{\mu\nu} = \frac{1}{8\pi}\sum_X\int\!{\rm d}\phi_X H_{\mu\nu}$
(the factor of $\frac{1}{8\pi}$ is included here by convention).
As it depends on two momenta $P^\mu$ and $q^\mu$, the most general gauge
invariant combination of the Lorentz tensor can be written as
\begin{equation*}
W_{\mu\nu}(P,q) = 
\left(-g_{\mu\nu} + \frac{q_\mu q_\nu}{q^2}\right) W_1(x,Q^2) +
\left(P_{\mu} - q_\mu \frac{P \cdot q}{q^2}\right)
\left(P_{\nu} - q_\nu \frac{P \cdot q}{q^2}\right)
\frac{W_2(x,Q^2)}{P\cdot q}
\,,
\end{equation*}
where the structure functions $W_i(x,Q^2)$ are dimensionless functions of
the scaling variable and the momentum transfer.

For the lepton part we express the kinematical relations 
$E'=(1-y)E$, $\cos\vartheta = 1-\frac{x y M} {(1-y) E}$
to change variables to scaling variable and relative energy loss:
\begin{equation*}
\frac{{\rm d}^3 k'}{(2\pi)^3 2 E'} =
\frac{{\rm d}\varphi}{2\pi} \frac{E'}{8\pi^2}\,{\rm d}E'\,{\rm
d}\cos\vartheta
= \frac{{\rm d}\varphi}{2\pi} \, \frac{yM E}{8\pi^2} \,{\rm d}y\,{\rm d}x
\,,
\end{equation*}
and compute the trace 
$
L^{\mu\nu} = \frac12 Tr[\slashed{k}\gamma^\mu\slashed{k}'\gamma^n] =
k^\mu k^{'\nu} + k^\nu k^{'\mu} -g^{\mu\nu} k\cdot k'
\,.
$
Then the differential cross section in $x$ and $y$ is obtained from 
\eqn{eq:dsigmaDIS} as
\begin{equation*}
\frac{{\rm d}^2 \sigma}{{\rm d} x\,{\rm d} y} =
\frac{4\pi \alpha^2}{y\,Q^2}\left[y^2 W_1(x,Q^2)+\left(\frac{1-y}{x}-x y
\frac{M^2}{Q^2}\right) W_2(x,Q^2)\right]
\,,
\end{equation*}
which we rewrite in the scaling limit, defined by $Q^2 \to \infty$ with $x$
fixed, as
\begin{equation}
\frac{{\rm d}^2 \sigma}{{\rm d} x\,{\rm d} y} =
\frac{4\pi \alpha^2}{y\,Q^2}\left[\big(1+(1-y)^2\big) F_1+\frac{1-y}{x}
\big(F_2-2 x F_1\big)\right]
\,.
\label{eq:DIShadron}
\end{equation}

\begin{wrapfigure}{l}{0.6\linewidth}
\includegraphics[width=1.0\linewidth]{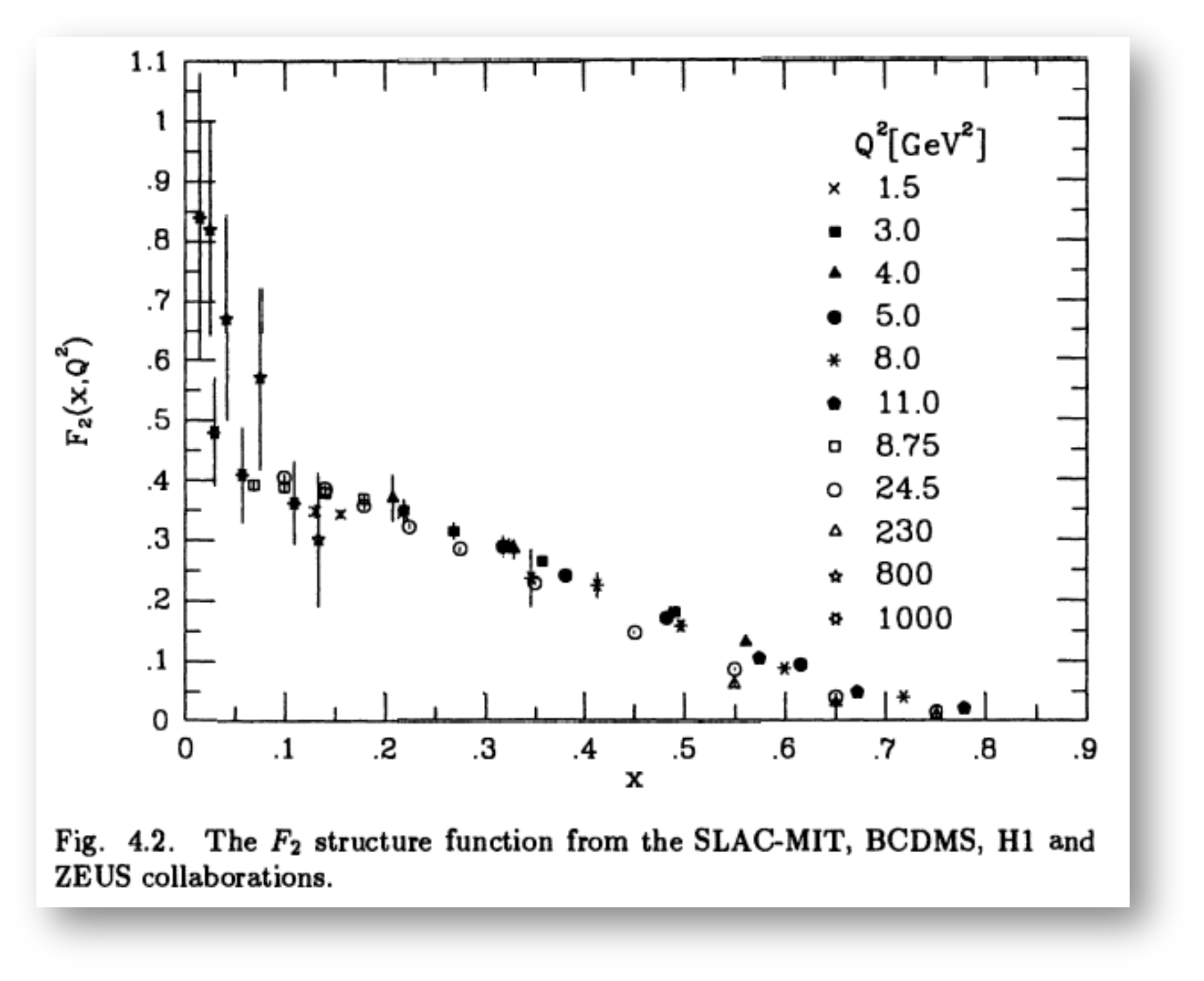}
\vspace*{-15pt}
\caption{Measured value of the $F_2$ structure function at several
different values of $Q^2$
~}
\label{fig:F2}
\end{wrapfigure}
The dimensionless functions $F_1$ and $F_2$ were first measured by the
SLAC-MIT experiment \cite{Miller:1971qb}. The result of that measurement
supplemented by some later ones is shown in \fig{fig:F2}. The interesting
feature is that in the scaling limit $F_2$ becomes independent of $Q^2$,
$F_2(x,Q^2) \to F_2(x)$ (in fact, the independence starts at quite low
values of $Q^2$).

\subsection{DIS in the parton model} 
\label{sec:partonmodel}

Let us now describe the same scattering process by assuming the proton is
a bunch of free flying quarks and the lepton exchanges a hard virtual
photon with one of those quarks as shown in \fig{fig:DIS}(b). The struck
quark carries a momentum $p^\mu$, which is a fraction of the proton
momentum, $p^\mu = \xi P^\mu$, so we consider the process $e(k) + q(p) \to
e(k') + q(p')$. The corresponding cross section is 
\begin{equation*}
{\rm d}\hat{\sigma} = \frac{1}{2 \hat{s}}\int\!{\rm d}\phi_2\,
\frac14\sum_{{\rm spin}}|{\cal M}|^2
\,,
\end{equation*}
with $\hat{s} = (p+k)^2$.
The SME is proportional to the product of the lepton tensor $L^{\mu\nu}$
and a similar quark tensor
$ Q_{\mu\nu} = \frac12 Tr[\slashed{q}\gamma^\mu\slashed{q}'\gamma^n] =
q^\mu q^{'\nu} + q^\nu q^{'\mu} -g^{\mu\nu} q\cdot q' $,
\ie $ L^{\mu\nu}Q_{\mu\nu} = 2 (\hat{s}^2+\hat{u}^2) $,
where $\hat{u} = (p-k')^2 = - 2 p \cdot k'$. As
$y = P\cdot q/P\cdot k = 2 p\cdot q/2 p\cdot k = (\hat{s}+\hat{u})/\hat{s}$,
momentum conservation, $p'_\mu = p_\mu+q_\mu$, implies for the
on-shell condition of the scattered quark
$0 = p^{'2} = (p+q)^2 = 2p\cdot q +q^2 = \hat{s}+\hat{u} - Q^2$. We have
$y = Q^2/\hat{s}$ and $\hat{u} = (y-1) \hat{s}$, so 
\[
\frac14\sum_{{\rm spin}}|{\cal M}|^2 =
\frac{e_q^2 e^4}{Q^4}L^{\mu\nu}Q_{\mu\nu} =
2e_q^2 e^4 \frac{\hat{s}^2}{Q^4} \big(1+(1-y)^2\big)
\,.
\]
Also $Q^2 = 2p\cdot q = 2 \xi P\cdot q$, so $p^{'2} = Q^2 (\xi/x -1)$.
Then the two-particle phase space is
\[
{\rm d}\phi_2 =
\frac{{\rm d}^3 k'}{(2\pi)^3 2 E_{k'}}\,
\frac{{\rm d}^4 p'}{(2\pi)^4} \,2\pi\delta_+\big(p^{'2}\big)\,
{(2\pi)^4\,\delta^4(k+p-k'-p')} = 
\frac{{\rm d}\varphi}{2\pi} \frac{E'}{4\pi}\,{\rm d}E'
\,{\rm d}\cos\vartheta \,\frac{x}{Q^2}\,\delta(\xi-x)
\,,
\]
or using $E'=\frac{\displaystyle\sqrt{\hat{s}}}{\displaystyle 2} (1-y)$ and
$\cos\vartheta = 1-\frac{\displaystyle 2y x} {\displaystyle\xi (1-y)}$,
we obtain
$
{\rm d}\phi_2 = \frac{\displaystyle{\rm d}\varphi}{\displaystyle(4\pi)^2}
\frac{\displaystyle y\,\hat{s}}{\displaystyle Q^2}\,{\rm d}y
\,{\rm d}x\,\delta(\xi-x)
\,.
$
The differential cross section in $x$ and $y$ 
\begin{equation}
\frac{{\rm d}^2 \hat{\sigma}}{{\rm d} x\,{\rm d} y} =
\frac{4\pi \alpha^2}{Q^2}\big[1+(1-y)^2\big]\,\frac12 e_q^2 \delta(\xi-x)
\,.
\label{eq:DISparton}
\end{equation}

Comparing \eqns{eq:DIShadron}{eq:DISparton}, we find the parton model
predictions
\begin{equation}
F_1(x) \propto e_q^2 \delta(\xi-x)
\,,\qquad 
F_2 - 2 x F_1 = 0
\,,\quad \mbox{called Callan-Gross relation.}
\label{eq:CallanGross}
\end{equation}
Thus $F_2$ probes the quark constituent of the proton with $\xi = x$.
However, this prediction for $F_2$ cannot be correct because
$F_2(x)$ is not a $\delta$ function as seen from \fig{fig:F2}, which
leads us to formulate the na\"\i ve parton model in the following way:
\begin{quote}
\emph{the virtual photon scatters incoherently off the constituents
(partons) of the proton;\\
the probability that a quark $q$ carries momentum fraction of the proton
between $\xi$ and $\xi+\delta \xi$ is $f_q(\xi)\rd \xi$.}
\end{quote}

\vspace{-4pt}
\noindent\rule{\textwidth}{1pt}


\begin{exe}

Compute the contribution to the DIS cross section in \eqn{eq:DIShadron}
with the exchange of a transversely polarized photon. Hint: Use
\eqn{eq:polsum} for the numerator in the propagator of the transversely
polarized photon and the Callan-Gross relation in \eqn{eq:CallanGross}.
Can you identify the result with any of the terms in \eqn{eq:DIShadron}?
What is the source of the remainder?

\end{exe}
\rule{\textwidth}{1pt}

\subsection{Measuring the proton structure} 
\label{sec:protonstructure}

\begin{figure}[ht]
\begin{center}
\includegraphics[width=0.48\textwidth]{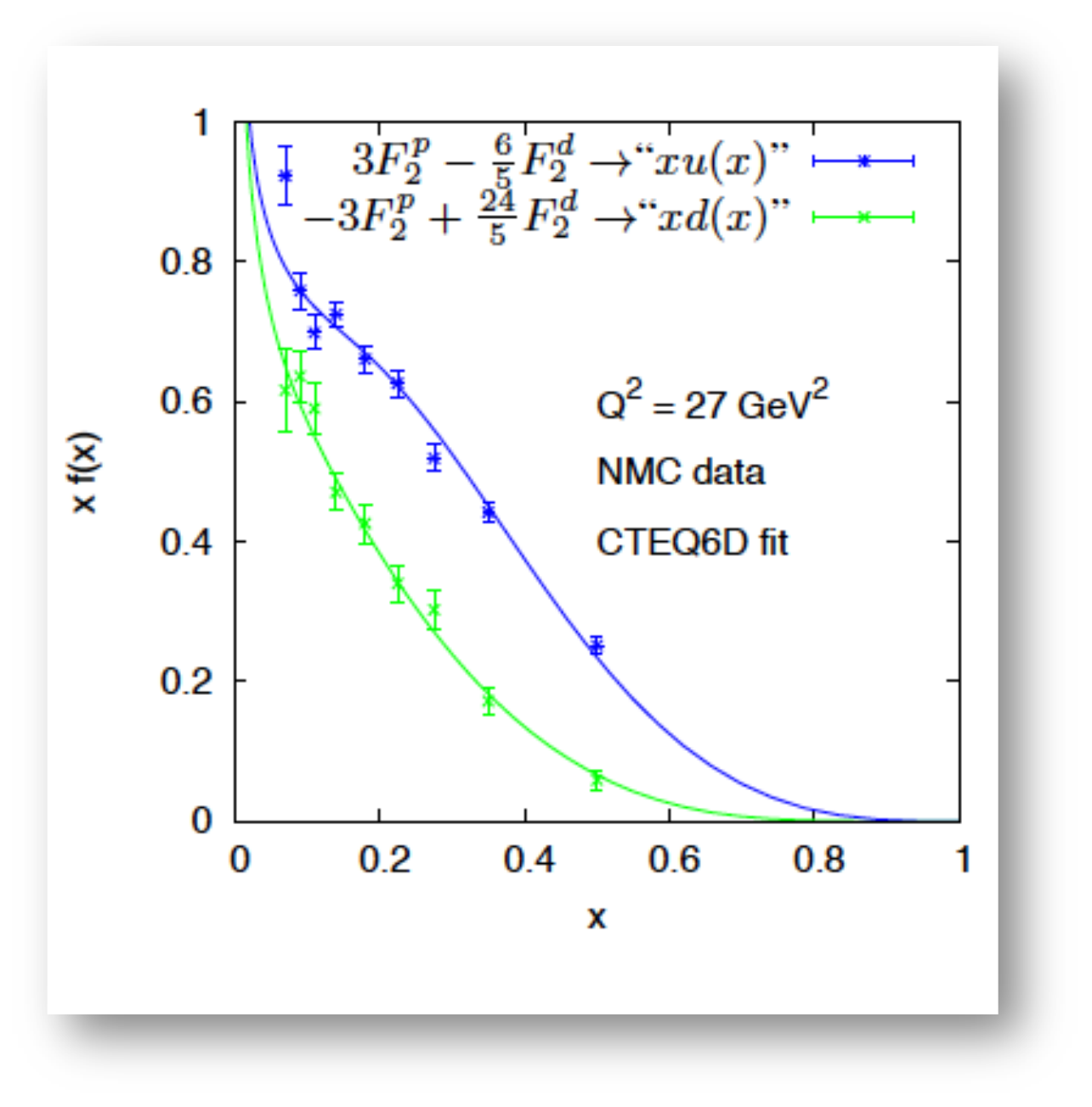}
\includegraphics[width=0.4\textwidth]{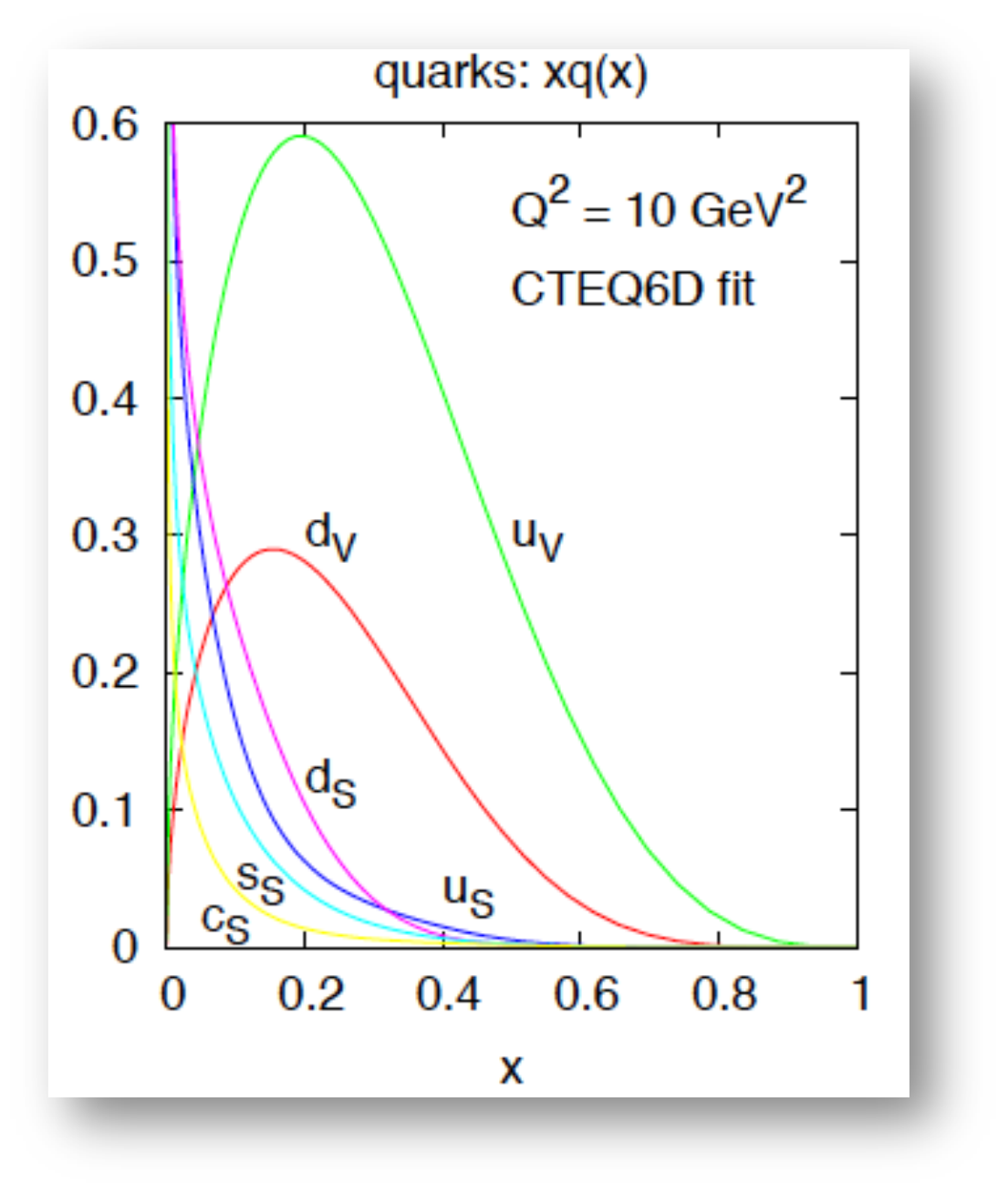}
\caption{(a) Measurement of combination of $F_2$ structure functions on
proton and deuteron targets by the NMC collaboration fit by CTEQ6D PDF
set, (b) CTEQ6D valence and sea quark distributions
}
\label{fig:pdf}
\end{center}
\end{figure}

With the assumptions of the na\"\i ve parton model the Callan-Gross
relation predicts
\begin{equation}
F_2(x) = 2xF_1(x)
= \sum_q \int_0^1\!{\rm d}\xi\,f_q(\xi)\,x\,e_q^2\,\delta(x-\xi)
= x\sum_q \,e_q^2\,f_q(x)
\,.
\label{eq:CallanGrossparton}
\end{equation}
Taking into account four flavours and simplifying the notation by using
$f_q(x) \equiv q(x)$, we obtain a prediction for the structure function
measured in scattering of charged-lepton off proton (neutral current
interaction):
\[
F_2^{\rm em}(x) = x
\bigg[\frac49\big(u(x)+\bar{u}(x)+c(x)+\bar{c}(x)\big)
+\frac19\big(d(x)+\bar{d}(x)+s(x)+\bar{s}(x)\big)\bigg]
\,.
\]
Similarly, in charged current interactions the prediction is
\[
F_2^{\bar{\nu}}(x) = 2x \big[u(x)+\bar{d}(x)+c(x)+\bar{s}(x)\big]
\;(\mbox{with }W^-)\,,\quad
F_2^{\nu}(x) = 2x \big[d(x)+\bar{u}(x)+s(x)+\bar{c}(x)\big]
\;(\mbox{with }W^+)\,.
\]
Further information can be obtained if we use different targets. Assuming
two flavours and isospin symmetry, the proton (with uud valence quarks)
structure is
\begin{equation}
F_2^{\rm proton}(x) = x
\bigg[\frac49\big(u_p(x)+\bar{u}_p(x)\big)
+\frac19\big(d_p(x)+\bar{d}_p(x)\big)\bigg]
\,,
\label{eq:protontarget}
\end{equation}
and that of the neutron (with udd valence quarks) is
\begin{equation}
F_2^{\rm neutron}(x) = x
\bigg[\frac49\big(u_n(x)+\bar{u}_n(x)\big)
+\frac19\big(d_n(x)+\bar{d}_n(x)\big)\bigg]
 = x
\bigg[\frac19\big(u_p(x)+\bar{u}_p(x)\big)
+\frac49\big(d_p(x)+\bar{d}_p(x)\big)\bigg]
\,.
\label{eq:neutrontarget}
\end{equation}
The measurements are supplemented by sum rules. For instance, as the
proton consists of uud valence quarks, we have
\[
\int_0^1\!{\rm d}x\,\big(u_p(x)-\bar{u}_p(x)\big) = 2
\,,\quad
\int_0^1\!{\rm d}x\,\big(d_p(x)-\bar{d}_p(x)\big) = 1
\,,\quad
\int_0^1\!{\rm d}x\,\big(s_p(x)-\bar{s}_p(x)\big) = 0
\,.
\]
The combination of the measurements and sum rules gives separate
information on the quark distributions in the proton $f_q(x)$. The
result of such measurements performed by the NMC collaboration
\cite{Arneodo:1996qe} is shown in \fig{fig:pdf}(a) together with a fit
to the data by the CTEQ collaboration \cite{Pumplin:2002vw}. The parton
distributions deduced from the fit are shown in \fig{fig:pdf}(b).

We can infer the proton momentum from the measurements. The surprising
result is that {\em quarks give only about half of the momentum of the
proton}, $\sum_q \int_0^1\!{\rm d}x\,x f_{q/p}(x) \simeq 0.5$. 
By now we know that the other half is carried by gluons, but clearly the
na\"\i ve parton model is not sufficient to interpret the gluon
distribution in the proton. With our experience in pQCD we try to compute
radiative corrections to the quark process to see if that helps to find
the role of the gluon distribution.

\newpage
\vspace*{-30pt}
\vspace{-4pt}
\noindent\rule{\textwidth}{1pt}


\begin{exe}

It is not feasible to use a neutron target experimentally. Instead
deuteron is used which is the bound state of a proton and a neutron. The
corresponding structure function is
$F_2^{\rm deuteron}(x) = \frac12 (F_2^{\rm proton}(x)+F_2^{\rm
neutron}(x))$, with $F_2^{\rm proton}$ and $F_2^{\rm neutron}$ given in
\eqns{eq:protontarget}{eq:neutrontarget}, respectively. Which combination
of the structure function on proton and deuteron targets gives the u- and
d-quark distributions?

\end{exe}
\rule{\textwidth}{1pt}

\subsection{Improved parton model: pQCD} 
\label{sec:DISpQCD}

Using the relations
${\rm d}y = {\rm d}Q^2/\hat{s}$ and
$\delta(\xi-x) = \frac{1}{\xi}\delta\left(1-\frac{x}{\xi}\right)$,
we rewrite the differential cross section (\ref{eq:DISparton}) in a more
usual notation,
\begin{equation}
\frac{{\rm d}^2 \sigma}{{\rm d} x\,{\rm d} Q^2} =
\int_0^1\frac{{\rm d}\xi}{\xi}\sum_i f_i(\xi)
\frac{{\rm d}^2 \hat{\sigma}}{{\rm d} x\,{\rm d}
Q^2}\left(\frac{x}{\xi},Q^2\right)
\,,
\label{eq:DISfactorization}
\end{equation}
which gives the cross section as a convolution of a long-distance
component (the PDF) and a short-distance component (the hard scattering
cross section). This form of the cross section is the main content of
the {\em factorization theorem}, which we derived heuristically, but a
rigorous proof, based on QFT exists.

The factorization formula (\ref{eq:DISfactorization}) raises some
questions. Knowing that the quarks do not give the total momentum of
the proton, it is natural to include the contribution of gluons in
\eqn{eq:DISfactorization}. However, we do not yet know the
corresponding hard scattering cross section. We also do not know how
we can apply PT. Furthermore, the scaling was exact in the parton model.
Is it so in QCD? There is a common answer to these questions: DIS in
pQCD. 

To develop pQCD for DIS, let us revisit the IR singularities once more.
Let us denote the hard scattering cross section for some final state by
$\sigma_h$. Then the cross section in the collinear approximation for
the same final state with an extra gluon of relative transverse
momentum $k_\perp = E \theta$, carrying momentum fraction $(1-z)$ is

\includegraphics[width=0.2\linewidth]{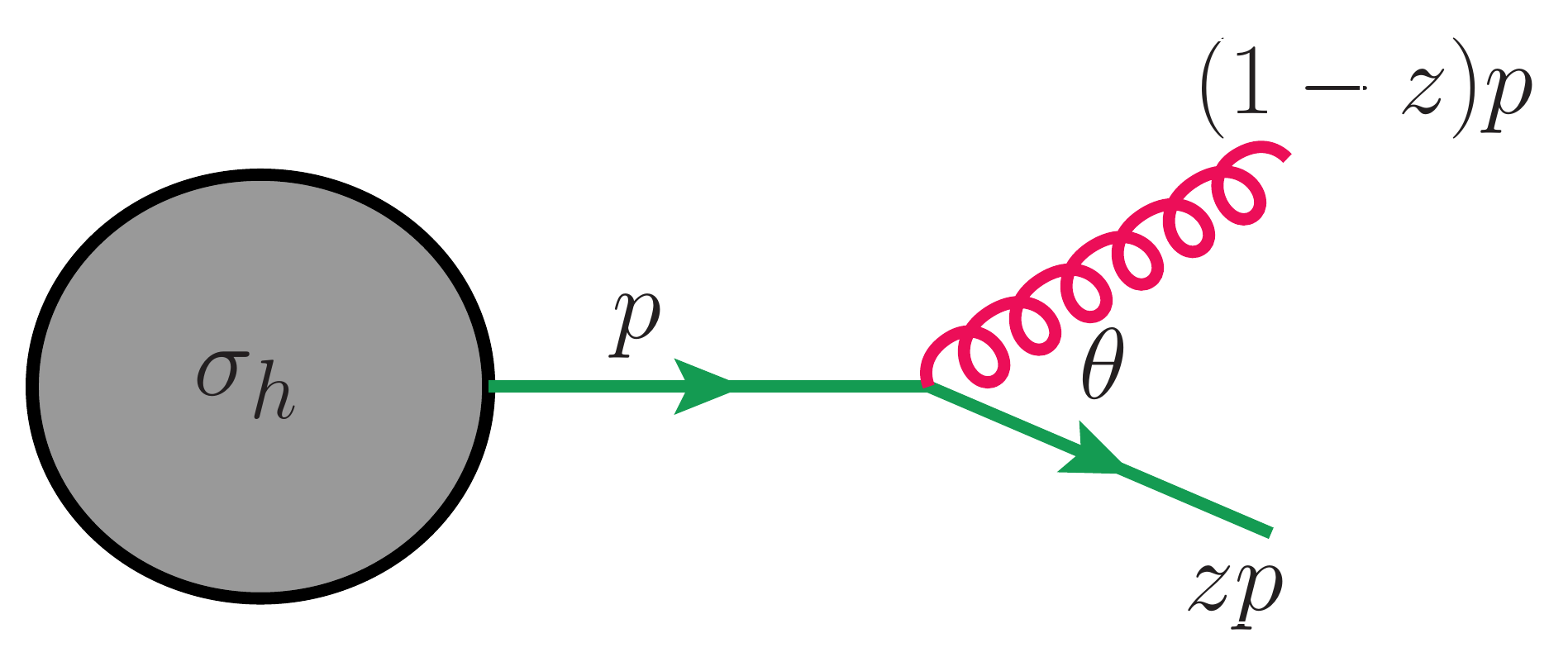}

\vspace*{-50pt} 
\begin{equation*}
:\qquad \sigma_{h+g} \simeq
\sigma_h 2 C_F \frac{\alpha_{\rm s}}{\pi}
\frac{{\rm d} E}{E} \frac{{\rm d} \theta}{\theta} =
\sigma_h C_F \frac{\alpha_{\rm s}}{\pi}
\frac{{\rm d} z}{1-z} \frac{{\rm d} k_\perp^2}{k_\perp^2}
\,.
\end{equation*}
Integrating over $z$ up to one and over $k_\perp$ we find soft and
collinear divergence, respectively.  In studying pQCD we found that
these IR singularities in the final state cancel against IR divergences
in the virtual correction for IR safe quantities:

\includegraphics[width=0.18\linewidth]{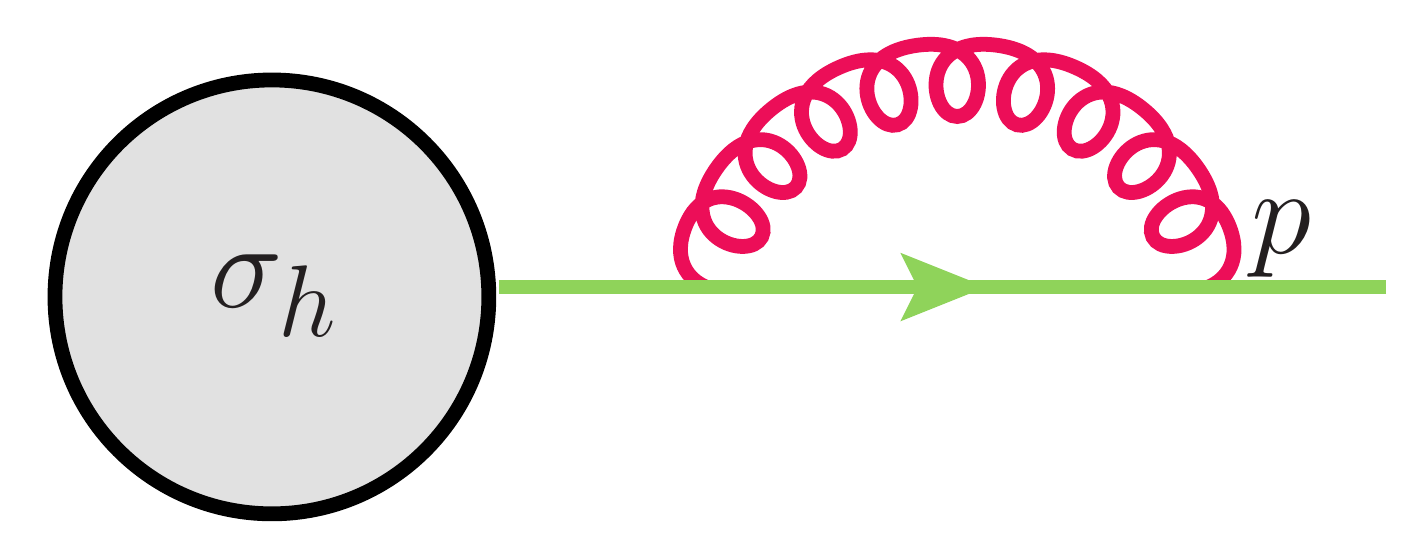}

\vspace*{-50pt} 
\begin{equation*}
:\qquad
\sigma_{h+V} \simeq
- \sigma_h C_F \frac{\alpha_{\rm s}}{\pi}
\frac{{\rm d} z}{1-z} \frac{{\rm d} k_\perp^2}{k_\perp^2}
\,.
\qquad\qquad \qquad\quad~
\end{equation*}

If there is a coloured parton in the initial state, then the splitting
may occur before the hard scattering and the momentum of the parton that
enters the hard process is reduced to $z p^\mu$, so

\includegraphics[width=0.2\linewidth]{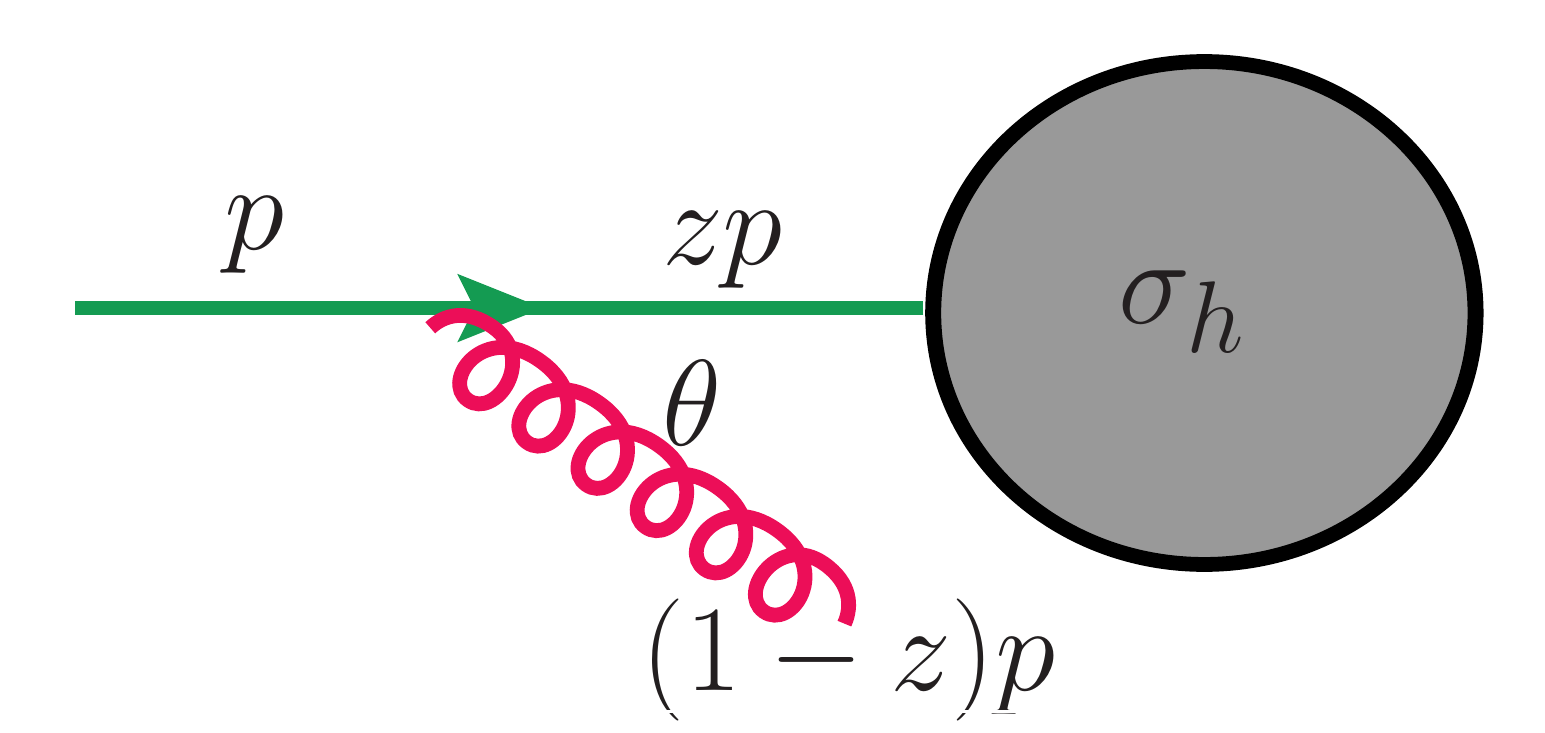}

\vspace*{-50pt} 
\begin{equation*}
:\qquad \sigma_{h+g}(p) \simeq
\sigma_h(zp) 2 C_F \frac{\alpha_{\rm s}}{\pi}
\frac{{\rm d} E}{E} \frac{{\rm d} \theta}{\theta} =
\sigma_h C_F \frac{\alpha_{\rm s}}{\pi}
\frac{{\rm d} z}{1-z} \frac{{\rm d} k_\perp^2}{k_\perp^2}
\,.
\end{equation*}
Integrating over $z$ up to one and over $k_\perp$ we again find soft and
collinear divergence, respectively. The corresponding $\epsilon$ poles
multiply $\sigma_h(zp)$, while in the virtual correction the poles
multiply $\sigma_h(p)$, irrespective whether the IR divergence is in the
initial or final state:

\includegraphics[width=0.18\linewidth]{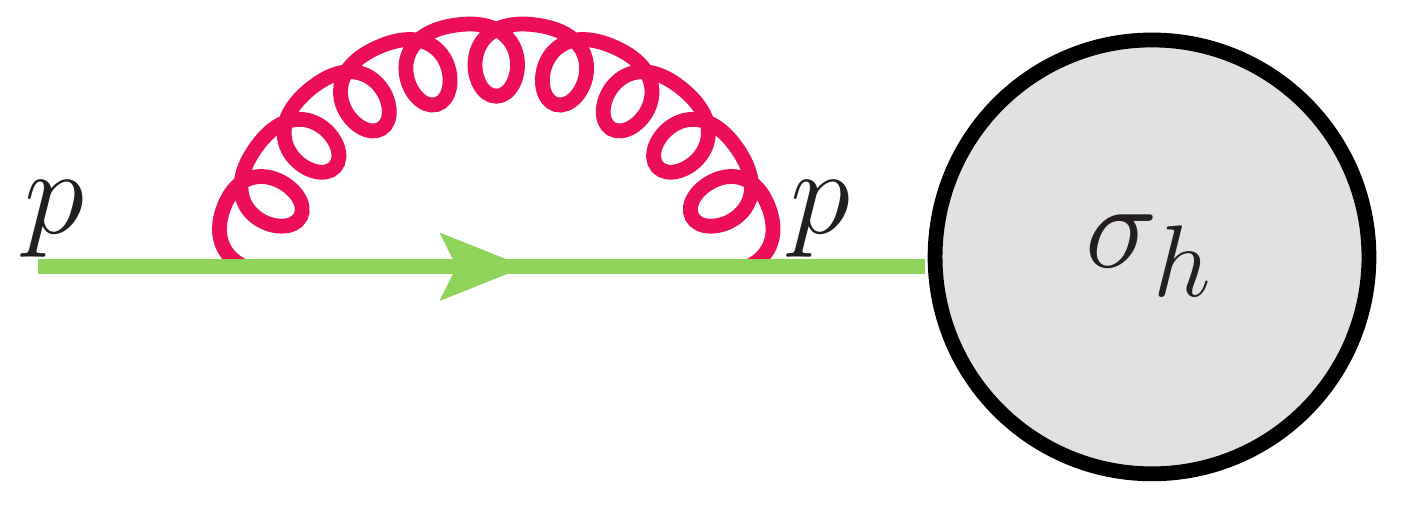}

\vspace*{-50pt} 
\begin{equation*}
:\qquad
\sigma_{h+V} \simeq
- \sigma_h C_F \frac{\alpha_{\rm s}}{\pi}
\frac{{\rm d} z}{1-z} \frac{{\rm d} k_\perp^2}{k_\perp^2}
\,.
\qquad\qquad \qquad\quad~
\end{equation*}
The sum of the real and virtual corrections then contains an uncancelled
singularity,
\begin{equation*}
\sigma_{h+g} + \sigma_{h+V} \simeq
C_F \frac{\alpha_{\rm s}}{\pi}
\underbrace{\int_{{\color[rgb]{1.000000,0.000000,0.000000} m_g^2}}^{Q^2}
\!\frac{{\rm d} k_\perp^2}{k_\perp^2}}_
{{\rm infinite\:if\:} m_g = 0}
\underbrace{\int_0^1\! \frac{{\rm d} z}{1-z} [\sigma_h(z p) -
\sigma_h(p)]}_
{{\rm finite}}
\,,
\end{equation*}
where we used a finite gluon mass to regulate the collinear divergence
(instead of dimensional regularization) to make manifest that the
collinear singularity remains, while the soft one (at $z\to 1$) vanishes
in the sum.

This uncancelled collinear singularity in the initial state is a
general feature of pQCD computations with incoming coloured partons and
its form is universal, so we can find its precise form studying the
structure function at NLO accuracy. We know that in the parton model (QCD
at LO) the prediction for hard scattering cross section $\hat{F}_2$ is 
finite:
\[
\hat{F}_{2,q}(x) =
\frac{{\rm d}^2 \hat{\sigma}}{{\rm d} x\,{\rm d} Q^2}\bigg|_{F_2} =
e_q^2 x\,\delta(1-x)
\,,\qquad
\hat{F}_{2,g}(x) =
\frac{{\rm d}^2 \hat{\sigma}}{{\rm d} x\,{\rm d} Q^2}\bigg|_{F_2} =
\sum_q e_q^2\:x\cdot 0
\]
\ie it is zero in the gluon channel because the virtual photon does not
interact with the gluon directly. At one order higher in $\alpha_{\rm s}$
we finds
\begin{equation}
\hat{F}_{2,q}(x) =
\frac{{\rm d}^2 \hat{\sigma}}{{\rm d} x\,{\rm d} Q^2}\bigg|_{F_2} =
e_q^2 x \bigg[ \delta(1-x)
+ \frac{\alpha_{\rm s}}{4\pi}\left(
P_{qg}(x) \ln\frac{Q^2}{m_g^2} +
C_2^q(x)\right)\bigg]\,,
\label{eq:F2q}
\end{equation}

\hspace*{180pt}
\includegraphics[width=0.1\linewidth]{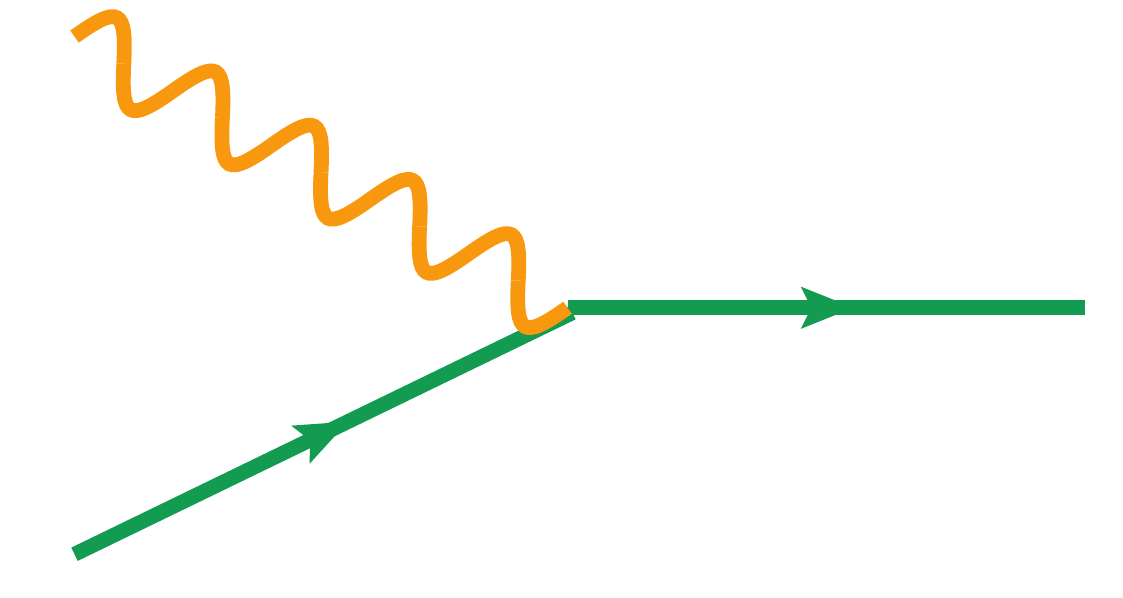}
\hspace*{20pt}
\includegraphics[width=0.1\linewidth]{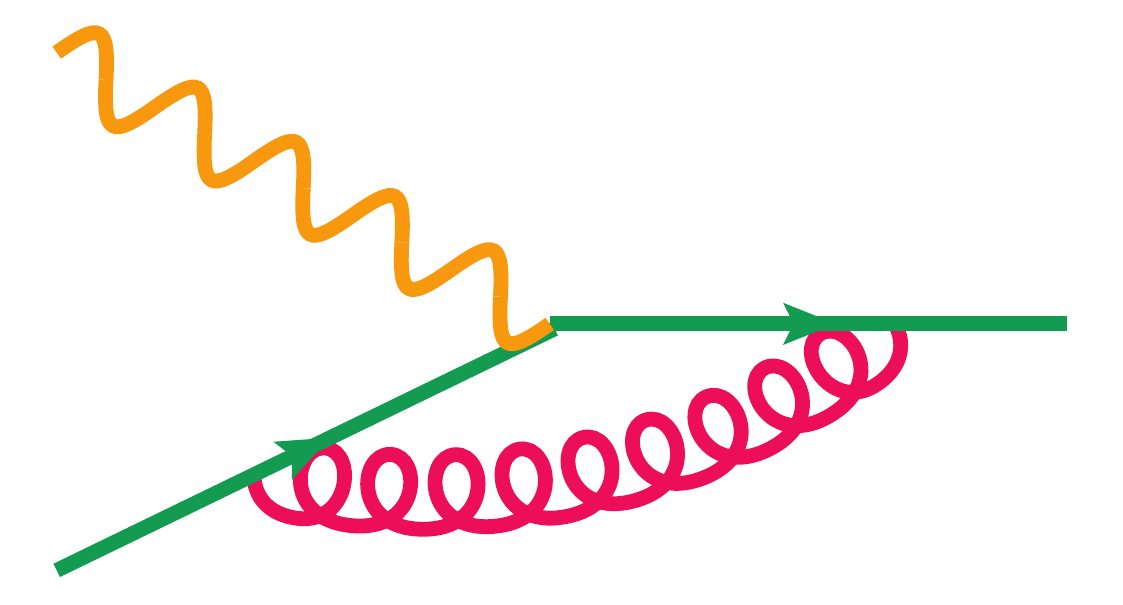}
\includegraphics[width=0.1\linewidth]{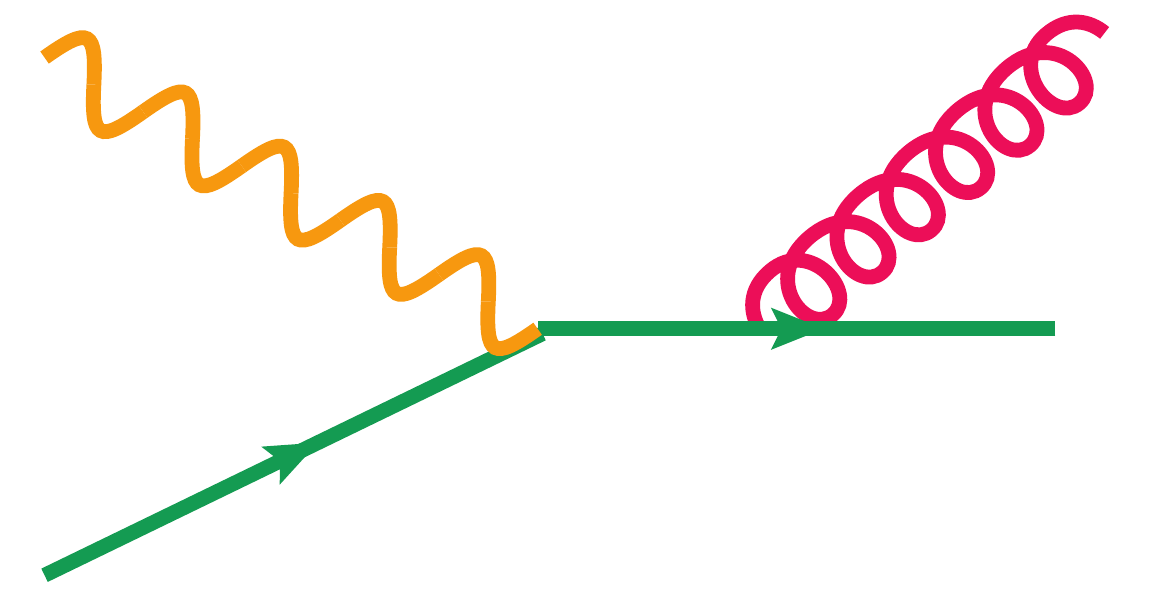}
\includegraphics[width=0.1\linewidth]{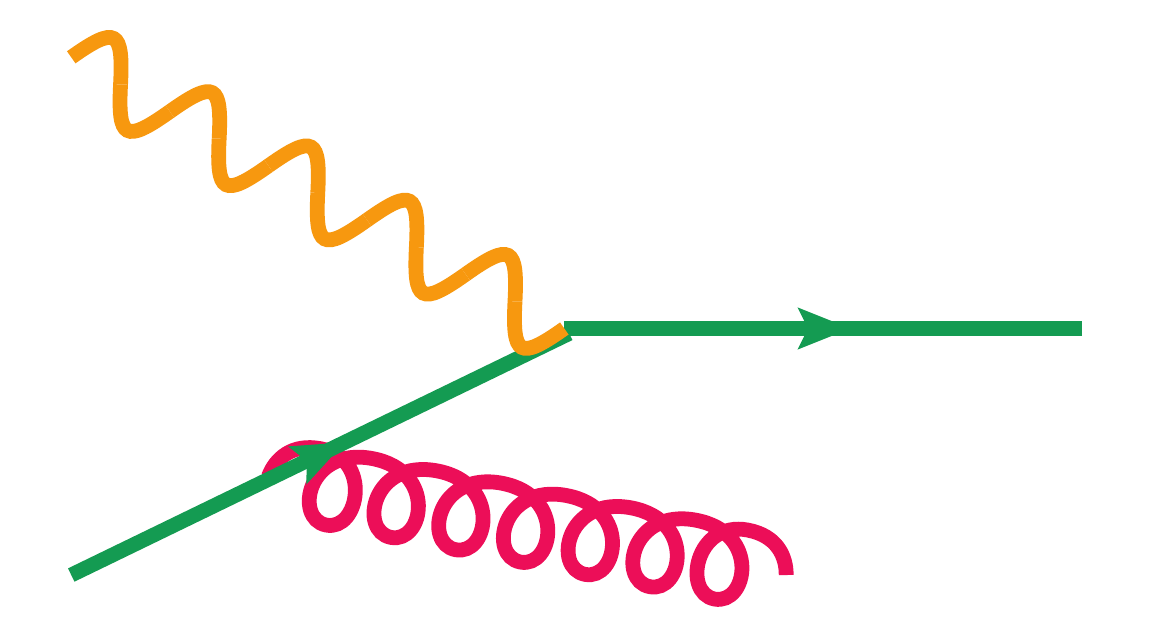}

and
\begin{equation}
\hat{F}_{2,g}(x) =
\frac{{\rm d}^2 \hat{\sigma}}{{\rm d} x\,{\rm d} Q^2}\bigg|_{F_2} =
\sum_q e_q^2 x
\bigg[ 0
+ \frac{\alpha_{\rm s}}{4\pi} \left(
P_{q\bar{q}}(x) \ln\frac{Q^2}{m_q^2} +
C_2^g(x)\right)\bigg]\,,
\quad~
\label{eq:F2g}
\end{equation}

\hspace*{260pt}
\includegraphics[width=0.1\linewidth]{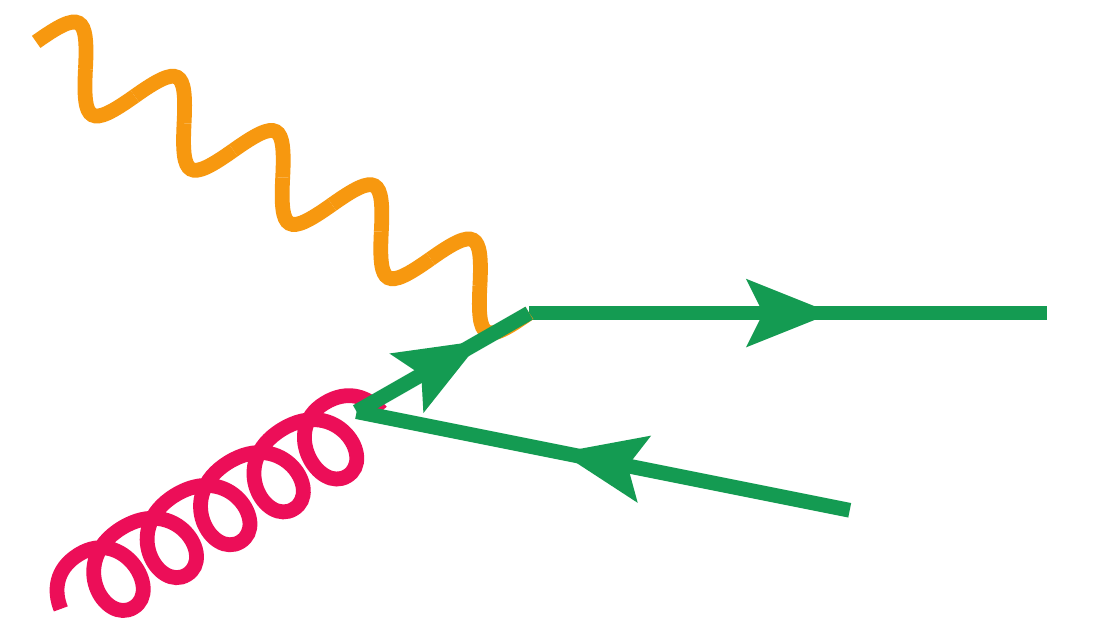}

\noindent where $P_{ij}(x)$ is the Altarelli-Parisi splitting function
(regularized at $x=1$), obtained from the splitting kernel $\hat{P}_{ij}$
in four dimensions by (i) averaging over the spin states of the splitting
parton and (ii) adding the contribution from the loop graphs, while
$C_2(x)$ is the remaining finite term, called coefficient function.
We see that at NLO the prediction for $\hat{F}_2$ is finite
in the UV and final state IR divergences cancel, but un-cancelled
singularity remains in the initial state IR, regularized with
a small mass here.

The hard scattering function is not measurable, only the structure
function is physical:
\begin{equation*}
F_{2,q}(x,Q^2) =
x \sum_i e_{q_i}^2\,
\bigg[f_{q_i}^{(0)}(x)
+ \frac{\alpha_{\rm s}}{2\pi}
\int_0^1\frac{{\rm d}\xi}{\xi}f_{q_i}^{(0)}(\xi) \left(
P_{qg}\left(\frac{x}{\xi}\right) \ln\frac{Q^2}{m_g^2} +
C_2^q\left(\frac{x}{\xi}\right)\right)\bigg]\,.
\end{equation*}
However, this function appears divergent if the regulator is removed,
$m_g \to 0$.
While $C_2(x)$ depends on the process under investigation, the divergence
does not because it is multiplied with universal splitting functions.

\vspace{-4pt}
\noindent\rule{\textwidth}{1pt}

 
\begin{exe} 
 
Compute the coefficient $C_2^g(x)$ in \eqn{eq:F2g}. 
 
\end{exe} 
\rule{\textwidth}{1pt}

\subsection{Factorization in DIS} 
\label{sec:DISfactorization}

If the remaining divergences are universal (and they are because do not
depend on the hard scattering), we can absorb the singularity into the
PDF's. For instance, defining
\begin{equation}
f_q(x,\mu_{\rm F}) =
f_q^{(0)}(x)
+ \frac{\alpha_{\rm s}}{2\pi}
\int_0^1\frac{{\rm d}\xi}{\xi}\bigg\{f_q^{(0)}(\xi)
\left[ P_{qg}\left(\frac{x}{\xi}\right) \ln\frac{\mu_{\rm F}^2}{m_g^2} +
z_{qq}\left(\frac{x}{\xi}\right)\right]\bigg\}
\,,
\label{eq:PDF}
\end{equation}
the structure function becomes
\begin{equation}
F_{2,q}(x,Q^2) =
x \sum_i e_{q_i}^2\,
\bigg\{f_i(x,\mu_{\rm F})
+ \frac{\alpha_{\rm s}(\mu_\rR)}{2\pi}
\int_0^1\frac{{\rm d}\xi}{\xi}
f_i(\xi,\mu_{\rm F}) \left[
P_{qg}\left(\frac{x}{\xi}\right)
\ln\frac{Q^2}{\mu_{\rm F}^2} +
\big(C_2^q-z_{qq}\big)\left(\frac{x}{\xi}\right)\right]\bigg\}
\,.
\label{eq:F2NLO}
\end{equation}
Defining the convolution in $x$-space, $f \otimes_x g \equiv
\int_0^1\!\frac{\displaystyle {\rm d}\xi}{\displaystyle\xi}
f(\xi)\,g\left(\frac{\displaystyle x}{\displaystyle\xi}\right)$, we see
that the structure function is `factorized' in the form of a convolution,
\begin{equation*}
F_{2,q}(x,Q^2) =
x \sum_i e_{q_i}^2\,
f_i(\mu_{\rm F})\otimes_x
\hat{F}_{2,i}( \mu_\rR,t)
\,,\qquad
{t = \ln\frac{Q^2}{\mu_{\rm F}^2}}
\,.
\end{equation*}
The long distance physics is factored into the PDF's that depend on the
{\em factorization scale} $\mu_{\rm F}$. The short distance physics is
factored into the hard scattering cross section that depends on both
the factorization and the renormalization scales. Both scales are
arbitrary, unphysical scales. The term $z$ defines the {\em factorization
scheme}. It is not unique, finite terms can be shifted between the short
and long distance parts, but it is important that it must be chosen the
same in all computations (the \msbar\ scheme is the standard).

\vspace{-4pt}
\noindent\rule{\textwidth}{1pt}

 
\begin{exe} 
 
The regularization of the splitting functions at $z=1$ is achieved by 
the +-prescription defined by  
\[ 
\int_0^1\!\rd x \frac{f(x)}{(1-x)_+} =  
\int_0^1\!\rd x \frac{f(x) - f(1)}{1-x}  
\] 
for any smooth test function $f(x)$. The contribution of the loop 
corrections has the same kinematics as the LO one, so it has to be 
proportional to $\delta(1-x)$. Thus the complete regularized splitting 
function has the form  
\[ 
P_{qg}(x) = C_{\rm F}\lp[\frac{1+x^2}{(1-x)_+} + K \delta(1-x)\rp] 
\,. 
\] 
We can obtain the parton distribution for quark in quark from \eqn{eq:PDF} 
by the substitution $f_q^{(0)}(x) \to \delta(1-x)$ 
\[ 
f_q(x,\mu_{\rm F}) = 
\delta(1-x) 
+ \frac{\alpha_{\rm s}}{2\pi} 
P_{qg}(x) \ln\frac{\mu_{\rm F}^2}{m_g^2} 
\,. 
\] 
Integration over $x$ gives the number of quarks in a quark that has to be 
one, independently of $\mu_F$. Thus we have the condition 
$\int_0^1\! \rd x P_{qg}(x) = 0$. Compute the regularized splitting function. 
\end{exe} 

\noindent\rule{\textwidth}{1pt}

\subsection{DGLAP equations} 
\label{sec:DGLAP}

The short-distance component of the factorized structure function in
\eqn{eq:F2NLO} can be computed in pQCD. It depends on the renormalization
scale, but recall that it has to satisfy the RGE.

We cannot compute the PDF's in PT, so it seems that this is the end of
the story: pQCD appears non-predictive for processes with hadrons in
the initial state. However, the arguments that lead to the RGE come to
the rescue. While the right hand side of \eqn{eq:F2NLO} depends on both
renormalization and factorization scales, the measurable quantity $F_2$
does not, which can be expressed by RGE. Of course, this statement has
to be understood perturbatively, namely at any order in PT, the right
hand side of the RGE is not exactly zero, but may contain terms that
are higher order in PT. Only infinite order is expected to give exact
independence of the scales. The RGE gives the missing piece of
information needed to make the theory predictive.

To write the RGE, we introduce Mellin transforms defined by $f(N) \equiv
\int_0^1\!{\rm d} x\,x^{N-1} f(x)$, which turns a convolution into a real
product:
\begin{equation*}
\begin{split}
\int_0^1\!{\rm d} x\,x^{N-1}
\int_0^1\!\frac{{\rm d}\xi}{\xi}
f(\xi)\,g\left(\frac{x}{\xi}\right) &=
\int_0^1\!{\rm d} x\,x^{N-1}
\int_0^1\!{\rm d}\xi \int_0^1\!{\rm d} y \delta(x-y\xi) f(\xi)\,g(y)
\\ &=
\int_0^1\!{\rm d}\xi \int_0^1\!{\rm d}y \,(\xi y)^{N-1} f(\xi)\,g(y)
=
f(N) g(N)
\,.
\end{split}
\end{equation*}
So $F_{2,q}(N,Q^2) =
x \sum_i e_{q_i}^2\,
f_i(N,{\color[rgb]{1.000000,0.000000,0.000000}\mu_{\rm F}})
\hat{F}_{2,i}(N,{\color[rgb]{1.000000,0.000000,0.000000} \mu_\rR,t})$ is
independent of $\mu_{\rm F}$, expressed as 
\[
\mu_{\rm F} \frac{{\rm d} F_2}{{\rm d} \mu_{\rm F}} = 0 \bigg
(= {\cal O}\big(\alpha_{\rm s}^{n+1}\big) \mbox{~in PT at~}
{\cal O}\big(\alpha_{\rm s}^{n}\big)\bigg)
\,.
\]

Let us explore the consequences of this RGE. For simplicity, let us assume
one quark flavour,
$F_{2,q}(N,Q^2) =
x e_{q_i}^2\,
f_q(N,{\color[rgb]{1.000000,0.000000,0.000000}\mu_F})
\hat{F}_{2,i}(N, \mu_R,{\color[rgb]{1.000000,0.000000,0.000000}t})$. Then
the RGE reads
\begin{equation*}
\hat{F}_{2,q}(N,t)\frac{{\rm d} f_q}{{\rm d} \mu_F}(N,\mu_F)
+ f_q(N,\mu_F)\frac{{\rm d} \hat{F}_{2,q}}{{\rm d} \mu_F}(N,t) = 0
\,.
\end{equation*}
Dividing with $f_q\,\hat{F}_{2,q}$, it turns into 
\begin{equation}
\mu_F \frac{{\rm d}\ln f_q}{{\rm d} \mu_F}(N,\mu_F)
=-\mu_F \frac{{\rm d} \ln \hat{F}_{2,q}}{{\rm d} \mu_F}(N,t)
\equiv -\gamma_{qg}(N)
\,,
\label{eq:Mellinevolution}
\end{equation}
where $\gamma_{qg}(N)$ is called the anomalous dimension because it
acts as a factor $\mu_{\rm F}^{-\gamma_{qg}(N)}$ in the dimensionless
function $\ln f_q(N,\mu_{\rm F})$.
Taking the Mellin moment of \eqn{eq:PDF} and then its derivative with
respect to $\mu_{\rm F}$, we obtain that the anomalous dimension is
\begin{equation}
\gamma_{qg}(N) = -\mu_F \frac{{\rm d}\ln f_q}{{\rm d} \mu_F}(N,\mu_F)
= -\frac{\alpha_{\rm s}(\mu_R)}{\pi} P_{qg}(N)
+ {\cal O}\big(\alpha_{\rm s}^{2}\big)
\,,
\label{eq:anomalousdim}
\end{equation}
\ie it is the Mellin transform of the splitting function, which can be
computed in PT. Equation (\ref{eq:Mellinevolution}) implies that {\em
the scale dependence of the PDF can be predicted in PT}. This together
with the universality of PDF's makes pQCD predictive: {\em we can
measure the PDF's in one process at a certain scale and then use it in
another process at another scale to make predictions}.

How shall we choose the renormalization and factorization scales? If we
want to avoid large logarithms that spoil the convergence of the
perturbative series, the scales should be chosen near the characteristic
physical scale of the process $Q$, \eg $\mu_\rR^2 = \mu_{\rm F}^2 = Q^2$.
Then the RGE becomes 
\begin{equation}
Q^2 \frac{{\rm d}\ln f_q}{{\rm d} Q^2}(N,Q^2)
= -\frac12 \gamma_{qg}\bigg(N, \alpha_{\rm s}\big(Q^2\big)\bigg)
\,,
\label{eq:DGLAPMellin}
\end{equation}
which is the Mellin transform of 
\begin{equation}
Q^2 \frac{{\rm d} f_q}{{\rm d} Q^2}(x,Q^2)
= \frac{\alpha_{\rm s}\big(Q^2\big)}{2\pi} P_{qg} \otimes_x f_q\big(Q^2\big)
\,.
\label{eq:DGLAP1flavour}
\end{equation}
Our discussion was highly simplified by considering only one quark flavour
and neglecting the mixing of partons. If we make the full computation we
obtain the gold-plated formula
\begin{equation}
Q^2 \frac{{\rm d} f_{(ij)}}{{\rm d} Q^2}(x,Q^2)
= \frac{\alpha_{\rm s}\big(Q^2\big)}{2\pi}\sum_i P_{ij} \otimes_x f_i\big(Q^2\big)
\,,
\label{eq:DGLAP}
\end{equation}
called DGLAP (for Dokshitzer \cite{Dokshitzer:1977sg}, Gribov-Lipatov
\cite{Gribov:1972ri} and Altarelli-Parisi \cite{Altarelli:1977zs}) equation.

Let us now solve the (simplified) DGLAP equation in Mellin space,
\eqn{eq:DGLAPMellin}. It is a simple first order differential equation
whose solution is
\[
f_q(N,Q^2) = f_q(N,Q^2_0) \,
\exp\left[-\int_ {t_0}^{t}
\!\! {\rm d} t\,\gamma_{qg}\bigg(N, \alpha_{\rm s}\big(\Lambda^2 {\rm
e}^t\big)\bigg) \right]
\,.
\]
Let us recall the one-loop formula in \eqn{eq:asymptoticfreedom}, 
$\alpha_{\rm s}\big(Q^{2}\big) = \frac{1}{b_0 t}\,,\quad
t = \ln\frac{Q^2}{\Lambda^2}$ and introduce the abbreviation 
$d_{qg}(N) = -\frac{\gamma_{qg}(N)}{2 \pi b_0} \leq 0$.
Then
\begin{equation}
f_q(N,Q^2) = f_q(N,Q^2_0) \,
\exp\left[d_{qg}(N) \int_ {t_0}^{t}
\!\!\frac{{\rm d} t}{t} \right]
\,,\quad {\rm or}\quad
f_q(N,Q^2) = f_q(N,Q^2_0) \, \left( \frac{t}{t_0} \right)^{d_{qg}(N)}
\,,
\label{eq:scalingviolation}
\end{equation}
called {\em scaling violation}.

\begin{figure}[ht]
\begin{center}
\includegraphics[width=0.43\textwidth]{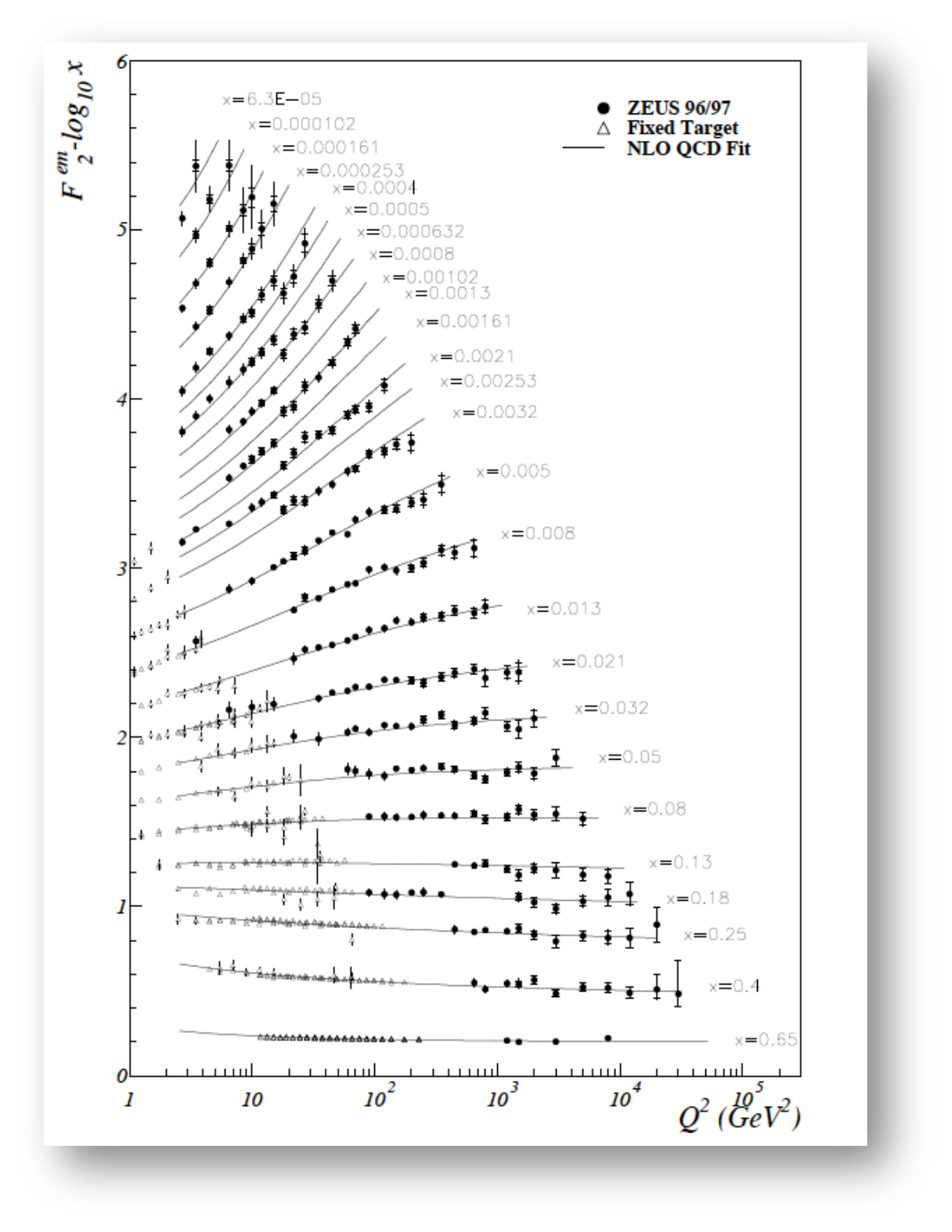}
\includegraphics[width=0.56\textwidth]{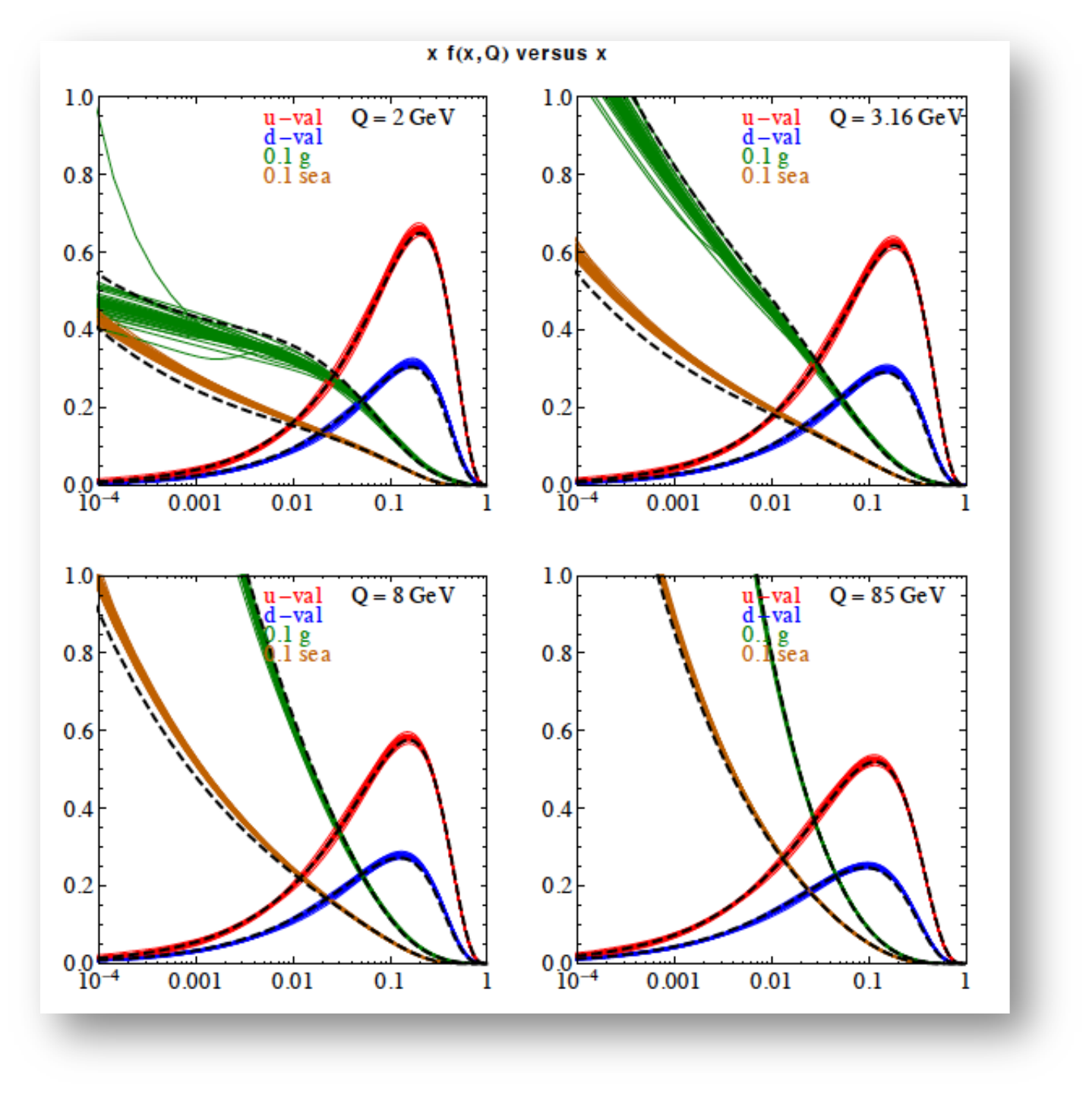}
\caption{(a) Measurement of $F_2$ structure function at different $Q^2$
as a function of $x$,
(b) evolution of valence quark, sea quark and gluon distributions
}
\label{fig:scalingviolation}
\end{center}
\end{figure}

As $\gamma_{qg}(1) = 0$, the valence q-quark in the proton, given by the
integral $\int_0^1\!{\rm d} x\, f_q(x,Q^2)$, is independent of $Q^2$.
Higher moments vanish more rapidly, therefore, the average $x$ decreases
as $Q^2$ increases. Thus we predict that $f_q(x,Q^2)$ increases at small
$x$ and decreases at large $x$. This prediction is seen to be valid from
the measurements shown in \fig{fig:scalingviolation}(a).

\vspace{-4pt}
\noindent\rule{\textwidth}{1pt}


\begin{exe}

Compute the anomalous dimension $\gamma_{qg}(x)$ using
\eqn{eq:anomalousdim}.

\end{exe}
\rule{\textwidth}{1pt}

\section{Hadron collisions} 
\label{sec:ppcollisions}

While electron-positron annihilation and DIS played very important role
in establishing pQCD for understanding high-energy scattering
experiments, presently and in the mid-term future the experiments at the
energy frontier can be found at the Large Hadron Collider (LHC). Thus we
are most interested in the theoretical tools needed to understand
high-energy proton-proton collisions.

\subsection{Factorization theorem} 
\label{sec:factorization}

Fortunately, the tools we have developed so far can be generalized
straightforwardly to hadron collisions. The most general form of the
factorization theorem includes convolution with two PDF's, one for each
colliding parton, the hard scattering cross section, and possibly a
convolution with a fragmentation function (FF) of a parton into an
identified hadron in the final state.  Thus, the differential cross
section for a hypothetical process $pp\to Z+\pi+X$ has the form
\begin{equation}
\begin{split}
{\rm d}\sigma_{pp\to Z+\pi+X}(s,x,\alpha_{\rm s},\mu_\rR,\mu_{\rm F}) =
\sum_{i,j,k}
\int_0^1\!{\rm d}x_1 f_{i/p}(x_1,\alpha_{\rm s}, \mu_{\rm F})
& \int_0^1\!{\rm d}x_2 f_{j/p}(x_2,\alpha_{\rm s}, \mu_{\rm F})
\\ \qquad\times
\int_x^1\!\frac{{\rm d}z}{z}
{\rm d}\hat{\sigma}_{ij\to Z+k+X}
(\hat{s},z,\alpha_{\rm s}(\mu_\rR),\mu_\rR,\mu_{\rm F})
D_{\pi/k}\left(\frac{x}{z},\hat{s}\right)
& + {\cal O}\left(\frac{\Lambda}{Q}\right)^p
\,.
\end{split}
\label{eq:ppfactorization}
\end{equation}
In \eqn{eq:ppfactorization} $s$ is the total centre-of-mass energy
squared, $x/z$ is the longitudinal momentum fraction of the pion in the
parton $k$, $\mu_{\rm R}$ and $\mu_{\rm F}$ are the
renormalization and factorization scales, $f_{i/p}(x)$ is the PDF for
parton $i$ in the proton with momentum fraction $x$, 
${\rm d}\hat{\sigma}_{ij\to Z+k+X}(\hat{s})$ is the hard scattering cross
section for the partonic process, $D_{\pi/k}(x)$ is the FF for the
process parton $k \to \pi$. The last term shows that contributions
suppressed at high $Q^2$ are neglected ($p>1$). Substituting the PDF's
and FF's with $\delta$ functions (in momentum and flavour) we obtain the
cross section formulae in DIS and electron-positron annihilation.

The PDF's and FF's constitute the non-perturbative, long-distance
components of the cross section that cannot be computed in pQCD, only
extracted from measurements. Thus, it is a natural question whether or
not the factorization theorem is predictive. The answer is a clear yes
for the following reasons.

We can compute the hard scattering cross section in PT, which involves
(i) renormalization of UV divergences (order by order in PT),
(ii) cancellation of IR ones for IR safe observables using a subtraction
method,
(iii) absorbing initial state collinear divergences into renormalization
of PDF's (and possibly uncancelled final state ones into that of the FF).
The non-perturbative components are universal, so can be measured in
one process and used to make prediction in another one. Furthermore, the
evolution of these with $Q^2$ can be predicted in PT (DGLAP equations),
shown in \fig{fig:scalingviolation}(b).

In summary, we are prepared to make predictions for any high-energy
scattering process. The theoretical framework for such predictions relies
on pQCD and the factorization theorem. In PT we can compute the hard
scattering cross section and the evolution of the PDF's. There are
universal elements, such as the PDF's and FF's, as well as the
subtraction method for computing radiative corrections.

\subsection{Are we happy?} 

At this point theorists can make precision predictions for distributions
of IR safe observables. The main bottleneck to make such predictions is
the algebraic complexity of computing amplitudes and the analytic
complexity of evaluating loop integrals. The state of the art considers
the computation of NLO corrections a solved problem with automated
implementations for processes up to about five partons in the final state
(at tree level). The exact number depends on the process being considered
because the numerical integrations become too expensive eventually.
Nevertheless, all processes listed in the `Les Houches wishlist (2011)'
are known by now. Furthermore, there is also a computer code to compute
seven-jet production in electron-positron annihilation \cite{Becker:2011vg}.

For experimenters the situation is less satisfactory. While pQCD
predictions are based on a solid theoretical ground, those lack important
features. On the one hand pQCD gives predictions for final states with
few partons, detectors detect hadrons. A tool that can simulate real
events with hadrons at correct rates would be much more handy. To finish
these lectures we look into modeling events in a qualitative way. A more
detailed description can be found in Ref.~\cite{Skands:2012ts}.

\subsection{Modelling events} 
\label{sec:events}

\begin{wrapfigure}{r}{0.45\linewidth}
\vspace*{-20pt}
\includegraphics[width=1.0\linewidth]{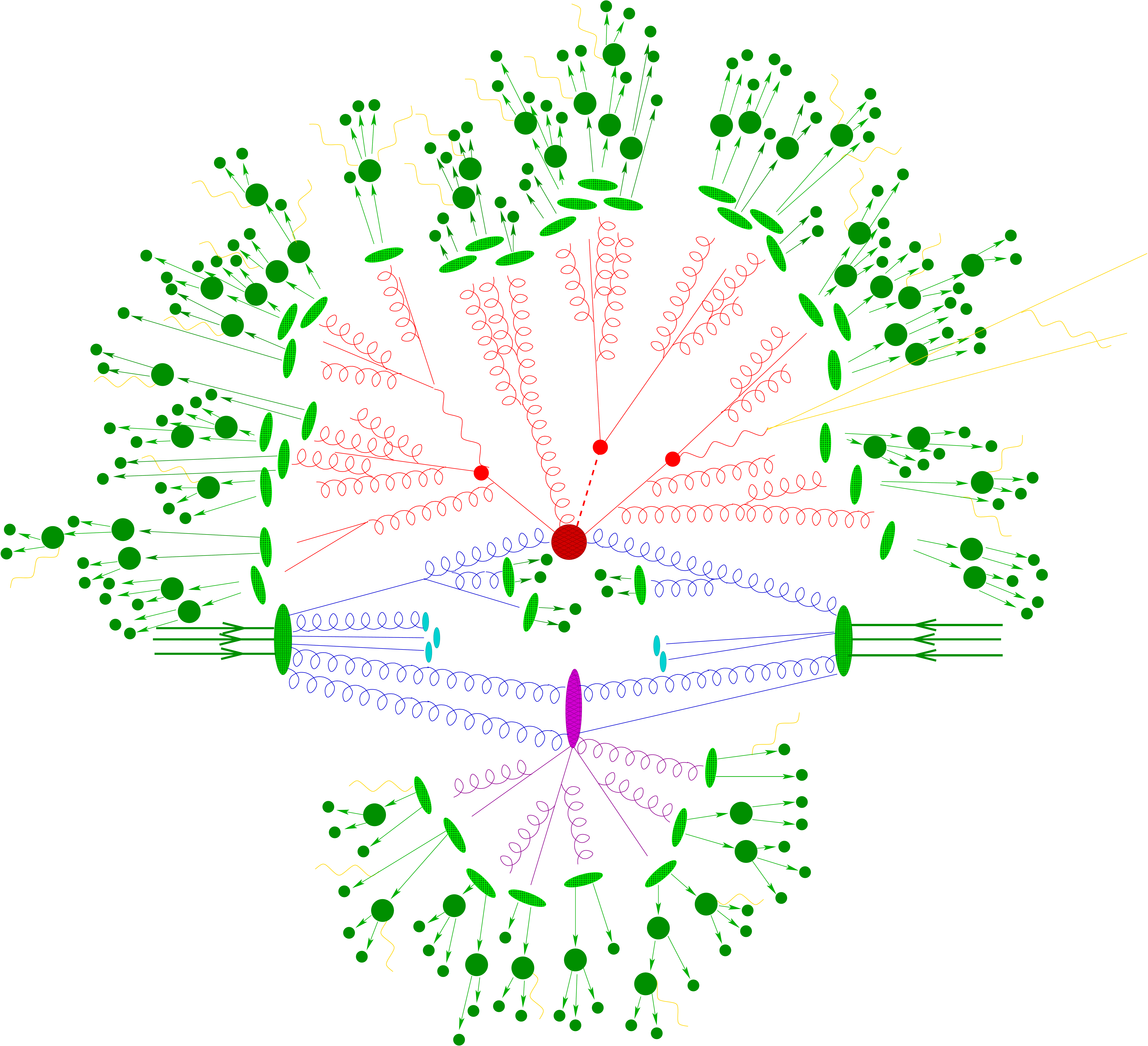}
\vspace*{-15pt}
\caption{Artistic view of a proton-proton scattering event
at high energy (curtesy of F.~Krauss)
~}
\label{fig:event}
\end{wrapfigure}
Figure \ref{fig:event} shows our view of a proton-proton scattering event
at high energy. The three parallel lines ending in discs from both sides
represent the two incoming protons. At high energies these protons
consist of (almost) free-flying partons, two of which (one from each)
collide at high centre-of-mass energy and produce the hard scattering,
with perturbatively computable cross section. This is where signs of new
physics may appear. The hard scattering cross section is process
dependent. We have discussed how it can be computed from first
principles, which can be improved systematically by computing the
radiative corrections.  

Before collision the colliding partons may emit other partons collinear
with the beam. These collinear emissions in the initial state give rise
to divergences that can be factored into the renormalized parton
distribution functions. After collision few energetic partons appear
that may emit less energetic partons and each develops showers of
partons. Emissions into almost the same direction as the original
parton occur with enhanced probability (due to the collinear
divergence) as well as emissions of soft gluon. This is represented in
Figure \ref{fig:event} by red quark and gluon lines.  Both
factorization and parton showering can be described from first
principles based upon known physics of QCD, and are universal, \ie
independent of the process and observable. We have seen how
factorization works, but have not discussed how parton showers are
modelled with shower Monte Carlo (SMC) programs
\cite{Sjostrand:2006za,Corcella:2000bw}. We mentioned
marginally how the large logarithms emerging in the final state
splittings can be resummed, which gives improved prediction for the
cross section (as seen in \fig{fig:3jetnnlo}), but does not simulate
events.  

Parton showers still only give a description of events in terms of
quarks and gluons, whereas detectors detect only hadrons. We do not
know how to compute hadronization, the transition from quarks and gluon
to hadrons, from first principles. Yet the idea of local parton-hadron
duality (LPHD) provides some sort of theoretical understanding (see, \eg
Ref.~\cite{BasicspQCD}). It states that
{\em after accounting for all gluon and quark production down to scales
$\simeq \Lambda_{\rm QCD}$, the transition from partons to hadrons is
essentially local in phase space.}
Thus the hadron directions and momenta will be closely related to
that of the partons, and the hadron multiplicity will reflect the parton
multiplicity, too. This is illustrated by the green lines with dots.

In addition to the energetic partons in the initial state, there are also
low-energy ones that may collide, which is energy and process dependent.
This low-energy physics is described in models of underlying event,
which are also part of modern SMC's. 
The underlying event produces low-energy partons. Also at the end of
the shower low-energy partons emerge. As QCD confines partons,
these partons turn into hadrons before detection, a process called
hadronization. We do not have a theory of hadronization based on first
principles. Instead, SMC's include models that describe hadronization
in a process independent way. These models contain parameters that are
fixed experimentally.  

\section{Conclusions}

In these lectures we discussed the theoretical basis of interpreting the
results of high-energy collider experiments. We discussed how pQCD can be
made predictive and also the main uncertainties in the predictions.
We used the following key ingredients in this tour:
(i) gauge invariance that allows us to write down the
Lagrangian and which predicts many important features of the
theory;
(ii) renormalization that cancels ultraviolet divergences systematically
order-by-order in perturbation theory and introduces a dimensionful scale into
even the scaleless Lagrangian of massless QCD, leading to scaling
violations of one-scale observables that would be scale independent in
the classical theory;
(iii) asymptotic freedom at high energies emerging from the quantum
structure of the theory and the non-Abelian nature of the gauge group;
(iv) need for infrared safety, emerging from asymptotic freedom, to ensure
that the IR divergences, associated with unresolved parton emission,
cancel between real and virtual contributions, allowing the
perturbative calculation of jet cross sections, without a detailed
understanding of the mechanism by which partons become jets;
(v) factorization that makes possible to use perturbative QCD to
calculate the interactions of hadrons, since all the non-perturbative
physics gets factorized, into parton distribution functions;
(vi) evolution and universality of PDF's that allows us to extract those
measuring cross sections in one process, like DIS, and then used to
predict the cross sections for any other process. Again, this
factorization introduces a scale dependence into the parton model so
that the structure functions of DIS, and other one-scale observables
become scale dependent.
These features make pQCD predictive, without forcing us to solve the
theory at all possible scales: unknown or uncalculable high- and
low-energy effects can be renormalized, factorized and cancelled away.

Of course, in four double lectures, it was impossible to give
full treatment of any of the topics we encountered. For that I refer
to any of the classic textbooks about QCD at colliders 
\cite{EllisStrilingWebber,DissertoriKnowlesSchmelling,Brock:1993sz}.

\section*{Acknowledgements}
My lectures grew partly out of research conducted during the past 20
years and supported in part by Hungarian Scientific Research Fund
grants, most recently under grant OTKA K-101482, and partly out of
lectures given at Universities of Debrecen and Z\"urich where I
received useful feedback from students. I benefited greatly from the
lectures of previous schools by G.~Salam \cite{Salam:2011bj} and F.~Maltoni 
(\underline{http://physicschool.web.cern.ch/PhysicSchool/ESHEP/ESHEP2011/programme.html}).
I am grateful to the organizers, discussion leaders and students of the
2013 European School of High-Energy Physics for providing a pleasant and
stimulating atmosphere.


\providecommand{\href}[2]{#2}\begingroup\raggedright
\endgroup

\end{document}